\documentclass[aps,prb,twocolumn,superscriptaddress,notitlepage]{revtex4-2}
\usepackage{graphicx}
\usepackage[caption=false]{subfig}
\usepackage{float}
\usepackage{natbib}
\usepackage{xcolor}
\usepackage{xr}
\usepackage{amsmath}
\usepackage{amssymb}
\usepackage{array}
\usepackage{wasysym}
\usepackage{color,soul}
\usepackage{braket}
\usepackage{verbatim}
\usepackage{hyperref}
\usepackage{gensymb}
\usepackage{url}

\usepackage[overload]{textcase} 
\newcommand\Alpha{\mathrm{A}}

\newcommand{\aKTH}{$\alpha$-KTi(C$_2$O$_4$)$_2\cdot$\textit{x}H$_2$O}

\usepackage{filecontents}
\begin{document}

\title{Realising square and diamond lattice $S=1/2$ Heisenberg antiferromagnet models in the $\alpha$ and $\beta$ phases of the coordination framework, KTi(C$_2$O$_4$)$_2\cdot$\textit{x}H$_2$O}

\author{Aly~H.~Abdeldaim}
\affiliation{Department of Chemistry and Materials Innovation Factory, University of Liverpool, 51 Oxford Street, Liverpool, L7 3NY, UK}
\affiliation{School of Chemistry, University of Birmingham, Edgbaston, Birmingham B15 2TT, UK}
\affiliation{ISIS Neutron and Muon Source, Science and Technology Facilities Council, Didcot OX11 0QX, UK}
\author{Teng~Li}
\affiliation{School of Chemistry, University of St. Andrews, St. Andrews KY16 9ST, UK}
\author{Lewis~Farrar}
\affiliation{Department of Chemistry and Materials Innovation Factory, University of Liverpool, 51 Oxford Street, Liverpool, L7 3NY, UK}
\author{Alexander~A.~Tsirlin}
\affiliation{Theoretical Physics and Applied Mathematics Department, Ural Federal University, 620002 Yekaterinburg, Russia}
\affiliation{Experimental Physics VI, Center for Electronic Correlations and Magnetism, Institute of Physics, University of Augsburg, 86135 Augsburg, Germany}
\author{Wenjiao~Yao}
\affiliation{School of Chemistry, University of St. Andrews, St. Andrews KY16 9ST, UK}
\affiliation{Functional Thin Films Research Center, Shenzhen Institutes of Advanced Technology, Chinese Academy of Sciences, Shenzhen 518055, China}
\author{Alexandra~S.~Gibbs}
\affiliation{ISIS Neutron and Muon Source, Science and Technology Facilities Council, Didcot OX11 0QX, UK}
\author{Pascal~Manuel}
\affiliation{ISIS Neutron and Muon Source, Science and Technology Facilities Council, Didcot OX11 0QX, UK}
\author{Philip~Lightfoot}
\affiliation{School of Chemistry, University of St. Andrews, St. Andrews KY16 9ST, UK}
\author{G\o{}ran~J.~Nilsen}
\affiliation{ISIS Neutron and Muon Source, Science and Technology Facilities Council, Didcot OX11 0QX, UK}
\author{Lucy~Clark}
\email[Email address:~]{l.m.clark@bham.ac.uk}
\affiliation{Department of Chemistry and Materials Innovation Factory, University of Liverpool, 51 Oxford Street, Liverpool, L7 3NY, UK}
\affiliation{School of Chemistry, University of Birmingham, Edgbaston, Birmingham B15 2TT, UK}

\begin{abstract}
\noindent We report the crystal structures and magnetic properties of two psuedo-polymorphs of the $S=1/2$~Ti\textsuperscript{3+} coordination framework, KTi(C$_2$O$_4$)$_2\cdot$\textit{x}H$_2$O. Single-crystal X-ray and powder neutron diffraction measurements on $\alpha$-KTi(C$_2$O$_4$)$_2\cdot$\textit{x}H$_2$O confirm its structure in the tetragonal $I4/mcm$ space group with a square planar arrangement of Ti\textsuperscript{3+} ions. Magnetometry and specific heat measurements reveal weak antiferromagnetic interactions, with $J_1\approx7$ K and $J_2/J_1=0.11$ indicating a slight frustration of nearest- and next-nearest-neighbor interactions. Below $1.8$ K, $\alpha$-KTi(C$_2$O$_4$)$_2\cdot$\textit{x}H$_2$O undergoes a transition to G-type antiferromagnetic order with magnetic moments aligned along the $c$ axis of the tetragonal structure. The estimated ordered moment of Ti\textsuperscript{3+} in \aKTH~is suppressed from its spin-only value to $0.62(3)~\mu_B$, thus verifying the two-dimensional nature of the magnetic interactions within the system. $\beta$-KTi(C$_2$O$_4$)$_2\cdot$2H$_2$O, on the other hand, realises a three-dimensional diamond-like magnetic network of  Ti\textsuperscript{3+} moments within a hexagonal $P6_222$ structure. An antiferromagnetic exchange coupling of $J\approx54$ K---an order of magnitude larger than in \aKTH---is extracted from magnetometry and specific heat data. $\beta$-KTi(C$_2$O$_4$)$_2\cdot$2H$_2$O undergoes N\'eel ordering at $T_N=28$ K, with the magnetic moments aligned within the $ab$ plane and a slightly reduced ordered moment of $0.79~\mu_B$ per Ti$^{3+}$. Through density-functional theory calculations, we address the origin of the large difference in the exchange parameters between the $\alpha$ and $\beta$ psuedo-polymorphs. Given their observed magnetic behaviors, we propose $\alpha$-KTi(C$_2$O$_4$)$_2\cdot$\textit{x}H$_2$O and $\beta$-KTi(C$_2$O$_4$)$_2\cdot$2H$_2$O as close to ideal model $S=1/2$ Heisenberg square and diamond lattice antiferromagnets, respectively.
\end{abstract}

\maketitle

\section{INTRODUCTION}
\noindent The discovery of new magnetic materials allows for the realization of theoretical ground state predictions as well as the identification of novel emergent phenomena. Both rely on a fine balance between several material ``ingredients'' that determine the magnetic properties, including exchange frustration\cite{Harrison2004}, low-dimensionality\cite{Vasiliev2018}, and spin-orbit coupling\cite{Rau2016}. Systems based on the $S=1/2$ Heisenberg frustrated square lattice (FSL) model, for example, are extensively studied as they provide a rich magnetic phase diagram depending on the degree of frustration between exchange interactions along the sides, $J_1$, and across the diagonal, $J_2$, of the square net. For the antiferromagnetic phase diagram\cite{Shannon2004}, theoretical predictions for the development of N\'eel and columnar antiferromagnetic orders within the respective dominant $J_1$ or $J_2$ regimes have been experimentally established in several materials\cite{Bombardi2004, Nath2008,Tsirlin2008,Tsirlin2011,  Tsirlin2013,Yang2017,Ishikawa2017, Mustonen2018a,Mustonen2019}. Intriguingly, at the border of these two regimes, materials in which the degree of frustration is maximized---\textit{i.e.} $0.4<J_2/J_1<0.6$---are predicted to map a region of the phase diagram within which a quantum spin liquid (QSL) ground state is realized \cite{ Chandra1988, Mezzacapo2012, Poilblanc2019}. Indeed, in a recent theoretical and experimental investigation of the solid-solution, Sr\textsubscript{2}Cu(Te\textsubscript{1-x}W\textsubscript{x})O\textsubscript{6} where $x=0$ and $x=1$ represent N\'eel and columnar ordered systems, respectively, QSL signatures have been observed for the $x=0.5$ compound\cite{Mustonen2018}. Interestingly, the behavior exhibited by complex materials derived from the FSL, such as BaCdVO(PO$_4$)$_2$\cite{Nath2008}, where signatures of a spin-nematic ground state have been observed in an applied field\cite{Skoulatos2019,Povarov2019}, has further enriched the phase diagram of FSL materials, leading to new theoretical predictions\cite{Smerald}. By extending the $J_1-J_2$ model to incorporate the effects of interplanar coupling, $J_3$, magnetic ground states inaccessible to the pure FSL model can also be realized\cite{Danu2016}.

While often sought within manifestations of low-dimensional models, the three-dimensional diamond lattice offers an alternative route towards unconventional magnetic ground states. Recent investigations of \textit{A}-site spinels have revived interest in the Heisenberg frustrated diamond lattice (FDL) model following the experimental observation of spin-spiral structures\cite{Ge2017} and spin-liquid regimes\cite{MacDougall2011, Gao2017}. As such structures indicate exchange competition, this behavior can be ascribed to the presence of a frustrated further-neighbor coupling, $J_2$, beyond the nearest-neighbor interaction, $J_1$. Beyond the theoretically predicted critical point at $J_2/J_1>0.125$, a set of  degenerate spin-spiral states were found to describe a novel spin-liquid regime\cite{Bergman2007}. Signatures of this dynamic behavior have been observed in material realizations of the $J_1$-$J_2$ FDL model, such as in MnSc$_2$S$_4$\cite{Gao2017} and CoAl$_2$O$_4$\cite{Tristan2005}, where theory combined with experimental diffuse neutron scattering studies conclude the presence of a continuous spin-spiral surface in momentum space\cite{Gao2017, MacDougall2011}. Perturbations further enrich the magnetism of such FDL systems, and spin-orbit coupling \cite{Chen2009,Chen2009a, Tsurkan2017}, further-neighbor couplings, and structural distortions\cite{Ge2017} can play a significant role in determining the magnetic ground state.

In this vein, alternative chemical realizations of highly sought magnetic models can be achieved through the synthesis of coordination framework materials. When compared to their inorganic counterparts, the versatility of possible organic linkers in coordination frameworks offers a wider command over the dimensionality and magnetic properties of the magnetic sublattice\cite{Tustain2019,Rao2008}. Indeed, a range of architectures, ranging from the star lattice\cite{Zheng2007}, to square\cite{Tsyrulin2010} and diamond networks\cite{English1993} have been prepared in coordination frameworks.

Motivated by the richness of intriguing behavior exhibited within the FDL and FSL models, we here explore a Ti\textsuperscript{3+}-based coordination framework realization of a diamond lattice, $\beta$-KTi(C$_2$O$_4$)$_2$.2H$_2$O~($\beta$), in addition to identifying its previously predicted \cite{Drew1977} square planar network pseudo-polymorph, \aKTH~($\alpha$). We report the crystal structures, thermodynamic properties, and magnetic structures of both pseudo-polymorphs using a combination of single-crystal X-ray diffraction, powder neutron diffraction, magnetic susceptibility, specific heat measurements, and \textit{ab initio} calculations. Beginning with Section.~\ref{sec:sec2}, we present our synthetic route for micro- and polycrystalline samples and summarize our experimental methods. In Sec.~\ref{sec:sec3}, we identify the crystal structure of $\alpha$ and provide evidence for the quasi-two-dimensional behavior of this non-stoichiometric hydrate. This is followed by a discussion of our structural and magnetic investigation of $\beta$ in Sec.~\ref{sec:sec4}. A discussion of the electronic structure calculations, single-ion properties, and exchange pathways of both compounds ensues in Sec.~\ref{sec:discuss}.  Finally, we conclude in Sec.~\ref{sec:sec6} and provide an outlook for future experiments.
\begin{figure}
\label{fig:hrpda}
\begin{center}
{\resizebox{0.99\columnwidth}{!}{\includegraphics{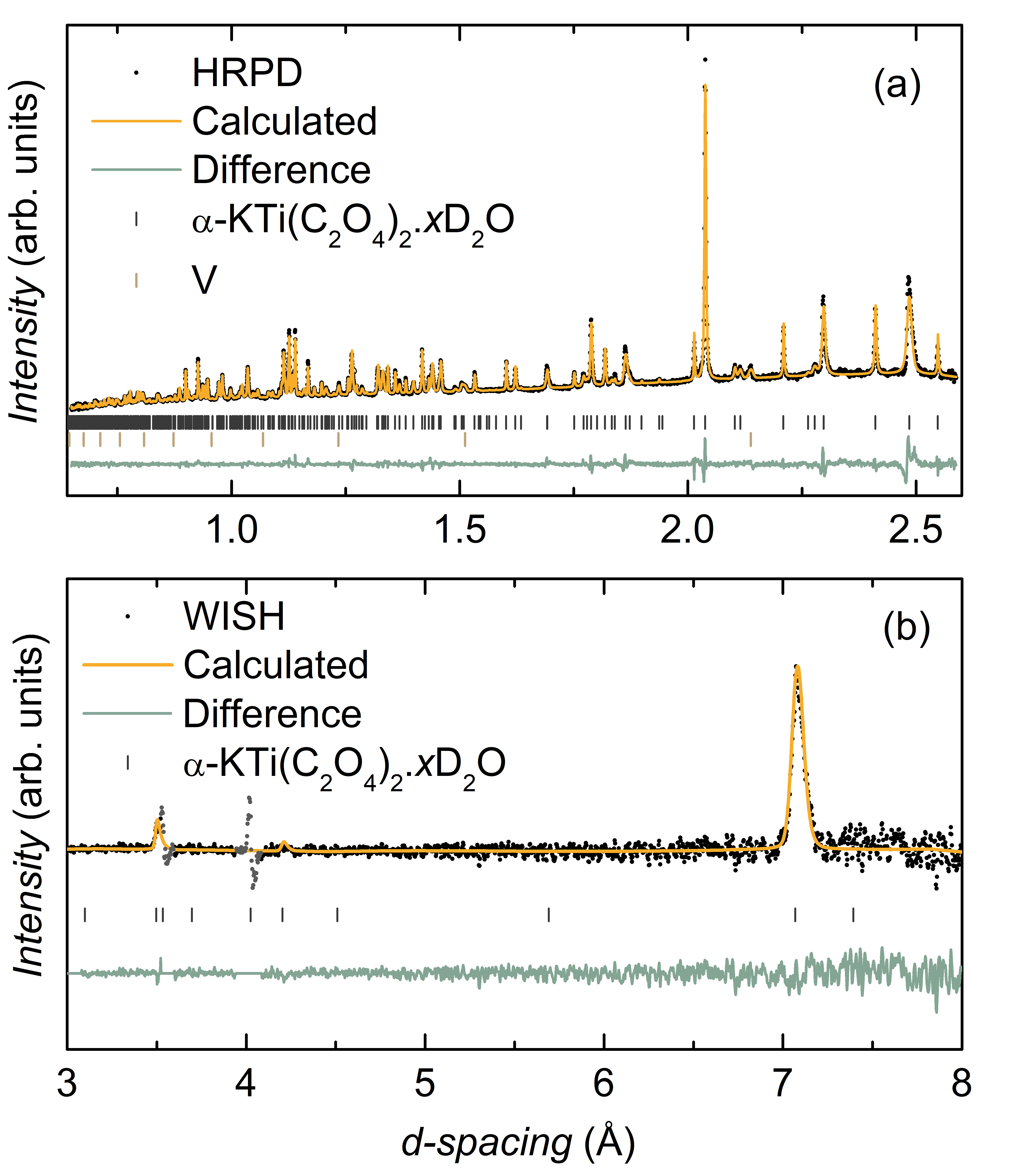}}}
\end{center}
\caption{(a) Rietveld refinement of the $I4/mcm$ model ($\chi^2=4.80$, $R_p=2.83\%$) describing the structure of deuterated $\alpha$-KTi(C$_2$O$_4$)$_2\cdot$1.48(4)D$_2$O using data collected on the HRPD instrument at $1.8$ K. (b) Rietveld plot ($\chi^2=1.81$ and $R_{mag}=3.16\%$) of the $I4/m^{\prime}cm$ magnetic structure to magnetic only scattering obtained by subtracting data collected on the WISH instrument at $15$ K from $1.2$ K data. In both plots, data points are shown in black, fitted curves in orange, difference curves in blue, Bragg reflection positions in grey and brown, and excluded data points in grey.}\label{fig:fig3}
\end{figure}
\section{EXPERIMENTAL METHODS}\label{sec:sec2}
\noindent Polycrystalline samples of $\alpha$ and $\beta$ were synthesized hydrothermally according to modified versions of previously published methods \cite{Stahler1905,Drew1977a}. In a typical synthesis of either sample, an aqueous solution of K$_2$CO$_3$ (Sigma Aldrich $99.99$\%)  and H$_2$C$_2$O$_4\cdot$2H$_2$O (Sigma Aldrich $>99.5$\%) was prepared in a round bottom flask and heated to $343$~K ($\alpha$) or $363$~K ($\beta$) under constant stirring. After $1$ hour of sparging with N$_2$ gas, TiCl$_3$ (Sigma Aldrich $\geq 12$\% TiCl$_3$ basis) was introduced to the solution, and either an orange or a maroon colored precipitate immediately formed for {$\alpha$} and {$\beta$}, respectively. The reaction proceeded for $30$ minutes, after which the round bottom flask was quenched in an ice bath. The resulting product was collected by filtration and washed with acetone. The molar ratios of K$_2$CO$_3$, H$_2$C$_2$O$_4\cdot2$H$_2$O, TiCl$_3$, and H$_2$O used were $1:3:5:277$ and $2:6:1:1000$ for $\alpha$ and $\beta$, respectively. It should be noted that the $\beta$ phase appears to be more thermodynamically stable as if not immediately collected from solution, samples of $\alpha$ slowly recrystallize into $\beta$. Deuterated samples of $\alpha$-KTi(C$_2$O$_4$)$_2\cdot$\textit{x}D$_2$O were synthesized using D$_2$O. Deuteration of $\beta$ was not achieved. While products are air-stable for short periods ($\approx 1$ day), handling in an inert environment is necessary to avoid H-D exchange. 

Single-crystal X-ray diffraction (SC-XRD) measurements were performed at $150$ K using a Bruker D8 VENTURE diffractometer equipped with a PHOTON-II detector and a Mo $K\alpha$ source with $\lambda = 0.71073$~\AA~ using $\omega$ scans.  The reduction and integration of the collected data were performed using the \texttt{APEX III} software package. Structure solutions were obtained using direct methods by utilizing \texttt{SHELXT-2013} \cite{Sheldrick2015a} followed by refinements using \texttt{SHELXL-2013}\cite{Sheldrick2015} on the \texttt{OLEX2} software with multi-scan absorption correction. Non-hydrogen atoms were modeled with anisotropic displacement parameters.

Powder neutron diffraction (PND) data\cite{HRPD_ex} were collected on the time-of-flight High-Resolution Powder Diffractometer (HRPD)\cite{HRPD} at the ISIS Neutron and Muon Source. Vanadium-windowed flat plate sample containers were used, and measurements were carried out on $2$~g samples of each compound at $1.8$~K and $300$~K. All three fixed angle detector banks (centered around $168$\degree, $90$\degree, and $30$\degree) were used for the nuclear structural refinement. To probe the magnetic structures of $\alpha$ and $\beta$, PND data\cite{WISH_ex} were collected for the same $2$~g samples using the long wavelength WISH diffractometer \cite{WISH} at the ISIS Neutron and Muon Source. Cylindrical vanadium containers were used, and measurements were carried out at 1.2~K, 2~K, 3~K, 4~K, 6~K, and 15~K for $\alpha$ and at 1.2~K and 35~K for $\beta$. Rietveld refinements of the nuclear and magnetic structures were conducted on the \texttt{GSAS}\cite{GSAS,GSAS2} and \texttt{FULLPROF}\cite{fullprof} software packages, respectively. Instrumental impurity phases (\textit{i.e.} steel and vanadium) were modeled using the Le Bail method.   

Temperature-dependent DC magnetic susceptibility data were measured on $13.21$~mg ({$\alpha$}) and $38.09$~mg ({$\beta$}) samples using both field cooled and zero-field cooled protocols in a $0.1$~T applied field between $1.8$~K and $300$~K using a Quantum Design MPMS 3 SQUID magnetometer. Specific heat data were recorded in a temperature range of $1.8$~K to $300$~K in zero-field on a Quantum Design PPMS DynaCool using $9.11$~mg and $6.71$~mg pressed powder samples of $\alpha$ and $\beta$, respectively. 

Quantum Monte Carlo (QMC) simulations were performed using the directed loop (\texttt{loop})\cite{Todo2001,Evertz2003} algorithm of the \texttt{ALPS}\cite{ALPS,ALPS2} simulation package. The temperature dependence of the magnetic susceptibility and specific heat were calculated for a $60\times60$ square lattice model ($2\times60^2$ spins) with $10^5$ thermalization and sweep steps. The high-temperature-series expansion (HTE) of the $S=1/2$ FSL and FDL models were calculated using the [4,6] Pad\'e approximants derived from the \texttt{HTE10}\cite{Lohmann} code.   

Fully relativistic density-functional theory (DFT) band-structure calculations were performed in the \texttt{FPLO} code\cite{Koepernik1999} using Perdew-Wang approximation for the exchange-correlation potential\cite{Perdew1992}. States near the Fermi level were used to construct Wannier functions, analyze their composition, and calculate hopping parameters $t_i$ that determine the antiferromagnetic superexchange $J_i=4t_i^2/U$, where $U$ is the on-site Coulomb repulsion parameter. A well-converged $\mathbf{k}$-mesh with $242$ irreducible $\mathbf{k}$-points for the $\alpha$ and $168$ $\mathbf{k}$-points for the $\beta$ were used. The experimentally determined crystallographic model was employed for $\beta$. In contrast, calculations for $\alpha$ required the construction of ordered models, owing to the partially disordered arrangement of K$^+$ and H$_2$O between the [Ti(C\textsubscript{2}O\textsubscript{4})\textsubscript{2}]$^-$ layers  (see Fig. S1(a) of the Supplemental Material \footnote{See Supplemental Material at http://link.aps.org/supplemental/
10.1103/PhysRevMaterials.xx.xxxxxx for further details on
DFT calculations and outputs, modeling of magnetic susceptibility
and heat capacity data, and fitting results of all symmetry
allowed magnetic models to powder neutron diffraction data.}). Three models were considered: (i) K$^+$ ions on the four-fold rotation axis and two water molecules per formula unit ($I4/m$)  [Fig. S1(b)], (ii) K$^+$ ions on the four-fold rotation axis, no water molecules ($I4/mcm$)  [Fig. S1(c)], and (iii) K$^+$ ions occupying half of their experimental positions, with no water molecules ($Ima2$) [Fig. S1(d)]. All three models produced very similar results for the bands around the Fermi level and led to essentially indistinguishable Ti$^{3+}$ Wannier functions, suggesting that the positions of K$^+$ ions and water molecules have no significant influence on the magnetism of the system.

\begin{figure}[h]
\begin{center}
{\resizebox{0.99\columnwidth}{!}{\includegraphics{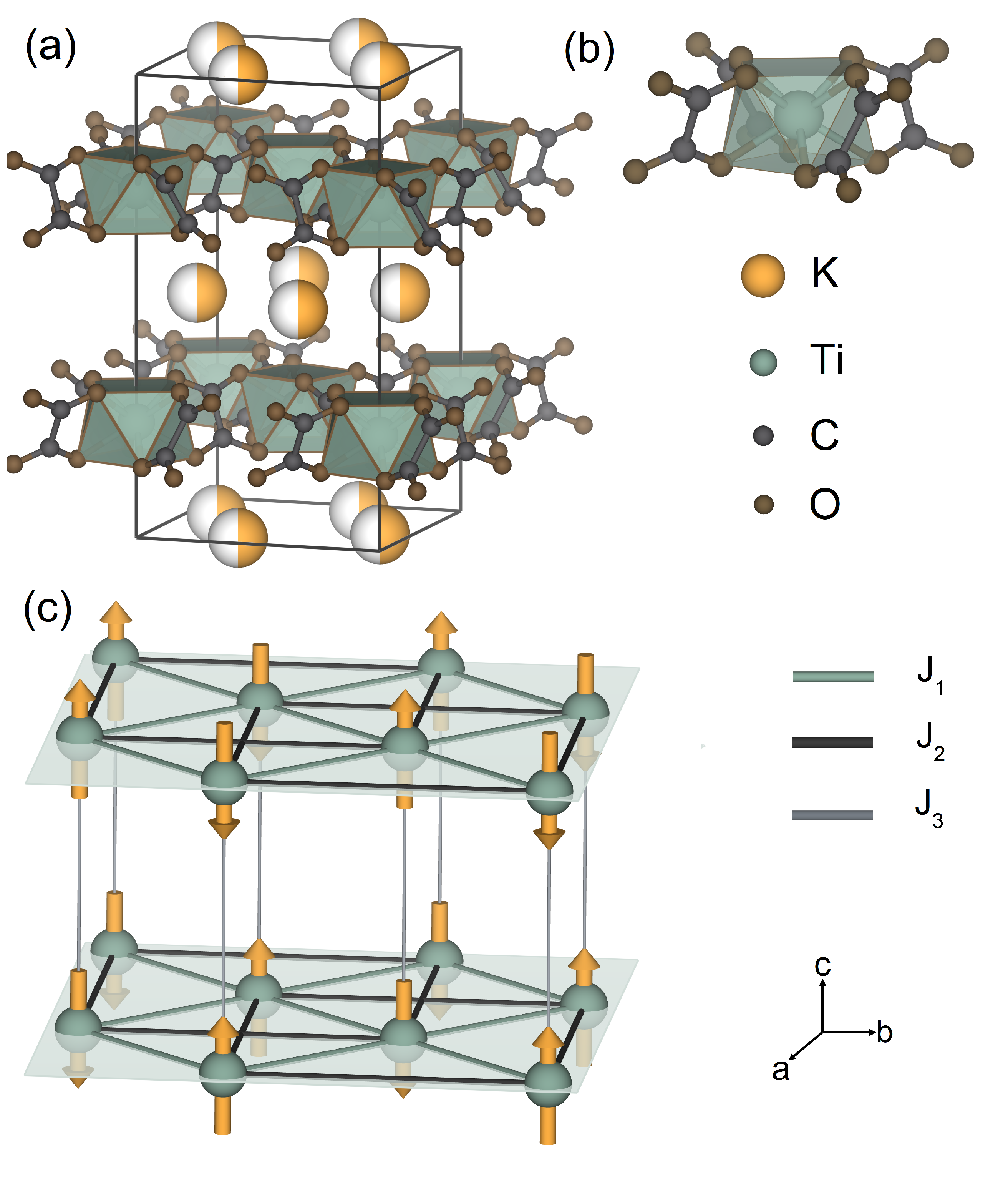}}}
\end{center}
\caption{(a) The crystal structure of $\alpha$-KTi(C$_2$O$_4$)$_2\cdot$\textit{x}H$_2$O as viewed along the $[111]$ direction where the (b) oxalate-bridged square antiprismatic Ti\textsuperscript{3+} ions form square planar sheets within the \textit{ab} plane separated by a disordered layer of orange K\textsuperscript{+} ions and water molecules which are omitted for clarity. (c) G-type antiferromagnetic ordering of the magnetic moments obtained from the analysis of PND data at $1.2$ K, where the orange Ti\textsuperscript{3+} magnetic moments lie along the crystallographic \textit{c} axis with nearest-neighbor ($J_1$), next-nearest-neighbour ($J_2$), and interplanar ($J_3$) exchange interactions highlighted in black, blue, and gray, respectively. Both structures were generated using the \texttt{VESTA}\cite{Vesta} visualization software.}\label{fig:fig1}
\end{figure}
\section{Results for \NoCaseChange{$\alpha$-KTi(C$_2$O$_4$)$_2\cdot$\textit{x}H$_2$O}}
\label{sec:sec3}
\subsection{Crystal Structure}

\noindent Analysis of SC-XRD data collected at $150$ K reveals that diffraction patterns of $\alpha$ can be indexed by the tetragonal space group $I4/mcm$. The refined parameters within this model are presented in Table.~\ref{SC_XRD}. Through a multibank Rietveld analysis of PND data collected for a deuterated sample of $\alpha$ on HRPD, the $I4/mcm$ model was verified at both $1.8$~K and $300$~K. A single deuterium site was located on a Fourier difference map, at a location similar to that obtained from SC-XRD, and refined according to an O-D bond distance restraint (0.97(3)~\AA). The refinement was initially carried out with the occupancy of the water site fixed to $0.5$, as obtained from SC-XRD, to represent a dihydrate, yielding $\chi^2=7.23$ and $R_{p}=3.45\%$. A significant improvement to the fit [Fig.~\ref{fig:fig3}(a)] is obtained, however, when the occupancy of the water site is varied resulting in $\chi^2=4.80$ and $R_{p}=2.83\%$. The refinement converges to a non-stoichiometric hydrate occupancy of 0.37(1) (\textit{i.e.} $\alpha$-KTi(C$_2$O$_4$)$_2\cdot$1.48(4)D$_2$O), with the rest of the refinement parameters presented in Table \ref{HRPD_Alpha}. 

The resulting crystal structure is consistent with that predicted by Drew \textit{et al.}\cite{Drew1977}, and is illustrated in Fig.~\ref{fig:fig1}(a). It is composed of two-dimensional layers of oxalate-bridged square antiprismatic Ti$^{3+}$ ions separated by a disordered layer of K${^+}$ ions and water molecules along the $c$ axis with an interplanar Ti-Ti distance of $7.36(1)$~\AA. Each square antiprism is coordinated by four oxalate molecules in the $ab$ plane [Fig.~\ref{fig:fig1}(b)] which results in a square planar arrangement with nearest-neighbor Ti-Ti superexchange, $J_1$, [Fig.~\ref{fig:fig1}(c)] mediated \textit{via} the Ti-O-C-O-Ti pathway at a distance of $5.69(1)$~\AA. The next-nearest-neighbor coupling, $J_{2}$ [Fig.~\ref{fig:fig1}(c)], runs along the $\langle$010$\rangle$ directions at a Ti-Ti distance of $8.05(1)$~\AA.

\begin{figure}
\begin{center}
{\resizebox{0.99\columnwidth}{!}{\includegraphics{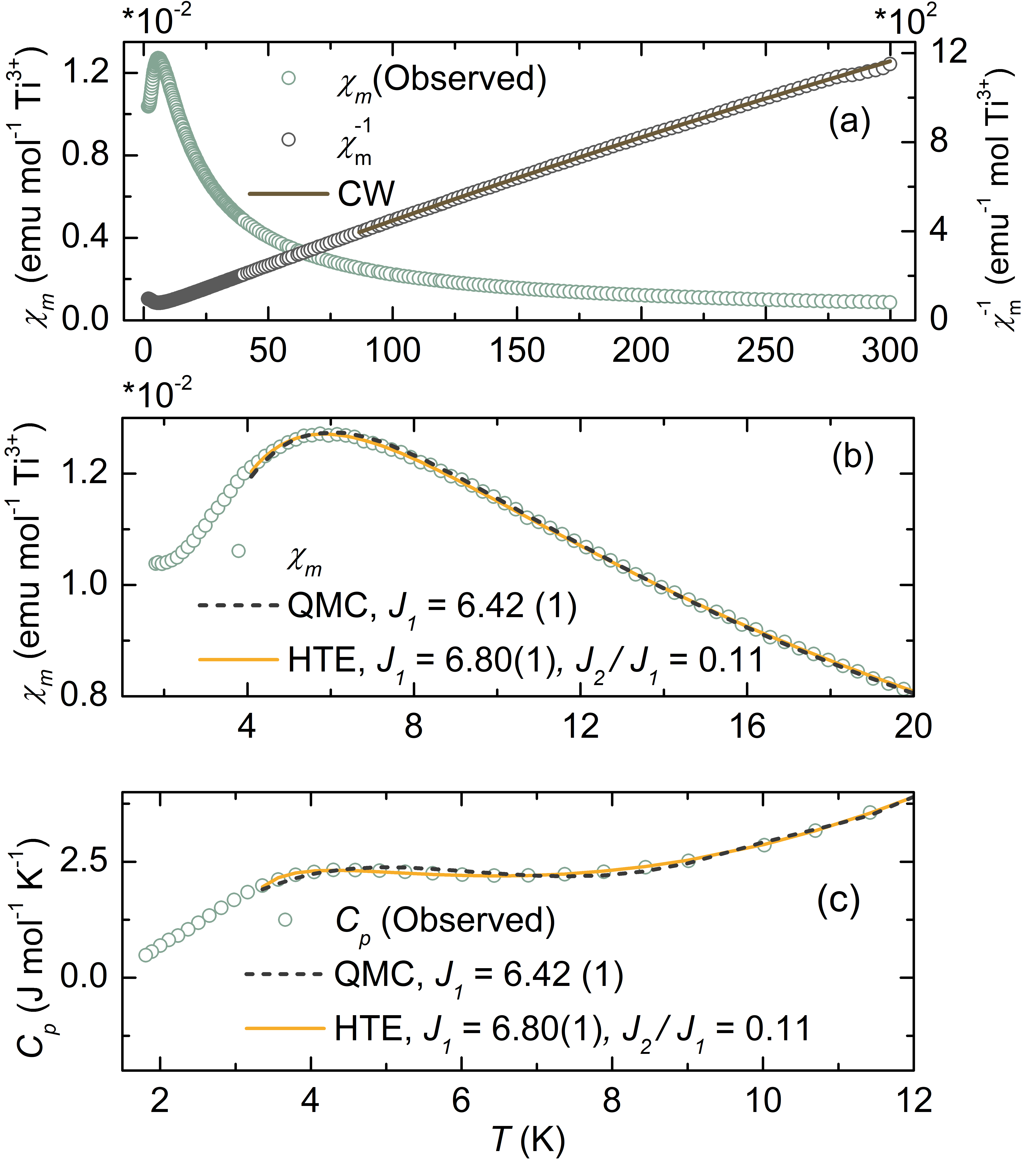}}}
\end{center}
\caption{(a) Temperature-dependent zero-field cooled magnetic susceptibility, $\chi_m$ (blue), of $\alpha$ measured in an applied field of $0.1$~T and its inverse, $\chi^{-1}_m$ (grey). A modified Curie-Weiss fit (brown) to $\chi^{-1}_m$ yields $C=0.233(1)$~emu~K~mol$^{-1}$, $\theta_{CW}=-7.86(1)$~K, and $\chi_0=1.19(2)\times10^{-4}$~emu~mol$^{-1}$. Quantum Monte Carlo (grey) and tenth-order high-temperature series expansion (orange) fits for a Heisenberg $S=1/2$ square lattice model and a $S=1/2$ frustrated square lattice model with interplanar coupling $J_3$, respectively, to (b) $\chi_m$ and (c) zero-field specific heat $C_p$, simultaneously. The resulting model parameters are $J_1=6.42(1)$ for the QMC $S=1/2$ square lattice model and $J_1=6.80(1)$ K, $J_2/J_1=0.11$, $J_3=0$, and $\chi_0=1.19\times10^{-4}$ for the HTE $S=1/2$ FSL model.}\label{fig:fig5}
\end{figure}

\begin{table*}[t]
\caption{\label{SC_XRD} Single crystal XRD data and structural refinement parameters for $\alpha$-KTi(C$_2$O$_4$)$_2\cdot$\textit{x}H$_2$O and $\beta$-KTi(C$_2$O$_4$)$_2\cdot$2H$_2$O. }
\begin{ruledtabular}
\hspace{-0.75cm}
\begin{tabular}{lll}
Compound                                     & $\alpha$-KTi(C$_2$O$_4$)$_2\cdot$\textit{x}H$_2$O                                    & $\beta$-KTi(C$_2$O$_4$)$_2\cdot$2H$_2$O                                     \\\hline 
Formula weight/g mol$^{-1}$                         & 299.03                                   & 299.03                                   \\
Temperature / K                                & 150                                      & 150                                      \\
Crystal system                               & Tetragonal                               & Hexagonal                                \\
Space group                                  & $I4/mcm$                                  & $P6_222$ (or $P6_422$)                                    \\
$a$ / \AA                                         & 8.0401(4)                                & 8.7926(4)                                \\
$b$ / \AA                                          & 8.0401(4)                                & 8.7926(4)                                \\
$c$ / \AA                                          & 14.8504(13)                              & 11.1247(5)                               \\
$\alpha$ / \degree                                          & 90                                       & 90                                       \\
$\beta$ / \degree                                         & 90                                       & 90                                       \\
$\gamma$ / \degree                                          & 90                                       & 120                                      \\
V / \AA$^3$                                    & 959.98(13)                               & 744.82(8)                                \\
Z                                            & 8                                        & 3                                        \\
Crystal size / mm\textsuperscript{3}                             & 0.07 $\times$ 0.07 $\times$ 0.03                    & 0.12 $\times$ 0.06 $\times$ 0.05                    \\
Color                                        & Orange                                   & Maroon                                   \\
$\rho$ / g cm$^{-3}$                                   & 2.069                                    & 2.000                                    \\
F(000)                                       & 596.0                                    & 447.0                                    \\
Reflections collected                        & 3279                                     & 5932                                     \\
Independent reflections                      & 287 {[}R\textsubscript{int} = 0.0614, R$_\sigma$ = 0.0252{]} & 512 {[}R\textsubscript{int}  = 0.0432, R$_\sigma$ = 0.0186{]} \\
Goodness-of-fit on F$^2$                        & 1.091                                    & 1.087                                    \\
Final R indexes {[}I\textgreater{}=2$\sigma$(I){]} & R$_1$ = 0.0453, wR$_2$ =  0.1112                 & R$_1$ = 0.0214, wR$_2$ = 0.0509                \\
\end{tabular}
\end{ruledtabular}
\hspace{-0.75cm}
\end{table*}

\begin{table}[]
\label{table:hrpda}
\caption{\label{HRPD_Alpha} Refined $I4/mcm$ structural model of  $\alpha$-KTi(C$_2$O$_4$)$_2\cdot$1.48(4)D$_2$O as obtained from the analysis of HRPD data ($\chi^2=4.80$, $R_p=2.83\%$) at $1.8$ K. The unit cell parameters are $a = b = 8.043(1)$~\AA~ and $~c=14.719(1)$~\AA.}
\begin{ruledtabular}

\begin{tabular}{lllllll}
\vspace{0.05cm}
Atom & Site                    & $x$           & $y$           & $z$           & $\mathit{Occ.}$ & $U_{\mathit{iso}}$~(\AA\textsuperscript{2}) 
\vspace{0.05cm}
\\\hline \vspace{0.1cm} 
Ti   & 4$\mathit{a}$                      & 0.0000     & 0.0000     & 0.2500     & 1.0  & 0.0066(8)                           \\
\vspace{0.05cm}
K    & 8$\mathit{h}$                      & 0.3753(5) & 0.1247(5) & 0.5000     & 0.5   & 0.0137(10)                           \\
\vspace{0.05cm}
O1   & 32$\mathit{m}$                     & 0.2098(1) & 0.0931(1) & 0.3277(6)  & 1.0  & 0.0065(2)                           \\
\vspace{0.05cm}
O2   & 16$\mathit{k}$                     & 0.2836(3) & 0.9232(3) & 0.5000     & 0.37(1)   & 0.0065(2)                           \\\vspace{0.05cm}
C1   & 16$\mathit{l}$                     & 0.2836(1) & 0.2164(1) & 0.2954(8)  & 1.0  & 0.0060(2)                           \\
\vspace{0.05cm}
D    & 32$\mathit{m}$ & 0.2531(1) & 0.9907(3) & 0.5514(2) & 0.37(1)   & 0.0363(1) 
\end{tabular}
\end{ruledtabular}
\end{table}

\subsection{Magnetic Susceptibility and Specific Heat}
\noindent The temperature dependencies of the molar magnetic susceptibility, $\chi_m$, the inverse molar susceptibility, $\chi^{-1}_m$, and the zero-field specific heat, $C_p$, of $\alpha$ are shown in Fig.~\ref{fig:fig5}. At 280 K, a broad feature is observed in the derivatives of both $\chi_m$ and $C_p$ [Fig.~S2]. Given its small amplitude and the porosity of the square network of $\alpha$, we ascribe this feature to the freezing of the lattice water position. A similar feature has been observed in the specific heat data of (CH$_3$NH$_3$)$_2$NaTi$_3$F$_{12}$\cite{Jiang} and has been attributed to the ordering of the methylammonium groups. Above 50 K, $\chi_m$ is well described by a modified Curie-Weiss (CW) law, $\chi_m(T)=C/(T-\theta_{CW})+\chi_0$, where $C=N_{A}\mu^2_{\mathit{eff}}/3k_B$ and $\theta_{CW}$ are the Curie and Weiss constants, respectively, and $\chi_0$ is a temperature independent background term. Relatively weak antiferromagnetic interactions are suggested as $\chi_m$ is consistently best described by $\theta_{CW}=-7.86(1)$~K. We also find a positive $\chi_0=1.19(2)\times10^{-4}$~emu~mol$^{-1}$, as seen in other Ti\textsuperscript{3+}-containing compounds\cite{Aggarwal1986,Eitel1986} which may be ascribed to a van Vleck contribution to the susceptibility\cite{Carr1960}. The fitted Curie constant, $C=0.233(1)$~emu~K~mol$^{-1}$, reveals a reduced effective magnetic moment, $\mu_{\mathit{eff}}=1.36(1)~\mu_\mathit{B}$ ($g=1.57(1))$, in comparison to the $S=1/2$ spin-only moment, $\mu_{\mathit{eff}}=1.73~\mu_\mathit{B}$. As discussed in Sec.~\ref{sec:discuss}, although a reduction of the effective moment in Ti\textsuperscript{3+}-containing compounds is usually a consequence of an orbital contribution, this picture is probably not applicable for $\alpha$. Instead, considering the instability of Ti\textsuperscript{3+} in air, it may reflect the presence of a minor impurity phase, for example an oxidized surface layer of the polycrystalline sample, or a data normalization problem arising from the non-stoichiometry of $\alpha$. Indeed, this is not uncommon for Ti\textsuperscript{3+}-containing compounds, where a similar reduction in $g$, unassociated with an orbital contribution\cite{ Kasinathan2013, Nilsen2015}, is observed for both KTi(SO\textsubscript{4})\textsubscript{2} \cite{ Bramwell1996} and KTi(SO\textsubscript{4})\textsubscript{2}$\cdot$H\textsubscript{2}O \cite{ Nilsen2008}. Finally, it should be noted that the possible presence of a magnetic ion deficiency was examined by allowing a free occupancy refinement of the Ti\textsuperscript{3+} and K\textsuperscript{+} sites in the models considered using both the PND and SC-XRD data. While it cannot be conclusively disregarded, it is highly unlikely that a deficiency exists as the resulting goodness-of-fit parameters worsen ($\chi^2=7.13$ and $R_{p}=3.23\%$), and only models with instead an insignificant Ti\textsuperscript{3+} excess of $1.4(1.3)\%$ were extracted. 

Upon cooling, the build up of short-range antiferromagnetic correlations is evidenced by a broad maximum in $\chi_m$ centered around $6.1$ K. This signature is also observed in the specific heat at a slightly lower temperature, as is typical for low-dimensional square lattice systems\cite{Tsirlin2011}. No features indicative of long-range magnetic order can be observed in either data-set down to 1.8 K. Because of the non-stoichiometric nature of $\alpha$, the values of the molar susceptibility and specific heat should be treated as approximate, and have been calculated with respect to the hydration of the deuterated sample obtained from the HRPD refinement.

To estimate the magnitude of the leading magnetic exchange parameter in $\alpha$, $J_1$, a QMC simulation was performed for the Heisenberg $S=1/2$ square lattice model using the \texttt{ALPS}\cite{ALPS,ALPS2} simulation package. By approximating the phonon contribution to the specific heat using $C_p=\alpha T^3+\beta T^5+\gamma T^7$, as done with other coordination frameworks\cite{Lancaster2007,Nath2015}, a simultaneous fit to $\chi_m$ and $C_p$ above $5$~K yields $J_1=6.42(1)$~K ($g=1.53(1)$) when $\chi_0$ is fixed to its CW fit value. Given that the QMC method is not generally applicable to frustrated systems, the \texttt{HTE10} code\cite{Lohmann} was used to calculate the [4,6] Pad\'e approximant of the high-temperature-series expansion of a $S=1/2$ FSL model, with two in-plane exchange interactions, $J_1$ and $J_2$, and an interplanar coupling, $J_3$ [Fig.~\ref{fig:fig1}(c)]. Using a similar simultaneous fitting procedure above $7$~K, the magnetic susceptibility and specific heat data were best described by a slightly frustrated system ($J_2/J_1=0.11$) with $J_1=6.80(1)$~K ($g=1.61$) and no interplanar coupling. The fitting parameters were $J_1$ and the coefficients of the phonon contribution, while $J_2$ and $J_3$ were varied in $0.01J_1$ steps between $-0.2J_1$ and $0.3J_1$, and $\chi_0$ was fixed to its CW fit value.  The absence of significant interplanar coupling, $J_3/J_1<1.9\times 10^{-3}$, is further implied by applying the empirical formula $T_N=(4\pi \rho_s)/(b-\textnormal{ln}(J_3/J_1))$\cite{Yasuda2005} for $T_N<1.8$ K (see Sec.~\ref{sec:mstruct}). Finally, as discussed in the methodology of the DFT calculations (Sec.~\ref{sec:sec2}) where a similar magnetic behavior is observed regardless of the model used to describe structural disorder, the extent of hydration and disorder in the potassium and water containing layers separating the [Ti(C\textsubscript{2}O\textsubscript{4})\textsubscript{2}]$^-$ planes indeed appear to be inconsequential to the magnetism of the titanium oxalate containing layers, as similar CW, QMC, and HTE parameters were extracted when fitting $\chi_m$ of another $\alpha$ sample [Fig.~S2].

At first glance, the solution extracted from the HTE fit is more consistent with the expected $\theta_{CW} = (J_1 + J_2)/k_B = 7.55(2)$ K for a FSL system than the QMC fit. This is further corroborated when considering the residual plots of both fits, as the HTE model better describes the experimental data over the whole fitting range [Fig. S4]. Our DFT calculations furthermore estimate that Ti-Ti exchanges beyond the nearest-neighbor are at least an order of magnitude smaller than $J_1$. As such, distinguishing between the two solutions provided by HTE and QMC will require probes such as inelastic neutron scattering, which are beyond the scope of this current study. Regardless, the dominant antiferromagnetic nearest-neighbor interactions and possible slight frustration indicate that $\alpha$ belongs to the N\'eel ordered region of the FSL phase diagram \cite{Shannon2004}.

\subsection{Magnetic Structure}
\noindent To investigate the nature of the low-temperature magnetic state of $\alpha$, PND data were collected at regular temperature intervals between $1.2$ K and $15$ K on the WISH instrument at ISIS. No magnetic features were present in the subtracted data above $2$~K, placing $1.2$~K~$<T_N<1.8$~K. Given the small ordered moment expected for a quasi-two-dimensional quantum spin system, the analysis was performed on the magnetic-only scattering obtained by subtracting the 15~K data set from that collected at 1.2~K. Two magnetic Bragg peaks were observed at positions corresponding to forbidden nuclear positions of the $I4/mcm$ space group [Fig.~\ref{fig:fig3}(b)] and were thus indexed by the commensurate propagation vector $\mathbf{k}=(0,0,0)$. To determine the magnetic structure, four irreducible representations---$m\Gamma_{3+}$, $m\Gamma_{3-}$, $m\Gamma_{5+}$, and $m\Gamma_{5-}$ in Miller-Love notation \cite{Miller}---were found to be compatible with the paramagnetic space group and propagation vector using the \texttt{BasIreps}\cite{fullprof} and \texttt{MAXMAGN}\cite{maxmagn} software packages. Of these, $m\Gamma_{3+}$ and $m\Gamma_{5+}$ correspond to ferromagnetic structures, and are therefore incompatible with the magnetic susceptibility of $\alpha$. Indeed, by refining the nuclear structural model to data collected at 15 K ($R_p=5.24\%$, $\mathit{Occ.}_{\mathsf{D_2O}}=0.40(2)$) and fixing the resulting structural and instrumental parameters for the magnetic model refinement, only the magnetic space group $I4/m^\prime cm$ (in Belov-Neronova-Smirnova notation \cite{belov}) belonging to $m\Gamma_{3-}$, correctly describes the data ($\chi^2=1.82, R_{mag}=3.16\%$) [Fig.~\ref{fig:fig3}(b)]. Comparative Rietveld plots for the other possible models are presented in Fig.~S5.

The resulting magnetic structure is shown in Fig.~\ref{fig:fig1}(c) and can be described as a G-type antiferromagnet with the magnetic moments aligned along the $c$ axis. The two-dimensional character of $\alpha$-KTi(C$_2$O$_4$)$_2\cdot$\textit{x}D$_2$O is corroborated by the observed ordered moment $\mu=0.62(3)~\mu_B$ per Ti\textsuperscript{3+} extracted from the Rietveld fit, similar to other square network systems\cite{Bettler2019, Bombardi2004, Koga2016}, and to the expected ordered moment of $0.6~\mu_B$ for a $S=1/2$ square lattice antiferromagnet \cite{Manousakis1991, Liu1990, Holt2011}. The discrepancy between the value extracted from the fit to PND data and  the expected $gS\mu_{B}=0.81~\mu_{B}$ extracted from the HTE fit is most likely associated with data renormalization issues as discussed in Sec.~\ref{Magsusc}. Also, given that the exact N\'eel temperature is unknown, thermal effects, resulting in the reduction of the ordered moment, cannot be excluded as the origin of the moment reduction.

\section{Results for \NoCaseChange{$\beta$-KTi(C$_2$O$_4$)$_2\cdot$2H$_2$O}}
\label{sec:sec4}
\subsection{Crystal Structure}
\begin{figure}
\begin{center}
{\resizebox{0.99\columnwidth}{!}{\includegraphics{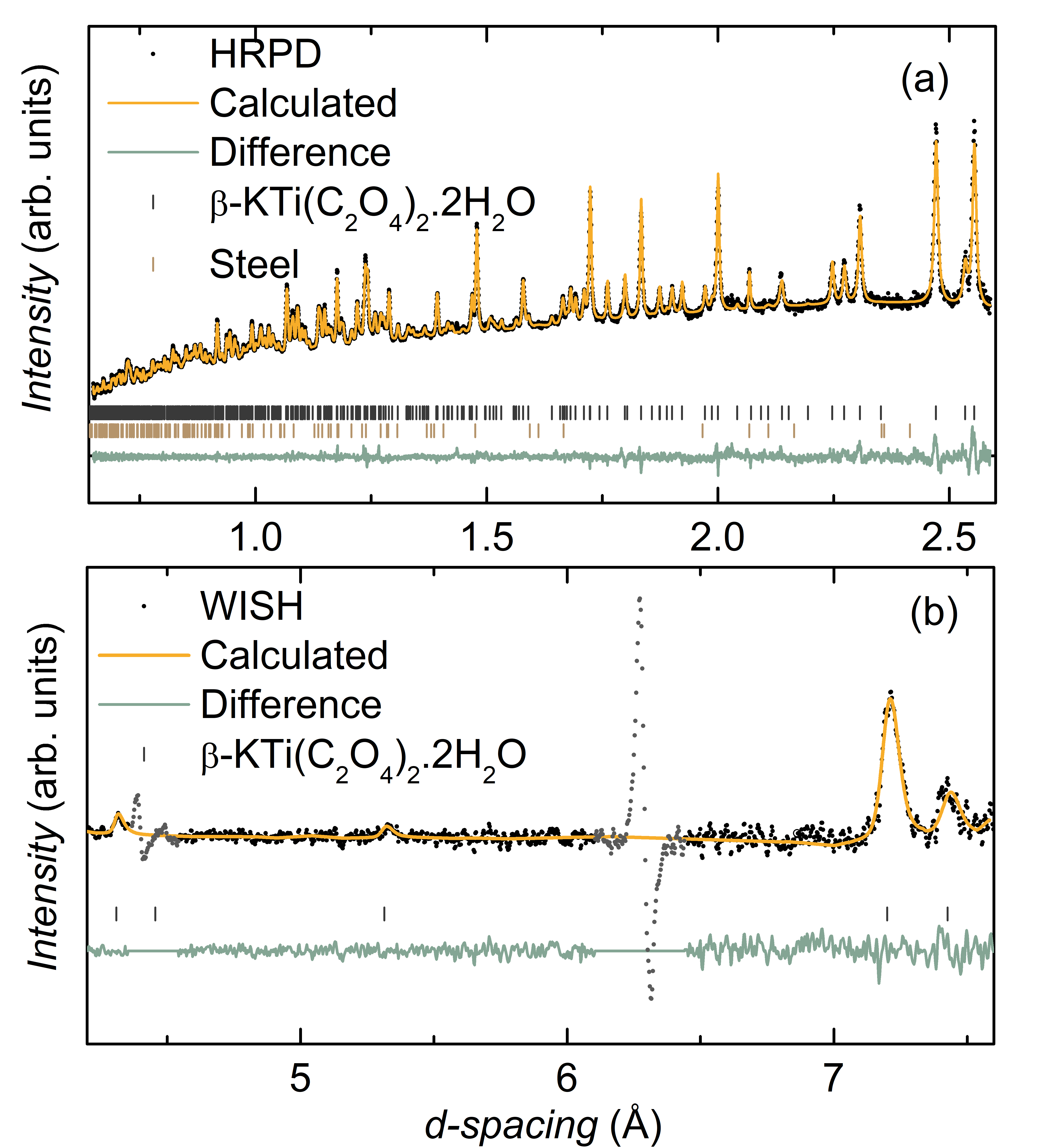}}}
\end{center}
\caption{(a) Rietveld plot for data collected on HRPD at $1.8$ K using the $P6_222$ structural model to describe $\beta$-KTi(C$_2$O$_4$)$_2\cdot$2H$_2$O with goodness-of-fit parameters $R_p=1.81\%$ and $\chi^2=3.06$. (b) Rietveld refinement ($R_{mag}=1.93\%$,~$\chi^2=3.91$) of the $P_b2_1$ magnetic space group model to  magnetic-only scattering obtained by subtracting data collected on the WISH instrument at $35$ K and $1.2$ K. In both plots, data points are shown in black, fitted curves in orange, difference curves in blue, Bragg reflection positions in grey and brown, and excluded data points in grey.}\label{fig:fig4}
\end{figure}

\begin{figure}
\begin{center}
{\resizebox{0.99\columnwidth}{!}{\includegraphics{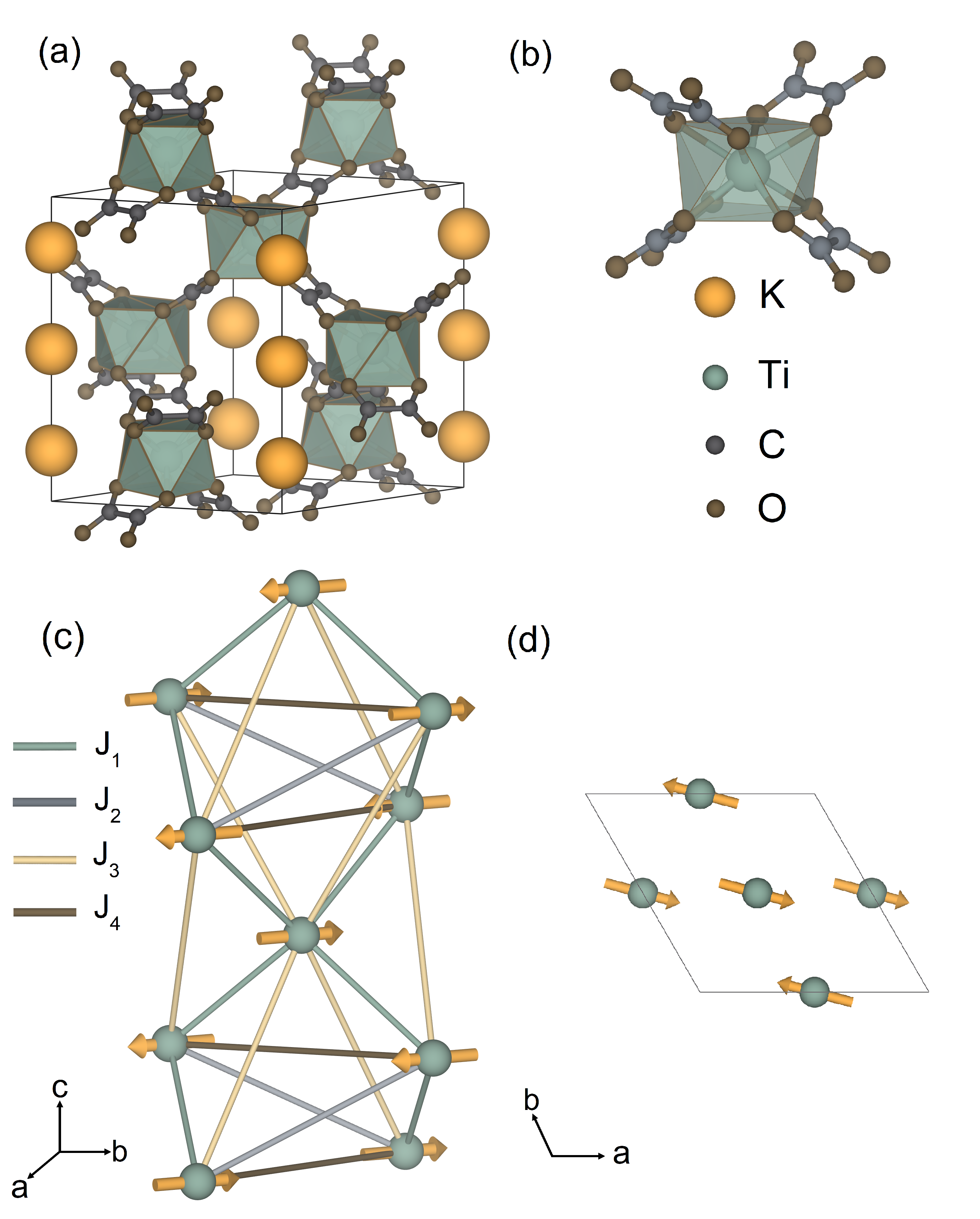}}}
\end{center}
\caption{(a) The crystal structure of $\beta$-KTi(C$_2$O$_4$)$_2\cdot$2H$_2$O as viewed along the $\langle 111\rangle$ directions where (b) distorted square antiprismatic Ti\textsuperscript{3+} ions form a diamond-like magnetic sublattice. (c) One of the possible magnetic moment arrangements for the $P_b2_1$ magnetic structure describing the coplanar antiferromagnetic ordering of $\beta$-KTi(C$_2$O$_4$)$_2\cdot$2H$_2$O with the nearest-neighbor and further neighbor exchange interactions shown in blue, gray, yellow, and brown, respectively. The possible spin arrangements only differ with respect to how the moment aligns within the \textit{ab} plane. The illustrated structure in (d) is representative of one of the possible moment directions.}\label{fig:fig2}
\end{figure}

    \noindent A chiral hexagonal $P6_222$ crystal structure of $\beta$ was determined through the analysis of SC-XRD data collected at 150~K, as summarised in Table.~\ref{SC_XRD}. This is consistent with the reported structure of the related $\beta$-NH$_4$Ti(C$_2$O$_4$)$_2\cdot$2H$_2$O, whereby the chirality of the crystal structure is such that both enantiomers, $P6_222$\cite{English1993} and $P6_422$\cite{Sheu1996}, are reported depending on the particular single crystal studied. A multi-bank refinement of this model to PND data collected on HRPD further confirms this structure at all measured temperatures [Fig.~\ref{fig:fig4}(a)] and gives an overall $\chi^2 = 3.06$ and $R_p=1.81\%$ at 1.8~K with the model shown in Table.~\ref{HRPD_Beta}. Because of the chirality of the crystal structure of $\beta$, standard powder diffraction measurements cannot distinguish between the enantiomers as polycrystalline samples are most likely comprised of a racemic mixture of the enantiomorphic space group pair\cite{Fujio2007}. As illustrated in Fig.~\ref{fig:fig2}(a), the crystal structure of $\beta$ forms a three-dimensional diamond-like network of Ti$^{3+}$ ions in a distorted square antiprismatic configuration linked by oxalate groups [Fig.~\ref{fig:fig2}(b)]. The distortion arises from the presence of two oxalate oxygen sites, giving rise to four Ti-O1 and two Ti-O2 bonds with distances of 2.11(3)~\AA~and 2.27(2)~\AA, respectively. Each oxalate group coordinates two Ti$^{3+}$ ions, thus providing a pathway for nearest-neighbour superexchange, $J_1$, with a Ti-Ti distance of 5.75(2)~\AA. Next-nearest-neighbour exchanges $J_2$, $J_3$, and $J_4$ run along the $\langle$111$\rangle$, $\langle100\rangle$, and $\langle$010$\rangle$ directions with similar Ti-Ti distances of 8.46(2)~\AA, 8.63(1)~\AA, and 8.78(3)~\AA, respectively [Fig.~\ref{fig:fig2}(c)]. Along the $c$ axis, K$^+$ ions and water molecules are packed in a column-like manner within the cavities surrounding the diamond-like sublattice [Fig. S6]. Water molecules are present on the O3 site and form hydrogen bonds with the O2 site of the oxalate groups at a distance of 1.92(2)~\AA.  

\begin{table}[]
\caption{\label{HRPD_Beta} Crystallographic data for $\beta$-KTi(C$_2$O$_4$)$_2\cdot$2H$_2$O as obtained by fitting the $P6_222$ model to data collected on HRPD at $1.8$ K. The unit cell parameters are $a = b = 8.784(1)$~\AA~and $~c=11.148(2)$~\AA~and the goodness-of-fit parameters are $\chi^2=3.06$ and $R_p=1.81\%$.}
\begin{ruledtabular}
\begin{tabular}{llllll}
Atom & Site                   & $x$           & $y$           & $z$           & $U_{\mathit{iso}}$~(\AA\textsuperscript{2}) 
\\ \hline 
\vspace{0.05cm} Ti   & 3$\mathit{d}$                     & 0.5000     & 0.0000      & 0.5000      & 0.0012(6)                           \\ \vspace{0.05cm}
K    & 3$\mathit{b}$                     & 0.0000      & 0.0000      & 0.5000      & 0.0019(2)                           \\\vspace{0.05cm}
O1   & 12$\mathit{k}$                    & 0.2656(2) & 0.9210(2) & 0.4001(1) & 0.0068(2)                           \\\vspace{0.05cm}
O2   & 12$\mathit{k}$                    & 0.4494(2) & 0.7615(2) & 0.3923(1) & 0.0063(2)                           \\\vspace{0.05cm}
O3   & 6$\mathit{g}$                     & 0.7662(3) & 0.7662(3) & 0.3333     & 0.0093(5)                           \\\vspace{0.05cm}
C1   & 6$\mathit{i}$ & 0.2094(1) & 0.7906(1) & 0.3333     & 0.0055(5)                           \\\vspace{0.05cm}
C2   & 6$\mathit{i}$ & 0.3083(1) & 0.6917(1) & 0.3333     & 0.0050(5)                           \\\vspace{0.05cm}
H    & 12$\mathit{k}$                    & 0.7493(5) & 0.6574(4) & 0.3152(3) & 0.0330(8)             
\end{tabular}
\end{ruledtabular}
\end{table}
\subsection{Magnetic Susceptibility and Specific Heat}
\label{Magsusc}
\begin{figure}
\begin{center}
{\resizebox{0.99\columnwidth}{!}{\includegraphics{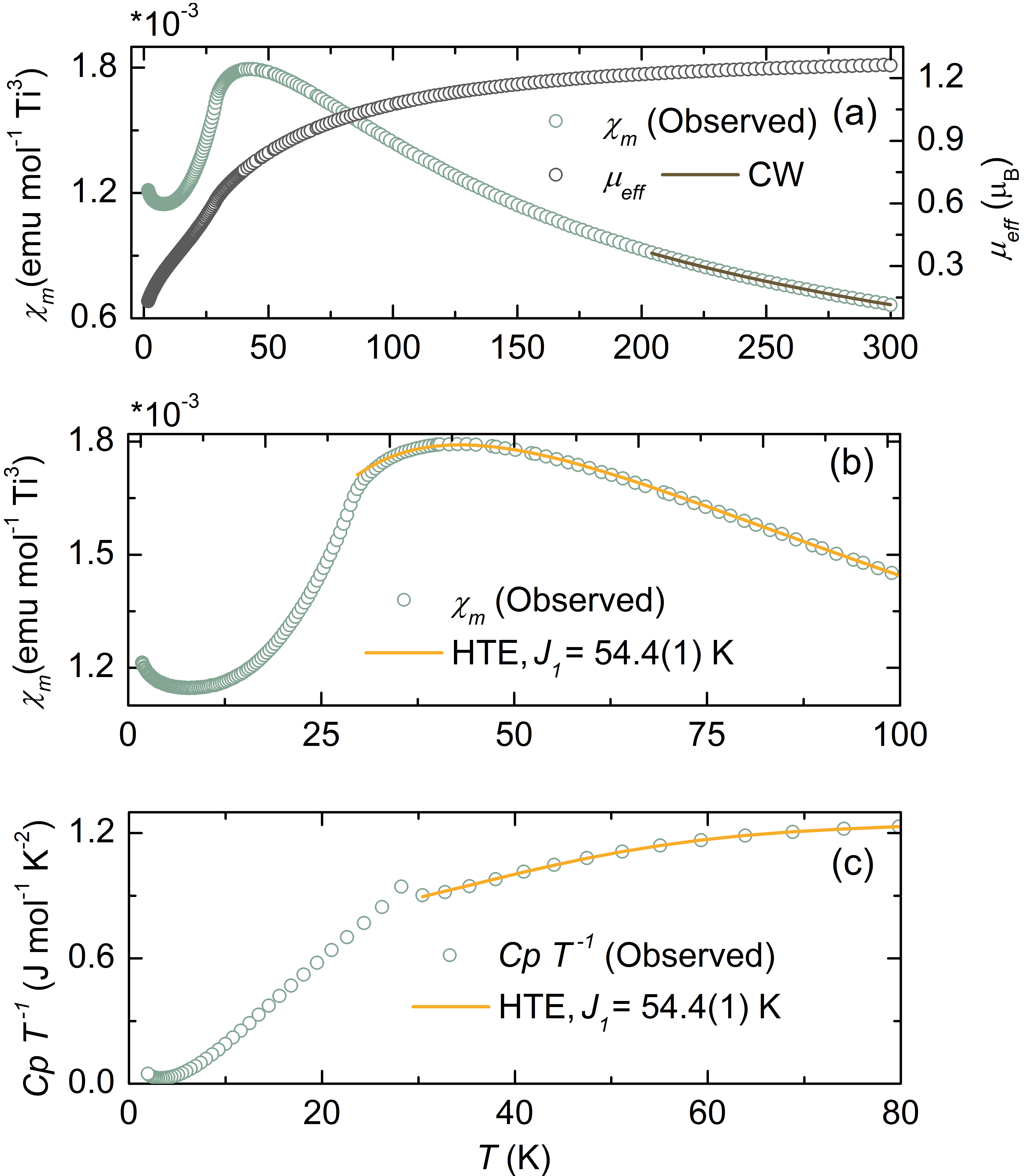}}}
\end{center}
\caption{(a) Zero-field cooled magnetic susceptibility of $\beta$-KTi(C$_2$O$_4$)$_2\cdot$2H$_2$O, $\chi_m$ (blue), of measured between $1.8$~K and $300$~K in a $0.1$~T applied magnetic field and corresponding calculated effective magnetic moment (grey). A Curie-Weiss fit to $\chi_m$ (brown) yields $\theta_\mathsf{CW}=-109.6(1)$~K, $C=0.329(2)$~emu~K~mol$^{-1}$, and $\chi_0=-4.88(3)\times10^{-5}$~emu~mol$^{-1}$. (b) and (c) Simultaneous fit to $\chi_m$ and the zero-field specific heat $C_p T^{-1}$ using the diamond lattice model with a nearest-neighbor coupling $J_1$ calculated by a tenth-order high-temperature series expansion. The model yields $J=54.4(1)$~K and $\chi_0=-4.11(8)\times10^{-5}$~emu~mol$^{-1}$.}\label{fig:fig6}
\end{figure}

\noindent The temperature-dependent molar magnetic susceptibility, $\chi_m$, effective magnetic moment, $\mu_{\mathit{eff}}$, and specific heat, $C_p$, of $\beta$ are shown in Fig.~\ref{fig:fig6}. Fitting $\chi_m$ over the temperature range $230$~K to $300$~K with the same modified CW model as applied to $\alpha$ yields $\theta_{CW}=-109.6(1)$~K and $C=0.329(1)$~emu~K~mol$^{-1}$ ($g=1.87$). The resulting negative $\theta_{CW}$ indicates dominant antiferromagnetic interactions an order of magnitude larger than those of $\alpha$. A reduced $\mu_{\mathit{eff}}=1.62(1)~\mu_\mathit{B}$~is calculated from $C$, which is consistent with a previous report \cite{Kalinnikov1969}, and its deviation from the spin-only moment---combined with the observed temperature dependence of the magnetic moment and relatively large negative $\theta_{CW}$---likely reflects the presence of antiferromagnetic correlations above $200$~K. We also find a small and negative $\chi_0=-4.88(3)\times10^{-5}$~emu~mol$^{-1}$, indicating the contribution of the sample holder and core diamagnetism of $\beta$ to the total magnetic susceptibility. The presence of short-range correlations is evidenced by a broad maximum in $\chi_m$ centered about $43$~K, followed by an inflection point at $28$~K that can be attributed to long-range antiferromagnetic ordering. This ordering transition is also present as an anomaly in $C_p$ at the same temperature [Fig.~\ref{fig:fig6}(c)]. 

To estimate the nearest-neighbor exchange interaction in $\beta$, $J_1$, the diamond lattice model was used to simultaneously fit $\chi_m$ and $C_p$ using the same HTE fitting procedure applied for $\alpha$. Above $60$~K, the data were consistently best described by a leading exchange $J_1=54.4(1)$~K ($g=1.74$) and a temperature independent background term $\chi_0=-4.11(8)\times10^{-5}$~emu~mol$^{-1}$, similar to that extracted from the CW fit. Interestingly, the resulting $T_N/J_1=0.51$ ratio is similar to the theoretically predicted ratio, $T_N/J_1=0.531(1)$\cite{Oitmaa2018}, expected for a $S=1/2$ Heisenberg diamond lattice model system. A frustrated diamond lattice model extended to account for further near-neighbor couplings was also used to simultaneously fit $\chi_m$ and $C_p$ [Fig.~S5]. While the resulting parameters ($J_2/J_1=0.04$, $J_3/J_1=0.03, J_4/J_1=0.04$, and $g=1.78$) better describe $\chi_m$ [Fig.~S7], this model should be taken at best as suggestive of frustration given the number of exchange parameters involved. Together with $T_N$, the fitted parameters are broadly consistent with minimal frustration and place $\beta$ within the N\'{e}el ordered region of the FDL phase diagram \cite{Oitmaa2019}.
\subsection{Magnetic Structure}
\label{sec:mstruct}
\noindent Indeed, the onset of three-dimensional magnetic order in $\beta$ is verified by the presence of additional intensity in the subtracted PND data ($1.2$~K~$-$~$35$~K) [Fig.~\ref{fig:fig4}(b)]. The observed magnetic reflections can be indexed by a unit cell doubled along the $c$ axis with the propagation vector $\mathbf{k}=(0,0,1/2)$. Symmetry analysis using \texttt{ISODISTORT}\cite{Campbell2006} and \texttt{SARAh}\cite{Wills2000} reveal five irreducible representations, $m\Alpha_1$, $m\Alpha_3$, $m\Alpha_4$, $m\Alpha_5$, and $m\Alpha_6$, compatible with $\mathbf{k}$ and the nuclear space group, $P6_222$ (or $P6_422$). However, all magnetic models given by these irreducible representations, except for $m\Alpha_6$, conflict with the presence of ($00l$)-type reflections in the PND data.  

Accordingly, the best description of the data ($R_{mag}=1.93\%,~\chi^2=3.91$), shown in Fig.~\ref{fig:fig4}(b), is obtained with using the $P_b2_1$ magnetic space group of the $m\Alpha_6$ irreducible representation. The resulting ordered moment, $\mu=0.79(2)~\mu_B$, is similar to the theoretically expected value, $\mu=0.76~\mu_B$\cite{Oitmaa2018}, for the $S=1/2$ Heisenberg diamond lattice model, and is slightly suppressed from the expected $gS\mu_B=0.87~\mu_B$ from the susceptibility fits. Interestingly, and in contrast to previous studies on $S=1/2$ diamond lattice systems\cite{Marjerrison2016,Injac2019}, $\beta$ is the first system in which the ordered moment of the $S=1/2$ diamond lattice model can be extracted without any ambiguity related to possible orbital contributions and/or covalency effects. The reasons for this are further discussed in Sec.~\ref{sec:discuss}. Comparative fits with other possible models are presented in Fig.~S8. It should be noted that the magnetic moment direction of the $P_b2_1$ structure in the $ab$ plane cannot be uniquely determined from the present data alone, because the structure factor of the strongest $(101)$ peak at $d\sim 7.2$~\AA~is insensitive to the in-plane moment direction. A representative illustration of the magnetic structure, in which nearest-neighbor (\textit{J}) magnetic moments are antiferromagnetic and align along the $ab$ plane, is therefore shown in Figs.~\ref{fig:fig2} (c) and (d). The full determination of the magnetic ground state of $\beta$ would be challenging even with a deuterated single crystal due to the 12 domains created by the lowering of symmetry from $P6_2221^\prime$ (or $P6_4221^\prime$) to $P_b2_1$. Given that the structure of $\beta$ is chiral, it should also be noted that the analysis of polycrystalline PND data is insensitive to the enantiomorphic space group pair, and an enantiopure single crystal would be necessary for further analysis.

\section{Discussion}
\label{sec:discuss}
\noindent One of the most striking observations made in comparing the magnetic properties of $\alpha$ and $\beta$ is the substantial difference between the magnitudes of their nearest-neighbor exchange interactions. Indeed, the nearest-neighbor exchange parameter, $J_1$, in the diamond network of Ti$^{3+}$ ions in ${\beta}$ is an order of magnitude larger than the corresponding leading exchange for $\alpha$. To understand the origin of this behavior, one has to consider the nature of the orbitals involved in the respective superexchange pathways of $\alpha$ and $\beta$ and how they overlap. For a transition metal ion in the square antiprismatic coordination environment exhibited in $\alpha$ and $\beta$, the crystal field symmetry splits the degenerate \textit{d}-orbitals into a low-lying \textit{A\textsubscript{1}} ($d_{z^2}$) ground state term and two higher energy levels corresponding to the doubly degenerate \textit{E\textsubscript{2}} and \textit{E\textsubscript{3}} terms\cite{Randic1960}. In the case of a Ti$^{3+}$ ion in such an environment, the lone 3\textit{d}\textsuperscript{1} electron is thus expected to populate the $d_{z^2}$ orbital within the \textit{A\textsubscript{1}} ground state. Indeed, through our DFT band-structure calculations, we identify $d_{z^2}$ as the active magnetic orbital for both $\alpha$ and $\beta$, even in the presence of the distorted square antiprismatic coordination environment found in the latter, as also observed for $\beta$-NH$_4$Ti(C$_2$O$_4$)$_2\cdot$2H$_2$O\cite{Sheu1996}. 

\begin{figure}
\begin{center}
{\resizebox{0.99\columnwidth}{!}{\includegraphics{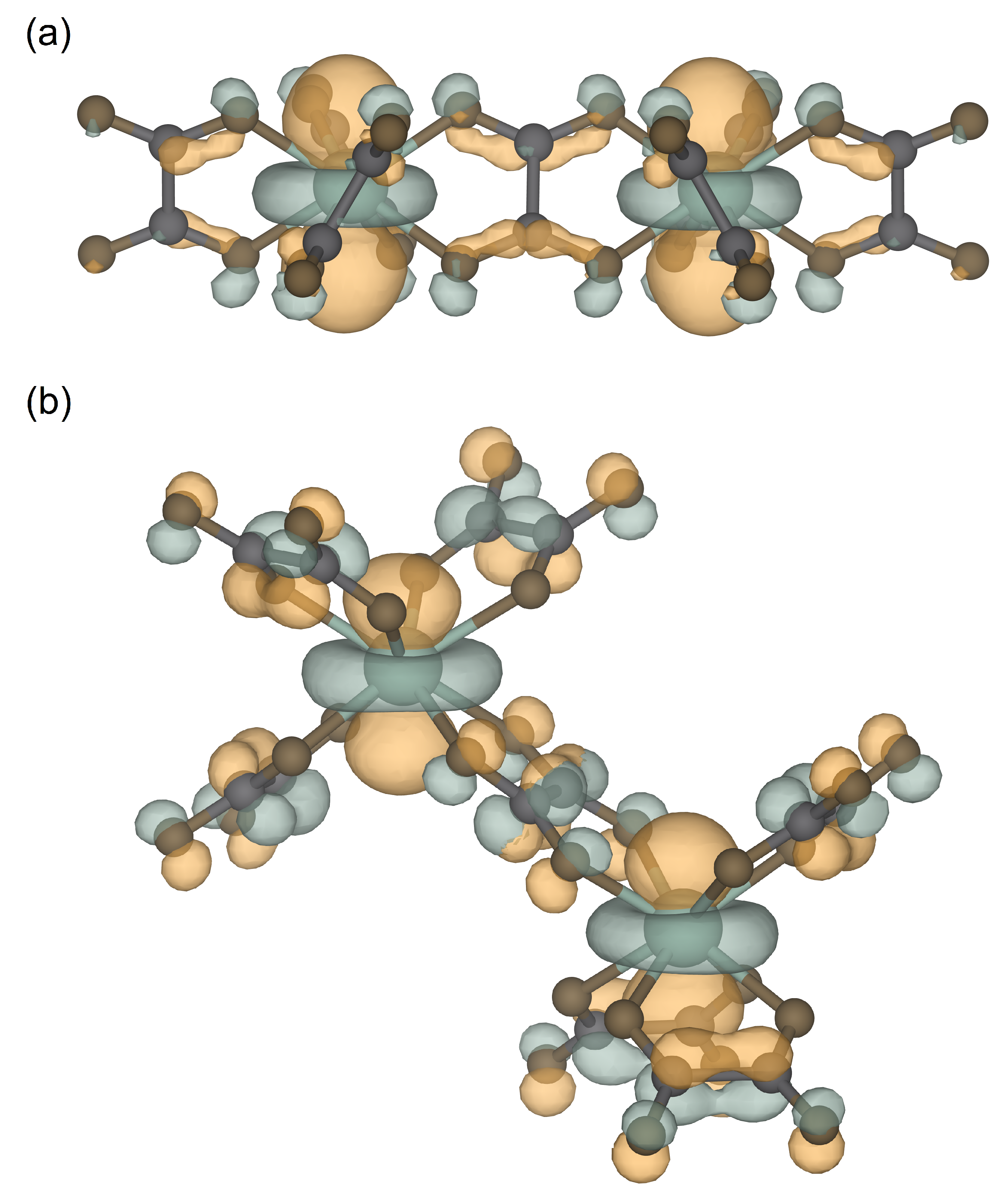}}}
\end{center}
\caption{Ti$^{3+}$ $d_{z^2}$-based Wannier functions showing the orbital overlaps for the $J_1$ superexchange pathways of (a) $\alpha$-KTi(C$_2$O$_4$)$_2\cdot$\textit{x}H$_2$O and (b) $\beta$-KTi(C$_2$O$_4$)$_2\cdot$2H$_2$O. }\label{fig:fig8}
\end{figure}

As described in the context of several other oxalate-based coordination frameworks, the interplay between the architecture of the oxalate-metal bridge and the active magnetic orbital plays a crucial role in determining the sign and strength of the resulting exchange\cite{Kahn1985, Cano1998}. In the case of $\alpha$, these oxalate bridges lie within the plane of the Ti$^{3+}$ ions, whereas the oxalate bridging in $\beta$ spans different planes. As illustrated through the calculated Wannier functions in Fig.~\ref{fig:fig8}, this leads to a significant superexchange in $\beta$, where the $d_{z^2}$ orbital overlaps directly with the oxalate $\pi$-bonding orbital. This results in a contribution to the exchange pathway at the C atoms within the oxalate bridge where the magnetic orbitals of the neighboring Ti\textsuperscript{3+} ions overlap. Conversely, when the active orbital is parallel to the oxalate group---as is more often discussed in the context of a $d_{x^2-y^2}$ active orbital in Cu\textsuperscript{2+}-containing compounds---a significantly reduced exchange interaction is expected\cite{Kahn1985,Cano1998}. This appears to be the case for $\alpha$, in which the Ti$^{3+}$ $d_{z^2}$ orbital overlaps with a different molecular orbital with minimal contribution at the C atoms, resulting in a reduced overlap with the neighboring Ti\textsuperscript{3+} ion, and thus a weaker superexchange interaction.  

While the effect of spin-orbit coupling is expected to be pronounced for octahedrally coordinated Ti\textsuperscript{3+}-containing compounds\cite{Carlin1986}, our DFT results also reveal a negligible contribution from orbitally excited states for the square antipristmatic coordination of Ti$^{3+}$ in $\alpha$ and $\beta$. This is unsurprising given the $d_{z^2}$ ($m_l = 0$) ground state. The combination of this result and our experimental observations indicate that $\alpha$ and $\beta$ should be considered close to ideal model $S=1/2$ Heisenberg square and diamond lattice antiferromagnets, respectively. 

Compared with other coordination frameworks, the magnetic response of $\alpha$ resembles that seen in the square lattice antiferromagnet, Cu(pz)$_2$(ClO$_4$)$_2$ (pz = pyrazine), where $J_2/J_1\approx0.02$\cite{Tsyrulin2009,Tsyrulin2010}. Given the relatively large distances between the magnetic ion centers in such systems, the weak frustration present in both $\alpha$ and Cu(pz)$_2$(ClO$_4$)$_2$ is to be expected. As a consequence, while some advantages are associated with coordination frameworks, especially related to tuning the energy scales present in the system through different bridging molecules, designing frustration into them is more difficult in comparison to their oxide counterparts. This is a result of the complex exchange pathways, which require several sets of orbitals to align favorably to generate an appreciable exchange. As for $\beta$, to the best of our knowledge, this is the first comprehensive study of the magnetic properties of a diamond lattice coordination framework, especially one in which the magnetic ion resides in an 8-coordinate environment. However, the extracted nearest-neighbour exchange parameter, $J_1$, is of the same order of magnitude as that of Ti$_2$(C$_2$O$_4$)$_3$(H$_2$O)$_5$ ($J=86$ K)\cite{WROBLESKI1980227}, and the magnetic response of $\beta$ bears resemblance to that of the $A$-site diamond lattice spinel, CoRh$_2$O$_4$\cite{Ge2017}. 

\section{Conclusions}
\label{sec:sec6}
\noindent In summary, we have presented a detailed investigation of the crystallographic and magnetic properties of the Ti\textsuperscript{3+}-containing coordination frameworks, $\alpha$-KTi(C$_2$O$_4$)$_2\cdot$\textit{x}H$_2$O and $\beta$-KTi(C$_2$O$_4$)$_2\cdot$2H$_2$O. Through analysis of SC-XRD and PND data, a quasi-two-dimensional square planar network of Ti$^{3+}$ ions was found within the crystal structure of $\alpha$. $\beta$, on the other hand, realises a three-dimensional diamond-like array of Ti$^{3+}$ moments. Bulk characterization indicates slight frustration of the antiferromagnetic interactions present in both systems, and simultaneous fits to the magnetic susceptibility and specific heat data demonstrate the two-dimensional character of the magnetic correlations in $\alpha$. Concomitantly, the order of magnitude difference in the extracted exchange parameters manifests the stark differences in the orbital interactions at play in each of the psuedo-polymorphs. Through supporting DFT calculations, these differences can be understood as arising from the different bridging geometries between neighboring Ti\textsuperscript{3+} ions, which result in  different orbital overlaps between the $d_{z^2}$-orbitals and the oxalate $\pi$ bonding orbitals. At low temperatures, the onset of long-range antiferromagnetic order was verified for both systems in PND data, with the resulting magnetic structures and ordered moments identifying $\alpha$ and $\beta$ as realizations of $S=1/2$ Heisenberg square and diamond lattice antiferromagnetic models, respectively.

Given the rarity of stable Ti\textsuperscript{3+}-based compounds and the versatility of coordination frameworks, we hope our results provide a pathway towards future experiments to further understand the observed magnetic behavior. Inelastic neutron scattering, for example, will provide a more detailed picture of the exchange Hamiltonian in both systems, and also access to exotic quantum phenomena like magnetic field-induced spontaneous decay of magnons in the square lattice antiferromagnet \cite{Zhitomirsky1999, Syljuasen2008}. For this purpose, the development of new synthetic routes for the growth of large single crystals will be necessary. Further understanding of the complex crystal structure of $\alpha$ may also be gleaned from single crystal neutron diffraction measurements combined with thermogravimetric analysis. As several crystallographic studies on \textit{AM}(C$_2$O$_4$)$\cdot$2H$_2$O (\textit{A} = NH$_4$\cite{English1993,Sheu1996, Kalinnikov1969}, Na\cite{Drew1977}, Rb\cite{Drew1977}, K\cite{Kalinnikov1969,Eve1966, Stahler1905}, Cs\cite{Drew1977a}, and \textit{M} = Ti, In\cite{Bulc1983}) already exist, it may prove fruitful to explore how the various energy scales involved in this family of materials can be tuned by varying the monovalent ion site. Finally, the roles of spin and orbital contributions to the magnetic properties of these models could be explored by varying the transition metal ions contained within the coordination framework from Ti$^{3+}$ ($S=1/2$) to V\textsuperscript{3+} ($S=1$), Mo\textsuperscript{3+} ($S=3/2$), or Fe\textsuperscript{3+} ($S=5/2$). 

\section{Acknowledgements}
\noindent Provision of a PhD studentship to A.H.A. by the University of Liverpool and the Science and Technology Facilities Council (STFC) is gratefully acknowledged. The work of T.L. was funded by the University of St Andrews and China Scholarship Council (CSC) joint scholarship (201606280032). A.T. was funded by the Federal Ministry of Education and Research through the Sofja Kovalevskaya Award of Alexander von Humboldt Foundation. Work at St Andrews was supported by the Leverhulme Trust (RPG-2013-343). The authors are also grateful to the STFC for access to neutron beamtime at ISIS and thank Dr.~Gavin~Stenning for aiding with SQUID and specific heat measurements at the Materials Characterization Laboratory, ISIS. We thank Dr.~Manh~Duc~Le (ISIS), Dr.~Ross~Stewart (ISIS), Dr.~Craig~Robertson (Uni. Liverpool), Dr.~Samantha~Chong (Uni. Liverpool), and Dr.~Hamish~Yeung (Uni. Birmingham)  for useful discussions.

\bibliography{bibliography}{}

\begin{thebibliography}{93}%
\makeatletter
\providecommand \@ifxundefined [1]{%
 \@ifx{#1\undefined}
}%
\providecommand \@ifnum [1]{%
 \ifnum #1\expandafter \@firstoftwo
 \else \expandafter \@secondoftwo
 \fi
}%
\providecommand \@ifx [1]{%
 \ifx #1\expandafter \@firstoftwo
 \else \expandafter \@secondoftwo
 \fi
}%
\providecommand \natexlab [1]{#1}%
\providecommand \enquote  [1]{``#1''}%
\providecommand \bibnamefont  [1]{#1}%
\providecommand \bibfnamefont [1]{#1}%
\providecommand \citenamefont [1]{#1}%
\providecommand \href@noop [0]{\@secondoftwo}%
\providecommand \href [0]{\begingroup \@sanitize@url \@href}%
\providecommand \@href[1]{\@@startlink{#1}\@@href}%
\providecommand \@@href[1]{\endgroup#1\@@endlink}%
\providecommand \@sanitize@url [0]{\catcode `\\12\catcode `\$12\catcode
  `\&12\catcode `\#12\catcode `\^12\catcode `\_12\catcode `\%12\relax}%
\providecommand \@@startlink[1]{}%
\providecommand \@@endlink[0]{}%
\providecommand \url  [0]{\begingroup\@sanitize@url \@url }%
\providecommand \@url [1]{\endgroup\@href {#1}{\urlprefix }}%
\providecommand \urlprefix  [0]{URL }%
\providecommand \Eprint [0]{\href }%
\providecommand \doibase [0]{https://doi.org/}%
\providecommand \selectlanguage [0]{\@gobble}%
\providecommand \bibinfo  [0]{\@secondoftwo}%
\providecommand \bibfield  [0]{\@secondoftwo}%
\providecommand \translation [1]{[#1]}%
\providecommand \BibitemOpen [0]{}%
\providecommand \bibitemStop [0]{}%
\providecommand \bibitemNoStop [0]{.\EOS\space}%
\providecommand \EOS [0]{\spacefactor3000\relax}%
\providecommand \BibitemShut  [1]{\csname bibitem#1\endcsname}%
\let\auto@bib@innerbib\@empty
\bibitem [{\citenamefont {Harrison}(2004)}]{Harrison2004}%
  \BibitemOpen
  \bibfield  {author} {\bibinfo {author} {\bibfnamefont {A.}~\bibnamefont
  {Harrison}},\ }\bibfield  {title} {\bibinfo {title} {{First catch your hare:
  the design and synthesis of frustrated magnets}},\ }\href
  {https://doi.org/10.1088/0953-8984/16/11/001} {\bibfield  {journal} {\bibinfo
   {journal} {J. Phys. Condens. Matter}\ }\textbf {\bibinfo {volume} {16}},\
  \bibinfo {pages} {S553} (\bibinfo {year} {2004})}\BibitemShut {NoStop}%
\bibitem [{\citenamefont {Vasiliev}\ \emph {et~al.}(2018)\citenamefont
  {Vasiliev}, \citenamefont {Volkova}, \citenamefont {Zvereva},\ and\
  \citenamefont {Markina}}]{Vasiliev2018}%
  \BibitemOpen
  \bibfield  {author} {\bibinfo {author} {\bibfnamefont {A.}~\bibnamefont
  {Vasiliev}}, \bibinfo {author} {\bibfnamefont {O.}~\bibnamefont {Volkova}},
  \bibinfo {author} {\bibfnamefont {E.}~\bibnamefont {Zvereva}},\ and\ \bibinfo
  {author} {\bibfnamefont {M.}~\bibnamefont {Markina}},\ }\bibfield  {title}
  {\bibinfo {title} {{Milestones of low-D quantum magnetism}},\ }\href
  {https://doi.org/10.1038/s41535-018-0090-7} {\bibfield  {journal} {\bibinfo
  {journal} {npj Quantum Mater.}\ }\textbf {\bibinfo {volume} {3}},\ \bibinfo
  {pages} {18} (\bibinfo {year} {2018})}\BibitemShut {NoStop}%
\bibitem [{\citenamefont {Rau}\ \emph {et~al.}(2016)\citenamefont {Rau},
  \citenamefont {Lee},\ and\ \citenamefont {Kee}}]{Rau2016}%
  \BibitemOpen
  \bibfield  {author} {\bibinfo {author} {\bibfnamefont {J.~G.}\ \bibnamefont
  {Rau}}, \bibinfo {author} {\bibfnamefont {E.~K.-H.}\ \bibnamefont {Lee}},\
  and\ \bibinfo {author} {\bibfnamefont {H.-Y.}\ \bibnamefont {Kee}},\
  }\bibfield  {title} {\bibinfo {title} {{Spin-orbit physics giving rise to
  novel phases in correlated systems: iridates and related materials}},\ }\href
  {https://doi.org/10.1146/annurev-conmatphys-031115-011319} {\bibfield
  {journal} {\bibinfo  {journal} {Annu. Rev. Condens. Matter Phys}\ }\textbf
  {\bibinfo {volume} {7}},\ \bibinfo {pages} {195} (\bibinfo {year}
  {2016})}\BibitemShut {NoStop}%
\bibitem [{\citenamefont {Shannon}\ \emph {et~al.}(2004)\citenamefont
  {Shannon}, \citenamefont {Schmidt}, \citenamefont {Penc},\ and\ \citenamefont
  {Thalmeier}}]{Shannon2004}%
  \BibitemOpen
  \bibfield  {author} {\bibinfo {author} {\bibfnamefont {N.}~\bibnamefont
  {Shannon}}, \bibinfo {author} {\bibfnamefont {B.}~\bibnamefont {Schmidt}},
  \bibinfo {author} {\bibfnamefont {K.}~\bibnamefont {Penc}},\ and\ \bibinfo
  {author} {\bibfnamefont {P.}~\bibnamefont {Thalmeier}},\ }\bibfield  {title}
  {\bibinfo {title} {{Finite temperature properties and frustrated
  ferromagnetism in a square lattice Heisenberg model}},\ }\href
  {https://doi.org/10.1140/epjb/e2004-00156-3} {\bibfield  {journal} {\bibinfo
  {journal} {Eur. Phys. J. B}\ }\textbf {\bibinfo {volume} {38}},\ \bibinfo
  {pages} {599} (\bibinfo {year} {2004})}\BibitemShut {NoStop}%
\bibitem [{\citenamefont {Bombardi}\ \emph {et~al.}(2004)\citenamefont
  {Bombardi}, \citenamefont {Rodriguez-Carvajal}, \citenamefont {{Di Matteo}},
  \citenamefont {de~Bergevin}, \citenamefont {Paolasini}, \citenamefont
  {Carretta}, \citenamefont {Millet},\ and\ \citenamefont
  {Caciuffo}}]{Bombardi2004}%
  \BibitemOpen
  \bibfield  {author} {\bibinfo {author} {\bibfnamefont {A.}~\bibnamefont
  {Bombardi}}, \bibinfo {author} {\bibfnamefont {J.}~\bibnamefont
  {Rodriguez-Carvajal}}, \bibinfo {author} {\bibfnamefont {S.}~\bibnamefont
  {{Di Matteo}}}, \bibinfo {author} {\bibfnamefont {F.}~\bibnamefont
  {de~Bergevin}}, \bibinfo {author} {\bibfnamefont {L.}~\bibnamefont
  {Paolasini}}, \bibinfo {author} {\bibfnamefont {P.}~\bibnamefont {Carretta}},
  \bibinfo {author} {\bibfnamefont {P.}~\bibnamefont {Millet}},\ and\ \bibinfo
  {author} {\bibfnamefont {R.}~\bibnamefont {Caciuffo}},\ }\bibfield  {title}
  {\bibinfo {title} {{Direct determination of the magnetic ground state in the
  square lattice $S = 1/2$ antiferromagnet Li$_2$VOSiO$_4$}},\ }\href
  {https://doi.org/10.1103/PhysRevLett.93.027202} {\bibfield  {journal}
  {\bibinfo  {journal} {Phys. Rev. Lett.}\ }\textbf {\bibinfo {volume} {93}},\
  \bibinfo {pages} {27202} (\bibinfo {year} {2004})}\BibitemShut {NoStop}%
\bibitem [{\citenamefont {Nath}\ \emph {et~al.}(2008)\citenamefont {Nath},
  \citenamefont {Tsirlin}, \citenamefont {Rosner},\ and\ \citenamefont
  {Geibel}}]{Nath2008}%
  \BibitemOpen
  \bibfield  {author} {\bibinfo {author} {\bibfnamefont {R.}~\bibnamefont
  {Nath}}, \bibinfo {author} {\bibfnamefont {A.~A.}\ \bibnamefont {Tsirlin}},
  \bibinfo {author} {\bibfnamefont {H.}~\bibnamefont {Rosner}},\ and\ \bibinfo
  {author} {\bibfnamefont {C.}~\bibnamefont {Geibel}},\ }\bibfield  {title}
  {\bibinfo {title} {{Magnetic properties of BaCdVO(PO$_4$)$_2$: A strongly
  frustrated spin-$1/2$ square lattice close to the quantum critical regime}},\
  }\href {https://doi.org/10.1103/PhysRevB.78.064422} {\bibfield  {journal}
  {\bibinfo  {journal} {Phys. Rev. B}\ }\textbf {\bibinfo {volume} {78}},\
  \bibinfo {pages} {64422} (\bibinfo {year} {2008})}\BibitemShut {NoStop}%
\bibitem [{\citenamefont {Tsirlin}\ \emph {et~al.}(2008)\citenamefont
  {Tsirlin}, \citenamefont {Belik}, \citenamefont {Shpanchenko}, \citenamefont
  {Antipov}, \citenamefont {Takayama-Muromachi},\ and\ \citenamefont
  {Rosner}}]{Tsirlin2008}%
  \BibitemOpen
  \bibfield  {author} {\bibinfo {author} {\bibfnamefont {A.~A.}\ \bibnamefont
  {Tsirlin}}, \bibinfo {author} {\bibfnamefont {A.~A.}\ \bibnamefont {Belik}},
  \bibinfo {author} {\bibfnamefont {R.~V.}\ \bibnamefont {Shpanchenko}},
  \bibinfo {author} {\bibfnamefont {E.~V.}\ \bibnamefont {Antipov}}, \bibinfo
  {author} {\bibfnamefont {E.}~\bibnamefont {Takayama-Muromachi}},\ and\
  \bibinfo {author} {\bibfnamefont {H.}~\bibnamefont {Rosner}},\ }\bibfield
  {title} {\bibinfo {title} {{Frustrated spin-$1/2$ square lattice in the
  layered perovskite PbVO$_3$}},\ }\href
  {https://doi.org/10.1103/PhysRevB.77.092402} {\bibfield  {journal} {\bibinfo
  {journal} {Phys. Rev. B}\ }\textbf {\bibinfo {volume} {77}},\ \bibinfo
  {pages} {92402} (\bibinfo {year} {2008})}\BibitemShut {NoStop}%
\bibitem [{\citenamefont {Tsirlin}\ \emph {et~al.}(2011)\citenamefont
  {Tsirlin}, \citenamefont {Nath}, \citenamefont {Abakumov}, \citenamefont
  {Furukawa}, \citenamefont {Johnston}, \citenamefont {Hemmida}, \citenamefont
  {{Krug von Nidda}}, \citenamefont {Loidl}, \citenamefont {Geibel},\ and\
  \citenamefont {Rosner}}]{Tsirlin2011}%
  \BibitemOpen
  \bibfield  {author} {\bibinfo {author} {\bibfnamefont {A.~A.}\ \bibnamefont
  {Tsirlin}}, \bibinfo {author} {\bibfnamefont {R.}~\bibnamefont {Nath}},
  \bibinfo {author} {\bibfnamefont {A.~M.}\ \bibnamefont {Abakumov}}, \bibinfo
  {author} {\bibfnamefont {Y.}~\bibnamefont {Furukawa}}, \bibinfo {author}
  {\bibfnamefont {D.~C.}\ \bibnamefont {Johnston}}, \bibinfo {author}
  {\bibfnamefont {M.}~\bibnamefont {Hemmida}}, \bibinfo {author} {\bibfnamefont
  {H.-A.}\ \bibnamefont {{Krug von Nidda}}}, \bibinfo {author} {\bibfnamefont
  {A.}~\bibnamefont {Loidl}}, \bibinfo {author} {\bibfnamefont
  {C.}~\bibnamefont {Geibel}},\ and\ \bibinfo {author} {\bibfnamefont
  {H.}~\bibnamefont {Rosner}},\ }\bibfield  {title} {\bibinfo {title} {{Phase
  separation and frustrated square lattice magnetism of
  Na$_{1.5}$VOPO$_4$F$_{0.5}$}},\ }\href
  {https://doi.org/10.1103/PhysRevB.84.014429} {\bibfield  {journal} {\bibinfo
  {journal} {Phys. Rev. B}\ }\textbf {\bibinfo {volume} {84}},\ \bibinfo
  {pages} {14429} (\bibinfo {year} {2011})}\BibitemShut {NoStop}%
\bibitem [{\citenamefont {Tsirlin}\ \emph {et~al.}(2013)\citenamefont
  {Tsirlin}, \citenamefont {Janson}, \citenamefont {Lebernegg},\ and\
  \citenamefont {Rosner}}]{Tsirlin2013}%
  \BibitemOpen
  \bibfield  {author} {\bibinfo {author} {\bibfnamefont {A.}~\bibnamefont
  {Tsirlin}}, \bibinfo {author} {\bibfnamefont {O.}~\bibnamefont {Janson}},
  \bibinfo {author} {\bibfnamefont {S.}~\bibnamefont {Lebernegg}},\ and\
  \bibinfo {author} {\bibfnamefont {H.}~\bibnamefont {Rosner}},\ }\bibfield
  {title} {\bibinfo {title} {{Square-lattice magnetism of diaboleite
  Pb$_2$Cu(OH)$_4$Cl$_2$}},\ }\href
  {https://link.aps.org/doi/10.1103/PhysRevB.87.064404} {\bibfield  {journal}
  {\bibinfo  {journal} {Phys. Rev. B}\ }\textbf {\bibinfo {volume} {87}}
  (\bibinfo {year} {2013})}\BibitemShut {NoStop}%
\bibitem [{\citenamefont {Yang}\ \emph {et~al.}(2017)\citenamefont {Yang},
  \citenamefont {Jeong}, \citenamefont {Babkevich}, \citenamefont {Katukuri},
  \citenamefont {N{\'{a}}fr{\'{a}}di}, \citenamefont {Shaik}, \citenamefont
  {Magrez}, \citenamefont {Berger}, \citenamefont {Schefer}, \citenamefont
  {Ressouche}, \citenamefont {Kriener}, \citenamefont {{\v{Z}}ivkovi{\'{c}}},
  \citenamefont {Yazyev}, \citenamefont {Forr{\'{o}}},\ and\ \citenamefont
  {R{\o}nnow}}]{Yang2017}%
  \BibitemOpen
  \bibfield  {author} {\bibinfo {author} {\bibfnamefont {L.}~\bibnamefont
  {Yang}}, \bibinfo {author} {\bibfnamefont {M.}~\bibnamefont {Jeong}},
  \bibinfo {author} {\bibfnamefont {P.}~\bibnamefont {Babkevich}}, \bibinfo
  {author} {\bibfnamefont {V.~M.}\ \bibnamefont {Katukuri}}, \bibinfo {author}
  {\bibfnamefont {B.}~\bibnamefont {N{\'{a}}fr{\'{a}}di}}, \bibinfo {author}
  {\bibfnamefont {N.~E.}\ \bibnamefont {Shaik}}, \bibinfo {author}
  {\bibfnamefont {A.}~\bibnamefont {Magrez}}, \bibinfo {author} {\bibfnamefont
  {H.}~\bibnamefont {Berger}}, \bibinfo {author} {\bibfnamefont
  {J.}~\bibnamefont {Schefer}}, \bibinfo {author} {\bibfnamefont
  {E.}~\bibnamefont {Ressouche}}, \bibinfo {author} {\bibfnamefont
  {M.}~\bibnamefont {Kriener}}, \bibinfo {author} {\bibfnamefont
  {I.}~\bibnamefont {{\v{Z}}ivkovi{\'{c}}}}, \bibinfo {author} {\bibfnamefont
  {O.~V.}\ \bibnamefont {Yazyev}}, \bibinfo {author} {\bibfnamefont
  {L.}~\bibnamefont {Forr{\'{o}}}},\ and\ \bibinfo {author} {\bibfnamefont
  {H.~M.}\ \bibnamefont {R{\o}nnow}},\ }\bibfield  {title} {\bibinfo {title}
  {{J$_1$-J$_2$ square lattice antiferromagnetism in the orbitally quenched
  insulator MoOPO$_4$}},\ }\href {https://doi.org/10.1103/PhysRevB.96.024445}
  {\bibfield  {journal} {\bibinfo  {journal} {Phys. Rev. B}\ }\textbf {\bibinfo
  {volume} {96}},\ \bibinfo {pages} {24445} (\bibinfo {year}
  {2017})}\BibitemShut {NoStop}%
\bibitem [{\citenamefont {Ishikawa}\ \emph {et~al.}(2017)\citenamefont
  {Ishikawa}, \citenamefont {Nakamura}, \citenamefont {Yoshida}, \citenamefont
  {Takigawa}, \citenamefont {Babkevich}, \citenamefont {Qureshi}, \citenamefont
  {R{\o}nnow}, \citenamefont {Yajima},\ and\ \citenamefont
  {Hiroi}}]{Ishikawa2017}%
  \BibitemOpen
  \bibfield  {author} {\bibinfo {author} {\bibfnamefont {H.}~\bibnamefont
  {Ishikawa}}, \bibinfo {author} {\bibfnamefont {N.}~\bibnamefont {Nakamura}},
  \bibinfo {author} {\bibfnamefont {M.}~\bibnamefont {Yoshida}}, \bibinfo
  {author} {\bibfnamefont {M.}~\bibnamefont {Takigawa}}, \bibinfo {author}
  {\bibfnamefont {P.}~\bibnamefont {Babkevich}}, \bibinfo {author}
  {\bibfnamefont {N.}~\bibnamefont {Qureshi}}, \bibinfo {author} {\bibfnamefont
  {H.~M.}\ \bibnamefont {R{\o}nnow}}, \bibinfo {author} {\bibfnamefont
  {T.}~\bibnamefont {Yajima}},\ and\ \bibinfo {author} {\bibfnamefont
  {Z.}~\bibnamefont {Hiroi}},\ }\bibfield  {title} {\bibinfo {title}
  {{J$_1$-J$_2$ square-lattice Heisenberg antiferromagnets with 4d$^1$ spins:
  \textit{A}MoOPO$_4$Cl (\textit{A}=K,Rb)}},\ }\href
  {https://doi.org/10.1103/PhysRevB.95.064408} {\bibfield  {journal} {\bibinfo
  {journal} {Phys. Rev. B}\ }\textbf {\bibinfo {volume} {95}},\ \bibinfo
  {pages} {64408} (\bibinfo {year} {2017})}\BibitemShut {NoStop}%
\bibitem [{\citenamefont {Mustonen}\ \emph
  {et~al.}(2018{\natexlab{a}})\citenamefont {Mustonen}, \citenamefont {Vasala},
  \citenamefont {Schmidt}, \citenamefont {Sadrollahi}, \citenamefont {Walker},
  \citenamefont {Terasaki}, \citenamefont {Litterst}, \citenamefont
  {Baggio-Saitovitch},\ and\ \citenamefont {Karppinen}}]{Mustonen2018a}%
  \BibitemOpen
  \bibfield  {author} {\bibinfo {author} {\bibfnamefont {O.}~\bibnamefont
  {Mustonen}}, \bibinfo {author} {\bibfnamefont {S.}~\bibnamefont {Vasala}},
  \bibinfo {author} {\bibfnamefont {K.~P.}\ \bibnamefont {Schmidt}}, \bibinfo
  {author} {\bibfnamefont {E.}~\bibnamefont {Sadrollahi}}, \bibinfo {author}
  {\bibfnamefont {H.~C.}\ \bibnamefont {Walker}}, \bibinfo {author}
  {\bibfnamefont {I.}~\bibnamefont {Terasaki}}, \bibinfo {author}
  {\bibfnamefont {F.~J.}\ \bibnamefont {Litterst}}, \bibinfo {author}
  {\bibfnamefont {E.}~\bibnamefont {Baggio-Saitovitch}},\ and\ \bibinfo
  {author} {\bibfnamefont {M.}~\bibnamefont {Karppinen}},\ }\bibfield  {title}
  {\bibinfo {title} {{Tuning the $S = 1/2$ square-lattice antiferromagnet
  Sr$_2$Cu(Te$_{\mathit{(1-x)}}$W$_{\mathit{x}}$)O$_6$ from N\'eel order to
  quantum disorder to columnar order}},\ }\href
  {https://doi.org/10.1103/PhysRevB.98.064411} {\bibfield  {journal} {\bibinfo
  {journal} {Phys. Rev. B}\ }\textbf {\bibinfo {volume} {98}},\ \bibinfo
  {pages} {64411} (\bibinfo {year} {2018}{\natexlab{a}})}\BibitemShut {NoStop}%
\bibitem [{\citenamefont {Mustonen}\ \emph {et~al.}(2019)\citenamefont
  {Mustonen}, \citenamefont {Vasala}, \citenamefont {Mutch}, \citenamefont
  {Thomas}, \citenamefont {Stenning}, \citenamefont {Baggio-Saitovitch},
  \citenamefont {Cussen},\ and\ \citenamefont {Karppinen}}]{Mustonen2019}%
  \BibitemOpen
  \bibfield  {author} {\bibinfo {author} {\bibfnamefont {O.}~\bibnamefont
  {Mustonen}}, \bibinfo {author} {\bibfnamefont {S.}~\bibnamefont {Vasala}},
  \bibinfo {author} {\bibfnamefont {H.}~\bibnamefont {Mutch}}, \bibinfo
  {author} {\bibfnamefont {C.~I.}\ \bibnamefont {Thomas}}, \bibinfo {author}
  {\bibfnamefont {G.~B.~G.}\ \bibnamefont {Stenning}}, \bibinfo {author}
  {\bibfnamefont {E.}~\bibnamefont {Baggio-Saitovitch}}, \bibinfo {author}
  {\bibfnamefont {E.~J.}\ \bibnamefont {Cussen}},\ and\ \bibinfo {author}
  {\bibfnamefont {M.}~\bibnamefont {Karppinen}},\ }\bibfield  {title} {\bibinfo
  {title} {{Magnetic interactions in the $S = 1/2$ square-lattice
  antiferromagnets Ba$_2$CuTeO$_6$ and Ba$_2$CuWO$_6$: parent phases of a
  possible spin liquid}},\ }\href {https://doi.org/10.1039/C8CC09479A}
  {\bibfield  {journal} {\bibinfo  {journal} {ChemComm}\ }\textbf {\bibinfo
  {volume} {55}},\ \bibinfo {pages} {1132} (\bibinfo {year}
  {2019})}\BibitemShut {NoStop}%
\bibitem [{\citenamefont {Chandra}\ and\ \citenamefont
  {Doucot}(1988)}]{Chandra1988}%
  \BibitemOpen
  \bibfield  {author} {\bibinfo {author} {\bibfnamefont {P.}~\bibnamefont
  {Chandra}}\ and\ \bibinfo {author} {\bibfnamefont {B.}~\bibnamefont
  {Doucot}},\ }\bibfield  {title} {\bibinfo {title} {{Possible spin-liquid
  state at large \textit{S} for the frustrated square Heisenberg lattice}},\
  }\href {https://doi.org/10.1103/PhysRevB.38.9335} {\bibfield  {journal}
  {\bibinfo  {journal} {Phys. Rev. B}\ }\textbf {\bibinfo {volume} {38}},\
  \bibinfo {pages} {9335} (\bibinfo {year} {1988})}\BibitemShut {NoStop}%
\bibitem [{\citenamefont {Mezzacapo}(2012)}]{Mezzacapo2012}%
  \BibitemOpen
  \bibfield  {author} {\bibinfo {author} {\bibfnamefont {F.}~\bibnamefont
  {Mezzacapo}},\ }\bibfield  {title} {\bibinfo {title} {{Ground-state phase
  diagram of the quantum J$_1$-J$_2$ model on the square lattice}},\ }\href
  {https://doi.org/10.1103/PhysRevB.86.045115} {\bibfield  {journal} {\bibinfo
  {journal} {Phys. Rev. B}\ }\textbf {\bibinfo {volume} {86}},\ \bibinfo
  {pages} {45115} (\bibinfo {year} {2012})}\BibitemShut {NoStop}%
\bibitem [{\citenamefont {Poilblanc}\ \emph {et~al.}(2019)\citenamefont
  {Poilblanc}, \citenamefont {Mambrini},\ and\ \citenamefont
  {Capponi}}]{Poilblanc2019}%
  \BibitemOpen
  \bibfield  {author} {\bibinfo {author} {\bibfnamefont {D.}~\bibnamefont
  {Poilblanc}}, \bibinfo {author} {\bibfnamefont {M.}~\bibnamefont
  {Mambrini}},\ and\ \bibinfo {author} {\bibfnamefont {S.}~\bibnamefont
  {Capponi}},\ }\bibfield  {title} {\bibinfo {title} {{Critical colored-RVB
  states in the frustrated quantum Heisenberg model on the square lattice}},\
  }\href {https://doi.org/10.21468/SciPostPhys.7.4.041} {\bibfield  {journal}
  {\bibinfo  {journal} {SciPost Phys.}\ }\textbf {\bibinfo {volume} {7}},\
  \bibinfo {pages} {41} (\bibinfo {year} {2019})}\BibitemShut {NoStop}%
\bibitem [{\citenamefont {Mustonen}\ \emph
  {et~al.}(2018{\natexlab{b}})\citenamefont {Mustonen}, \citenamefont {Vasala},
  \citenamefont {Sadrollahi}, \citenamefont {Schmidt}, \citenamefont {Baines},
  \citenamefont {Walker}, \citenamefont {Terasaki}, \citenamefont {Litterst},
  \citenamefont {Baggio-Saitovitch},\ and\ \citenamefont
  {Karppinen}}]{Mustonen2018}%
  \BibitemOpen
  \bibfield  {author} {\bibinfo {author} {\bibfnamefont {O.}~\bibnamefont
  {Mustonen}}, \bibinfo {author} {\bibfnamefont {S.}~\bibnamefont {Vasala}},
  \bibinfo {author} {\bibfnamefont {E.}~\bibnamefont {Sadrollahi}}, \bibinfo
  {author} {\bibfnamefont {K.~P.}\ \bibnamefont {Schmidt}}, \bibinfo {author}
  {\bibfnamefont {C.}~\bibnamefont {Baines}}, \bibinfo {author} {\bibfnamefont
  {H.~C.}\ \bibnamefont {Walker}}, \bibinfo {author} {\bibfnamefont
  {I.}~\bibnamefont {Terasaki}}, \bibinfo {author} {\bibfnamefont {F.~J.}\
  \bibnamefont {Litterst}}, \bibinfo {author} {\bibfnamefont {E.}~\bibnamefont
  {Baggio-Saitovitch}},\ and\ \bibinfo {author} {\bibfnamefont
  {M.}~\bibnamefont {Karppinen}},\ }\bibfield  {title} {\bibinfo {title}
  {{Spin-liquid-like state in a spin-$1/2$ square-lattice antiferromagnet
  perovskite induced by \textit{d}$^{10}$–\textit{d}$^0$ cation mixing}},\
  }\href {https://doi.org/10.1038/s41467-018-03435-1} {\bibfield  {journal}
  {\bibinfo  {journal} {Nat. Commun.}\ }\textbf {\bibinfo {volume} {9}},\
  \bibinfo {pages} {1085} (\bibinfo {year} {2018}{\natexlab{b}})}\BibitemShut
  {NoStop}%
\bibitem [{\citenamefont {Skoulatos}\ \emph {et~al.}(2019)\citenamefont
  {Skoulatos}, \citenamefont {Rucker}, \citenamefont {Nilsen}, \citenamefont
  {Bertin}, \citenamefont {Pomjakushina}, \citenamefont {Ollivier},
  \citenamefont {Schneidewind}, \citenamefont {Georgii}, \citenamefont
  {Zaharko}, \citenamefont {Keller}, \citenamefont {R{\"{u}}egg}, \citenamefont
  {Pfleiderer}, \citenamefont {Schmidt}, \citenamefont {Shannon}, \citenamefont
  {Kriele}, \citenamefont {Senyshyn},\ and\ \citenamefont
  {Smerald}}]{Skoulatos2019}%
  \BibitemOpen
  \bibfield  {author} {\bibinfo {author} {\bibfnamefont {M.}~\bibnamefont
  {Skoulatos}}, \bibinfo {author} {\bibfnamefont {F.}~\bibnamefont {Rucker}},
  \bibinfo {author} {\bibfnamefont {G.~J.}\ \bibnamefont {Nilsen}}, \bibinfo
  {author} {\bibfnamefont {A.}~\bibnamefont {Bertin}}, \bibinfo {author}
  {\bibfnamefont {E.}~\bibnamefont {Pomjakushina}}, \bibinfo {author}
  {\bibfnamefont {J.}~\bibnamefont {Ollivier}}, \bibinfo {author}
  {\bibfnamefont {A.}~\bibnamefont {Schneidewind}}, \bibinfo {author}
  {\bibfnamefont {R.}~\bibnamefont {Georgii}}, \bibinfo {author} {\bibfnamefont
  {O.}~\bibnamefont {Zaharko}}, \bibinfo {author} {\bibfnamefont
  {L.}~\bibnamefont {Keller}}, \bibinfo {author} {\bibfnamefont
  {C.}~\bibnamefont {R{\"{u}}egg}}, \bibinfo {author} {\bibfnamefont
  {C.}~\bibnamefont {Pfleiderer}}, \bibinfo {author} {\bibfnamefont
  {B.}~\bibnamefont {Schmidt}}, \bibinfo {author} {\bibfnamefont
  {N.}~\bibnamefont {Shannon}}, \bibinfo {author} {\bibfnamefont
  {A.}~\bibnamefont {Kriele}}, \bibinfo {author} {\bibfnamefont
  {A.}~\bibnamefont {Senyshyn}},\ and\ \bibinfo {author} {\bibfnamefont
  {A.}~\bibnamefont {Smerald}},\ }\bibfield  {title} {\bibinfo {title}
  {{Putative spin-nematic phase in BaCdVO(PO$_4$)$_2$}},\ }\href
  {https://doi.org/10.1103/PhysRevB.100.014405} {\bibfield  {journal} {\bibinfo
   {journal} {Phys. Rev. B}\ }\textbf {\bibinfo {volume} {100}},\ \bibinfo
  {pages} {14405} (\bibinfo {year} {2019})}\BibitemShut {NoStop}%
\bibitem [{\citenamefont {Povarov}\ \emph {et~al.}(2019)\citenamefont
  {Povarov}, \citenamefont {Bhartiya}, \citenamefont {Yan},\ and\ \citenamefont
  {Zheludev}}]{Povarov2019}%
  \BibitemOpen
  \bibfield  {author} {\bibinfo {author} {\bibfnamefont {K.~Y.}\ \bibnamefont
  {Povarov}}, \bibinfo {author} {\bibfnamefont {V.~K.}\ \bibnamefont
  {Bhartiya}}, \bibinfo {author} {\bibfnamefont {Z.}~\bibnamefont {Yan}},\ and\
  \bibinfo {author} {\bibfnamefont {A.}~\bibnamefont {Zheludev}},\ }\bibfield
  {title} {\bibinfo {title} {{Thermodynamics of a frustrated quantum magnet on
  a square lattice}},\ }\href {https://doi.org/10.1103/PhysRevB.99.024413}
  {\bibfield  {journal} {\bibinfo  {journal} {Phys. Rev. B}\ }\textbf {\bibinfo
  {volume} {99}},\ \bibinfo {pages} {24413} (\bibinfo {year}
  {2019})}\BibitemShut {NoStop}%
\bibitem [{\citenamefont {Smerald}(2020)}]{Smerald}%
  \BibitemOpen
  \bibfield  {author} {\bibinfo {author} {\bibfnamefont {A.}~\bibnamefont
  {Smerald}},\ }\bibfield  {title} {\bibinfo {title} {{Magnon binding in
  BaCdVO(PO$_4$)$_2$}},\ }\href {https://arxiv.org/abs/2003.12747} {\bibfield
  {journal} {\bibinfo  {journal} {arXiv:2003.12747 [cond-mat.str-el]}\ }
  (\bibinfo {year} {2020})}\BibitemShut {NoStop}%
\bibitem [{\citenamefont {Danu}\ \emph {et~al.}(2016)\citenamefont {Danu},
  \citenamefont {Nambiar},\ and\ \citenamefont {Ganesh}}]{Danu2016}%
  \BibitemOpen
  \bibfield  {author} {\bibinfo {author} {\bibfnamefont {B.}~\bibnamefont
  {Danu}}, \bibinfo {author} {\bibfnamefont {G.}~\bibnamefont {Nambiar}},\ and\
  \bibinfo {author} {\bibfnamefont {R.}~\bibnamefont {Ganesh}},\ }\bibfield
  {title} {\bibinfo {title} {{Extended degeneracy and order by disorder in the
  square lattice J$_1$-J$_2$-J$_3$ model}},\ }\href
  {https://doi.org/10.1103/PhysRevB.94.094438} {\bibfield  {journal} {\bibinfo
  {journal} {Phys. Rev. B}\ }\textbf {\bibinfo {volume} {94}},\ \bibinfo
  {pages} {94438} (\bibinfo {year} {2016})}\BibitemShut {NoStop}%
\bibitem [{\citenamefont {Ge}\ \emph {et~al.}(2017)\citenamefont {Ge},
  \citenamefont {Flynn}, \citenamefont {Paddison}, \citenamefont {Stone},
  \citenamefont {Calder}, \citenamefont {Subramanian}, \citenamefont
  {Ramirez},\ and\ \citenamefont {Mourigal}}]{Ge2017}%
  \BibitemOpen
  \bibfield  {author} {\bibinfo {author} {\bibfnamefont {L.}~\bibnamefont
  {Ge}}, \bibinfo {author} {\bibfnamefont {J.}~\bibnamefont {Flynn}}, \bibinfo
  {author} {\bibfnamefont {J.~A.~M.}\ \bibnamefont {Paddison}}, \bibinfo
  {author} {\bibfnamefont {M.~B.}\ \bibnamefont {Stone}}, \bibinfo {author}
  {\bibfnamefont {S.}~\bibnamefont {Calder}}, \bibinfo {author} {\bibfnamefont
  {M.~A.}\ \bibnamefont {Subramanian}}, \bibinfo {author} {\bibfnamefont
  {A.~P.}\ \bibnamefont {Ramirez}},\ and\ \bibinfo {author} {\bibfnamefont
  {M.}~\bibnamefont {Mourigal}},\ }\bibfield  {title} {\bibinfo {title} {{Spin
  order and dynamics in the diamond-lattice Heisenberg antiferromagnets
  CuRh$_2$O$_4$ and CoRh$_2$O$_4$}},\ }\href
  {https://doi.org/10.1103/PhysRevB.96.064413} {\bibfield  {journal} {\bibinfo
  {journal} {Phys. Rev. B}\ }\textbf {\bibinfo {volume} {96}},\ \bibinfo
  {pages} {64413} (\bibinfo {year} {2017})}\BibitemShut {NoStop}%
\bibitem [{\citenamefont {MacDougall}\ \emph {et~al.}(2011)\citenamefont
  {MacDougall}, \citenamefont {Gout}, \citenamefont {Zarestky}, \citenamefont
  {Ehlers}, \citenamefont {Podlesnyak}, \citenamefont {McGuire}, \citenamefont
  {Mandrus},\ and\ \citenamefont {Nagler}}]{MacDougall2011}%
  \BibitemOpen
  \bibfield  {author} {\bibinfo {author} {\bibfnamefont {G.~J.}\ \bibnamefont
  {MacDougall}}, \bibinfo {author} {\bibfnamefont {D.}~\bibnamefont {Gout}},
  \bibinfo {author} {\bibfnamefont {J.~L.}\ \bibnamefont {Zarestky}}, \bibinfo
  {author} {\bibfnamefont {G.}~\bibnamefont {Ehlers}}, \bibinfo {author}
  {\bibfnamefont {A.}~\bibnamefont {Podlesnyak}}, \bibinfo {author}
  {\bibfnamefont {M.~A.}\ \bibnamefont {McGuire}}, \bibinfo {author}
  {\bibfnamefont {D.}~\bibnamefont {Mandrus}},\ and\ \bibinfo {author}
  {\bibfnamefont {S.~E.}\ \bibnamefont {Nagler}},\ }\bibfield  {title}
  {\bibinfo {title} {{Kinetically inhibited order in a diamond-lattice
  antiferromagnet}},\ }\href {https://doi.org/10.1073/pnas.1107861108}
  {\bibfield  {journal} {\bibinfo  {journal} {Proc. Natl. Acad. Sci. U.S.A.}\
  }\textbf {\bibinfo {volume} {108}},\ \bibinfo {pages} {15693 LP } (\bibinfo
  {year} {2011})}\BibitemShut {NoStop}%
\bibitem [{\citenamefont {Gao}\ \emph {et~al.}(2017)\citenamefont {Gao},
  \citenamefont {Zaharko}, \citenamefont {Tsurkan}, \citenamefont {Su},
  \citenamefont {White}, \citenamefont {Tucker}, \citenamefont {Roessli},
  \citenamefont {Bourdarot}, \citenamefont {Sibille}, \citenamefont
  {Chernyshov}, \citenamefont {Fennell}, \citenamefont {Loidl},\ and\
  \citenamefont {R{\"{u}}egg}}]{Gao2017}%
  \BibitemOpen
  \bibfield  {author} {\bibinfo {author} {\bibfnamefont {S.}~\bibnamefont
  {Gao}}, \bibinfo {author} {\bibfnamefont {O.}~\bibnamefont {Zaharko}},
  \bibinfo {author} {\bibfnamefont {V.}~\bibnamefont {Tsurkan}}, \bibinfo
  {author} {\bibfnamefont {Y.}~\bibnamefont {Su}}, \bibinfo {author}
  {\bibfnamefont {J.~S.}\ \bibnamefont {White}}, \bibinfo {author}
  {\bibfnamefont {G.}~\bibnamefont {Tucker}}, \bibinfo {author} {\bibfnamefont
  {B.}~\bibnamefont {Roessli}}, \bibinfo {author} {\bibfnamefont
  {F.}~\bibnamefont {Bourdarot}}, \bibinfo {author} {\bibfnamefont
  {R.}~\bibnamefont {Sibille}}, \bibinfo {author} {\bibfnamefont
  {D.}~\bibnamefont {Chernyshov}}, \bibinfo {author} {\bibfnamefont
  {T.}~\bibnamefont {Fennell}}, \bibinfo {author} {\bibfnamefont
  {A.}~\bibnamefont {Loidl}},\ and\ \bibinfo {author} {\bibfnamefont
  {C.}~\bibnamefont {R{\"{u}}egg}},\ }\bibfield  {title} {\bibinfo {title}
  {{Spiral spin-liquid and the emergence of a vortex-like state in
  MnSc$_2$S$_4$}},\ }\href {https://doi.org/10.1038/nphys3914} {\bibfield
  {journal} {\bibinfo  {journal} {Nat. Phys.}\ }\textbf {\bibinfo {volume}
  {13}},\ \bibinfo {pages} {157} (\bibinfo {year} {2017})}\BibitemShut
  {NoStop}%
\bibitem [{\citenamefont {Bergman}\ \emph {et~al.}(2007)\citenamefont
  {Bergman}, \citenamefont {Alicea}, \citenamefont {Gull}, \citenamefont
  {Trebst},\ and\ \citenamefont {Balents}}]{Bergman2007}%
  \BibitemOpen
  \bibfield  {author} {\bibinfo {author} {\bibfnamefont {D.}~\bibnamefont
  {Bergman}}, \bibinfo {author} {\bibfnamefont {J.}~\bibnamefont {Alicea}},
  \bibinfo {author} {\bibfnamefont {E.}~\bibnamefont {Gull}}, \bibinfo {author}
  {\bibfnamefont {S.}~\bibnamefont {Trebst}},\ and\ \bibinfo {author}
  {\bibfnamefont {L.}~\bibnamefont {Balents}},\ }\bibfield  {title} {\bibinfo
  {title} {{Order-by-disorder and spiral spin-liquid in frustrated
  diamond-lattice antiferromagnets}},\ }\href
  {https://doi.org/10.1038/nphys622} {\bibfield  {journal} {\bibinfo  {journal}
  {Nat. Phys.}\ }\textbf {\bibinfo {volume} {3}},\ \bibinfo {pages} {487}
  (\bibinfo {year} {2007})}\BibitemShut {NoStop}%
\bibitem [{\citenamefont {Tristan}\ \emph {et~al.}(2005)\citenamefont
  {Tristan}, \citenamefont {Hemberger}, \citenamefont {Krimmel}, \citenamefont
  {{Krug von Nidda}}, \citenamefont {Tsurkan},\ and\ \citenamefont
  {Loidl}}]{Tristan2005}%
  \BibitemOpen
  \bibfield  {author} {\bibinfo {author} {\bibfnamefont {N.}~\bibnamefont
  {Tristan}}, \bibinfo {author} {\bibfnamefont {J.}~\bibnamefont {Hemberger}},
  \bibinfo {author} {\bibfnamefont {A.}~\bibnamefont {Krimmel}}, \bibinfo
  {author} {\bibfnamefont {H.-A.}\ \bibnamefont {{Krug von Nidda}}}, \bibinfo
  {author} {\bibfnamefont {V.}~\bibnamefont {Tsurkan}},\ and\ \bibinfo {author}
  {\bibfnamefont {A.}~\bibnamefont {Loidl}},\ }\bibfield  {title} {\bibinfo
  {title} {{Geometric frustration in the cubic spinels \textit{M}Al$_2$O$_4$
  (\textit{M}=Co, Fe, and Mn)}},\ }\href
  {https://doi.org/10.1103/PhysRevB.72.174404} {\bibfield  {journal} {\bibinfo
  {journal} {Phys. Rev. B}\ }\textbf {\bibinfo {volume} {72}},\ \bibinfo
  {pages} {174404} (\bibinfo {year} {2005})}\BibitemShut {NoStop}%
\bibitem [{\citenamefont {Chen}\ \emph
  {et~al.}(2009{\natexlab{a}})\citenamefont {Chen}, \citenamefont {Balents},\
  and\ \citenamefont {Schnyder}}]{Chen2009}%
  \BibitemOpen
  \bibfield  {author} {\bibinfo {author} {\bibfnamefont {G.}~\bibnamefont
  {Chen}}, \bibinfo {author} {\bibfnamefont {L.}~\bibnamefont {Balents}},\ and\
  \bibinfo {author} {\bibfnamefont {A.~P.}\ \bibnamefont {Schnyder}},\
  }\bibfield  {title} {\bibinfo {title} {{Spin-orbital singlet and quantum
  critical point on the diamond lattice: FeSc$_2$S$_4$}},\ }\href
  {https://doi.org/10.1103/PhysRevLett.102.096406} {\bibfield  {journal}
  {\bibinfo  {journal} {Phys. Rev. Lett.}\ }\textbf {\bibinfo {volume} {102}},\
  \bibinfo {pages} {96406} (\bibinfo {year} {2009}{\natexlab{a}})}\BibitemShut
  {NoStop}%
\bibitem [{\citenamefont {Chen}\ \emph
  {et~al.}(2009{\natexlab{b}})\citenamefont {Chen}, \citenamefont {Schnyder},\
  and\ \citenamefont {Balents}}]{Chen2009a}%
  \BibitemOpen
  \bibfield  {author} {\bibinfo {author} {\bibfnamefont {G.}~\bibnamefont
  {Chen}}, \bibinfo {author} {\bibfnamefont {A.~P.}\ \bibnamefont {Schnyder}},\
  and\ \bibinfo {author} {\bibfnamefont {L.}~\bibnamefont {Balents}},\
  }\bibfield  {title} {\bibinfo {title} {{Excitation spectrum and magnetic
  field effects in a quantum critical spin-orbital system: The case of
  FeSc$_2$S$_4$}},\ }\href {https://doi.org/10.1103/PhysRevB.80.224409}
  {\bibfield  {journal} {\bibinfo  {journal} {Phys. Rev. B}\ }\textbf {\bibinfo
  {volume} {80}},\ \bibinfo {pages} {224409} (\bibinfo {year}
  {2009}{\natexlab{b}})}\BibitemShut {NoStop}%
\bibitem [{\citenamefont {Tsurkan}\ \emph {et~al.}(2017)\citenamefont
  {Tsurkan}, \citenamefont {Prodan}, \citenamefont {Felea}, \citenamefont
  {Filippova}, \citenamefont {Kravtsov}, \citenamefont {G{\"{u}}nther},
  \citenamefont {Widmann}, \citenamefont {{Krug von Nidda}}, \citenamefont
  {Deisenhofer},\ and\ \citenamefont {Loidl}}]{Tsurkan2017}%
  \BibitemOpen
  \bibfield  {author} {\bibinfo {author} {\bibfnamefont {V.}~\bibnamefont
  {Tsurkan}}, \bibinfo {author} {\bibfnamefont {L.}~\bibnamefont {Prodan}},
  \bibinfo {author} {\bibfnamefont {V.}~\bibnamefont {Felea}}, \bibinfo
  {author} {\bibfnamefont {I.}~\bibnamefont {Filippova}}, \bibinfo {author}
  {\bibfnamefont {V.}~\bibnamefont {Kravtsov}}, \bibinfo {author}
  {\bibfnamefont {A.}~\bibnamefont {G{\"{u}}nther}}, \bibinfo {author}
  {\bibfnamefont {S.}~\bibnamefont {Widmann}}, \bibinfo {author} {\bibfnamefont
  {H.-A.}\ \bibnamefont {{Krug von Nidda}}}, \bibinfo {author} {\bibfnamefont
  {J.}~\bibnamefont {Deisenhofer}},\ and\ \bibinfo {author} {\bibfnamefont
  {A.}~\bibnamefont {Loidl}},\ }\bibfield  {title} {\bibinfo {title}
  {{Structure, magnetic susceptibility, and specific heat of the
  spin-orbital-liquid candidate FeSc$_2$S$_4$: influence of Fe
  off-stoichiometry}},\ }\href {https://doi.org/10.1103/PhysRevB.96.054417}
  {\bibfield  {journal} {\bibinfo  {journal} {Phys. Rev. B}\ }\textbf {\bibinfo
  {volume} {96}},\ \bibinfo {pages} {54417} (\bibinfo {year}
  {2017})}\BibitemShut {NoStop}%
\bibitem [{\citenamefont {Tustain}\ \emph {et~al.}(2019)\citenamefont
  {Tustain}, \citenamefont {Farrar}, \citenamefont {Yao}, \citenamefont
  {Lightfoot}, \citenamefont {da~Silva}, \citenamefont {Telling},\ and\
  \citenamefont {Clark}}]{Tustain2019}%
  \BibitemOpen
  \bibfield  {author} {\bibinfo {author} {\bibfnamefont {K.}~\bibnamefont
  {Tustain}}, \bibinfo {author} {\bibfnamefont {L.}~\bibnamefont {Farrar}},
  \bibinfo {author} {\bibfnamefont {W.}~\bibnamefont {Yao}}, \bibinfo {author}
  {\bibfnamefont {P.}~\bibnamefont {Lightfoot}}, \bibinfo {author}
  {\bibfnamefont {I.}~\bibnamefont {da~Silva}}, \bibinfo {author}
  {\bibfnamefont {M.~T.~F.}\ \bibnamefont {Telling}},\ and\ \bibinfo {author}
  {\bibfnamefont {L.}~\bibnamefont {Clark}},\ }\bibfield  {title} {\bibinfo
  {title} {{Materialization of a geometrically frustrated magnet in a hybrid
  coordination framework: a study of the iron(II) oxalate fluoride framework,
  KFe(C$_2$O$_4$)F}},\ }\href {https://doi.org/10.1021/acs.inorgchem.9b00571}
  {\bibfield  {journal} {\bibinfo  {journal} {Inorg. Chem.}\ }\textbf {\bibinfo
  {volume} {58}},\ \bibinfo {pages} {11971} (\bibinfo {year}
  {2019})}\BibitemShut {NoStop}%
\bibitem [{\citenamefont {Rao}\ \emph {et~al.}(2008)\citenamefont {Rao},
  \citenamefont {Cheetham},\ and\ \citenamefont {Thirumurugan}}]{Rao2008}%
  \BibitemOpen
  \bibfield  {author} {\bibinfo {author} {\bibfnamefont {C.~N.~R.}\
  \bibnamefont {Rao}}, \bibinfo {author} {\bibfnamefont {A.~K.}\ \bibnamefont
  {Cheetham}},\ and\ \bibinfo {author} {\bibfnamefont {A.}~\bibnamefont
  {Thirumurugan}},\ }\bibfield  {title} {\bibinfo {title} {{Hybrid
  inorganic-organic materials: a new family in condensed matter physics}},\
  }\href {https://doi.org/10.1088/0953-8984/20/15/159801} {\bibfield  {journal}
  {\bibinfo  {journal} {J. Phys.: Condens. Matter}\ }\textbf {\bibinfo {volume}
  {20}},\ \bibinfo {pages} {159801} (\bibinfo {year} {2008})}\BibitemShut
  {NoStop}%
\bibitem [{\citenamefont {Zheng}\ \emph {et~al.}(2007)\citenamefont {Zheng},
  \citenamefont {Tong}, \citenamefont {Xue}, \citenamefont {Zhang},
  \citenamefont {Chen}, \citenamefont {Grandjean},\ and\ \citenamefont
  {Long}}]{Zheng2007}%
  \BibitemOpen
  \bibfield  {author} {\bibinfo {author} {\bibfnamefont {Y.-Z.}\ \bibnamefont
  {Zheng}}, \bibinfo {author} {\bibfnamefont {M.-L.}\ \bibnamefont {Tong}},
  \bibinfo {author} {\bibfnamefont {W.}~\bibnamefont {Xue}}, \bibinfo {author}
  {\bibfnamefont {W.-X.}\ \bibnamefont {Zhang}}, \bibinfo {author}
  {\bibfnamefont {X.-M.}\ \bibnamefont {Chen}}, \bibinfo {author}
  {\bibfnamefont {F.}~\bibnamefont {Grandjean}},\ and\ \bibinfo {author}
  {\bibfnamefont {G.}~\bibnamefont {Long}},\ }\bibfield  {title} {\bibinfo
  {title} {{A “star” antiferromagnet: a polymeric iron(III) acetate that
  exhibits both spin frustration and long-range magnetic ordering}},\ }\href
  {https://doi.org/10.1002/anie.200701954} {\bibfield  {journal} {\bibinfo
  {journal} {Angew. Chem. Int. Ed.}\ }\textbf {\bibinfo {volume} {46}},\
  \bibinfo {pages} {6076} (\bibinfo {year} {2007})}\BibitemShut {NoStop}%
\bibitem [{\citenamefont {Tsyrulin}\ \emph {et~al.}(2010)\citenamefont
  {Tsyrulin}, \citenamefont {Xiao}, \citenamefont {Schneidewind}, \citenamefont
  {Link}, \citenamefont {R{\o}nnow}, \citenamefont {Gavilano}, \citenamefont
  {Landee}, \citenamefont {Turnbull},\ and\ \citenamefont
  {Kenzelmann}}]{Tsyrulin2010}%
  \BibitemOpen
  \bibfield  {author} {\bibinfo {author} {\bibfnamefont {N.}~\bibnamefont
  {Tsyrulin}}, \bibinfo {author} {\bibfnamefont {F.}~\bibnamefont {Xiao}},
  \bibinfo {author} {\bibfnamefont {A.}~\bibnamefont {Schneidewind}}, \bibinfo
  {author} {\bibfnamefont {P.}~\bibnamefont {Link}}, \bibinfo {author}
  {\bibfnamefont {H.~M.}\ \bibnamefont {R{\o}nnow}}, \bibinfo {author}
  {\bibfnamefont {J.}~\bibnamefont {Gavilano}}, \bibinfo {author}
  {\bibfnamefont {C.~P.}\ \bibnamefont {Landee}}, \bibinfo {author}
  {\bibfnamefont {M.~M.}\ \bibnamefont {Turnbull}},\ and\ \bibinfo {author}
  {\bibfnamefont {M.}~\bibnamefont {Kenzelmann}},\ }\bibfield  {title}
  {\bibinfo {title} {{Two-dimensional square-lattice $S = 1/2$ antiferromagnet
  Cu(pz)$_2$(ClO$_4$)$_2$}},\ }\href
  {https://link.aps.org/doi/10.1103/PhysRevB.81.134409} {\bibfield  {journal}
  {\bibinfo  {journal} {Phys. Rev. B}\ }\textbf {\bibinfo {volume} {81}}
  (\bibinfo {year} {2010})}\BibitemShut {NoStop}%
\bibitem [{\citenamefont {English}\ and\ \citenamefont
  {Eve}(1993)}]{English1993}%
  \BibitemOpen
  \bibfield  {author} {\bibinfo {author} {\bibfnamefont {R.~B.}\ \bibnamefont
  {English}}\ and\ \bibinfo {author} {\bibfnamefont {D.~J.}\ \bibnamefont
  {Eve}},\ }\bibfield  {title} {\bibinfo {title} {{Ammonium
  di-$\mu$-oxalatotitanate(III) dihydrate: an eight-coordinate, polymeric
  titanium(III) complex}},\ }\href
  {https://doi.org/https://doi.org/10.1016/S0020-1693(00)81660-6} {\bibfield
  {journal} {\bibinfo  {journal} {Inorg. Chim. Acta}\ }\textbf {\bibinfo
  {volume} {203}},\ \bibinfo {pages} {219} (\bibinfo {year}
  {1993})}\BibitemShut {NoStop}%
\bibitem [{\citenamefont {Drew}\ and\ \citenamefont
  {Eve}(1977{\natexlab{a}})}]{Drew1977}%
  \BibitemOpen
  \bibfield  {author} {\bibinfo {author} {\bibfnamefont {M.~G.~B.}\
  \bibnamefont {Drew}}\ and\ \bibinfo {author} {\bibfnamefont {D.~J.}\
  \bibnamefont {Eve}},\ }\bibfield  {title} {\bibinfo {title} {{Observations on
  the spectra and structures of seven-and eight-co-ordinate oxalato complexes
  of titanium(III)}},\ }\href
  {https://doi.org/https://doi.org/10.1016/S0020-1693(00)95667-6} {\bibfield
  {journal} {\bibinfo  {journal} {Inorg. Chim. Acta}\ }\textbf {\bibinfo
  {volume} {25}},\ \bibinfo {pages} {L111} (\bibinfo {year}
  {1977}{\natexlab{a}})}\BibitemShut {NoStop}%
\bibitem [{\citenamefont {St{\"{a}}hler}(1905)}]{Stahler1905}%
  \BibitemOpen
  \bibfield  {author} {\bibinfo {author} {\bibfnamefont {A.}~\bibnamefont
  {St{\"{a}}hler}},\ }\bibfield  {title} {\bibinfo {title} {{Zur Kenntniss des
  Titans. II. (Zum Theil gemeinsam mit Heinz Wirthwein)}},\ }\href
  {https://doi.org/10.1002/cber.19050380333} {\bibfield  {journal} {\bibinfo
  {journal} {Ber. Dtsch. Chem. Ges.}\ }\textbf {\bibinfo {volume} {38}},\
  \bibinfo {pages} {2619} (\bibinfo {year} {1905})}\BibitemShut {NoStop}%
\bibitem [{\citenamefont {Drew}\ and\ \citenamefont
  {Eve}(1977{\natexlab{b}})}]{Drew1977a}%
  \BibitemOpen
  \bibfield  {author} {\bibinfo {author} {\bibfnamefont {M.~G.~B.}\
  \bibnamefont {Drew}}\ and\ \bibinfo {author} {\bibfnamefont {D.~J.}\
  \bibnamefont {Eve}},\ }\bibfield  {title} {\bibinfo {title} {{Caesium
  triaquabis(oxalato)titanate(III) dihydrate}},\ }\href
  {https://doi.org/10.1107/S0567740877009819} {\bibfield  {journal} {\bibinfo
  {journal} {Acta Crystallogr. B}\ }\textbf {\bibinfo {volume} {33}},\ \bibinfo
  {pages} {2919} (\bibinfo {year} {1977}{\natexlab{b}})}\BibitemShut {NoStop}%
\bibitem [{\citenamefont {Sheldrick}(2015{\natexlab{a}})}]{Sheldrick2015a}%
  \BibitemOpen
  \bibfield  {author} {\bibinfo {author} {\bibfnamefont {G.}~\bibnamefont
  {Sheldrick}},\ }\bibfield  {title} {\bibinfo {title} {{SHELXT - Integrated
  space-group and crystal-structure determination}},\ }\href
  {https://doi.org/10.1107/S2053273314026370} {\bibfield  {journal} {\bibinfo
  {journal} {Acta Crystallogr. A}\ }\textbf {\bibinfo {volume} {71}},\ \bibinfo
  {pages} {3} (\bibinfo {year} {2015}{\natexlab{a}})}\BibitemShut {NoStop}%
\bibitem [{\citenamefont {Sheldrick}(2015{\natexlab{b}})}]{Sheldrick2015}%
  \BibitemOpen
  \bibfield  {author} {\bibinfo {author} {\bibfnamefont {G.}~\bibnamefont
  {Sheldrick}},\ }\bibfield  {title} {\bibinfo {title} {{Crystal structure
  refinement with SHELXL}},\ }\href {https://doi.org/10.1107/S2053229614024218}
  {\bibfield  {journal} {\bibinfo  {journal} {Acta Crystallogr. C}\ }\textbf
  {\bibinfo {volume} {71}},\ \bibinfo {pages} {3} (\bibinfo {year}
  {2015}{\natexlab{b}})}\BibitemShut {NoStop}%
\bibitem [{\citenamefont {Clark}\ \emph {et~al.}(2018)\citenamefont {Clark},
  \citenamefont {Lightfoot}, \citenamefont {Gibbs}, \citenamefont {Nilsen},\
  and\ \citenamefont {Farrar}}]{HRPD_ex}%
  \BibitemOpen
  \bibfield  {author} {\bibinfo {author} {\bibfnamefont {L.}~\bibnamefont
  {Clark}}, \bibinfo {author} {\bibfnamefont {P.}~\bibnamefont {Lightfoot}},
  \bibinfo {author} {\bibfnamefont {A.}~\bibnamefont {Gibbs}}, \bibinfo
  {author} {\bibfnamefont {G.}~\bibnamefont {Nilsen}},\ and\ \bibinfo {author}
  {\bibfnamefont {L.}~\bibnamefont {Farrar}},\ }\bibfield  {title} {\bibinfo
  {title} {{Structural, orbital and magnetic orders in the alpha- and
  beta-phases of $S = 1/2$ KTi(C$_2$O$_4$)$_2\cdot$2H$_2$O - a neutron
  diffraction study}},\ }\href {https://doi.org/10.5286/ISIS.E.RB1810582}
  {\bibfield  {journal} {\bibinfo  {journal} {STFC ISIS Neutron and Muon
  Source}\ } (\bibinfo {year} {2018})}\BibitemShut {NoStop}%
\bibitem [{\citenamefont {Ibberson}(2009)}]{HRPD}%
  \BibitemOpen
  \bibfield  {author} {\bibinfo {author} {\bibfnamefont {R.~M.}\ \bibnamefont
  {Ibberson}},\ }\bibfield  {title} {\bibinfo {title} {{Design and performance
  of the new supermirror guide on HRPD at ISIS}},\ }\href
  {https://doi.org/https://doi.org/10.1016/j.nima.2008.11.066} {\bibfield
  {journal} {\bibinfo  {journal} {Nucl. Instrum. Methods Phys. Res}\ }\textbf
  {\bibinfo {volume} {600}},\ \bibinfo {pages} {47 } (\bibinfo {year}
  {2009})}\BibitemShut {NoStop}%
\bibitem [{\citenamefont {Clark}\ \emph {et~al.}(2020)\citenamefont {Clark},
  \citenamefont {Abdeldaim}, \citenamefont {Tustain},\ and\ \citenamefont
  {Graham}}]{WISH_ex}%
  \BibitemOpen
  \bibfield  {author} {\bibinfo {author} {\bibfnamefont {L.}~\bibnamefont
  {Clark}}, \bibinfo {author} {\bibfnamefont {A.}~\bibnamefont {Abdeldaim}},
  \bibinfo {author} {\bibfnamefont {K.}~\bibnamefont {Tustain}},\ and\ \bibinfo
  {author} {\bibfnamefont {J.}~\bibnamefont {Graham}},\ }\bibfield  {title}
  {\bibinfo {title} {{Magnetic orders in the alpha- and beta-Phases of $S =
  1/2$ KTi(C$_2$O$_4$)$_2$·2H$_2$O}},\ }\href
  {https://doi.org/10.5286/ISIS.E.RB2010102} {\bibfield  {journal} {\bibinfo
  {journal} {STFC ISIS Neutron and Muon Source}\ } (\bibinfo {year}
  {2020})}\BibitemShut {NoStop}%
\bibitem [{\citenamefont {Chapon}\ \emph {et~al.}(2011)\citenamefont {Chapon},
  \citenamefont {Manuel}, \citenamefont {Radaelli}, \citenamefont {Benson},
  \citenamefont {Perrott}, \citenamefont {Ansell}, \citenamefont {Rhodes},
  \citenamefont {Raspino}, \citenamefont {Duxbury}, \citenamefont {Spill},\
  and\ \citenamefont {Norris}}]{WISH}%
  \BibitemOpen
  \bibfield  {author} {\bibinfo {author} {\bibfnamefont {L.~C.}\ \bibnamefont
  {Chapon}}, \bibinfo {author} {\bibfnamefont {P.}~\bibnamefont {Manuel}},
  \bibinfo {author} {\bibfnamefont {P.~G.}\ \bibnamefont {Radaelli}}, \bibinfo
  {author} {\bibfnamefont {C.}~\bibnamefont {Benson}}, \bibinfo {author}
  {\bibfnamefont {L.}~\bibnamefont {Perrott}}, \bibinfo {author} {\bibfnamefont
  {S.}~\bibnamefont {Ansell}}, \bibinfo {author} {\bibfnamefont {N.~J.}\
  \bibnamefont {Rhodes}}, \bibinfo {author} {\bibfnamefont {D.}~\bibnamefont
  {Raspino}}, \bibinfo {author} {\bibfnamefont {D.}~\bibnamefont {Duxbury}},
  \bibinfo {author} {\bibfnamefont {E.}~\bibnamefont {Spill}},\ and\ \bibinfo
  {author} {\bibfnamefont {J.}~\bibnamefont {Norris}},\ }\bibfield  {title}
  {\bibinfo {title} {{Wish: the new powder and single crystal magnetic
  diffractometer on the second target station}},\ }\href
  {https://doi.org/10.1080/10448632.2011.569650} {\bibfield  {journal}
  {\bibinfo  {journal} {Neutron News}\ }\textbf {\bibinfo {volume} {22}},\
  \bibinfo {pages} {22} (\bibinfo {year} {2011})}\BibitemShut {NoStop}%
\bibitem [{\citenamefont {Toby}(2001)}]{GSAS}%
  \BibitemOpen
  \bibfield  {author} {\bibinfo {author} {\bibfnamefont {B.~H.}\ \bibnamefont
  {Toby}},\ }\bibfield  {title} {\bibinfo {title} {Expgui, a graphical user
  interface for gsas},\ }\href {https://doi.org/10.1107/S0021889801002242}
  {\bibfield  {journal} {\bibinfo  {journal} {J. Appl. Crystallogr}\ }\textbf
  {\bibinfo {volume} {34}},\ \bibinfo {pages} {210} (\bibinfo {year}
  {2001})}\BibitemShut {NoStop}%
\bibitem [{\citenamefont {Larson}\ and\ \citenamefont {von
  Dreele}(2004)}]{GSAS2}%
  \BibitemOpen
  \bibfield  {author} {\bibinfo {author} {\bibfnamefont {A.~C.}\ \bibnamefont
  {Larson}}\ and\ \bibinfo {author} {\bibfnamefont {R.~B.}\ \bibnamefont {von
  Dreele}},\ }\href
  {http://www.ccp14.ac.uk/ccp/ccp14/ftp-mirror/gsas/public/gsas/mantemperatureual/GSASManual.pdf}
  {\bibfield  {journal} {\bibinfo  {journal} {Los Alamos National Laboratory
  Report LAUR 86-748}\ } (\bibinfo {year} {2004})}\BibitemShut {NoStop}%
\bibitem [{\citenamefont {Rodríguez-Carvajal}(1993)}]{fullprof}%
  \BibitemOpen
  \bibfield  {author} {\bibinfo {author} {\bibfnamefont {J.}~\bibnamefont
  {Rodríguez-Carvajal}},\ }\bibfield  {title} {\bibinfo {title} {Recent
  advances in magnetic structure determination by neutron powder diffraction},\
  }\href {https://doi.org/https://doi.org/10.1016/0921-4526(93)90108-I}
  {\bibfield  {journal} {\bibinfo  {journal} {Physica B}\ }\textbf {\bibinfo
  {volume} {192}},\ \bibinfo {pages} {55 } (\bibinfo {year}
  {1993})}\BibitemShut {NoStop}%
\bibitem [{\citenamefont {Todo}\ and\ \citenamefont {Kato}(2001)}]{Todo2001}%
  \BibitemOpen
  \bibfield  {author} {\bibinfo {author} {\bibfnamefont {S.}~\bibnamefont
  {Todo}}\ and\ \bibinfo {author} {\bibfnamefont {K.}~\bibnamefont {Kato}},\
  }\bibfield  {title} {\bibinfo {title} {{Cluster Algorithms for General-$S$
  Quantum Spin Systems}},\ }\href
  {https://doi.org/10.1103/PhysRevLett.87.047203} {\bibfield  {journal}
  {\bibinfo  {journal} {Phys. Rev. Lett.}\ }\textbf {\bibinfo {volume} {87}},\
  \bibinfo {pages} {47203} (\bibinfo {year} {2001})}\BibitemShut {NoStop}%
\bibitem [{\citenamefont {Evertz}(2003)}]{Evertz2003}%
  \BibitemOpen
  \bibfield  {author} {\bibinfo {author} {\bibfnamefont {H.~G.}\ \bibnamefont
  {Evertz}},\ }\bibfield  {title} {\bibinfo {title} {{The loop algorithm}},\
  }\href {https://doi.org/10.1080/0001873021000049195} {\bibfield  {journal}
  {\bibinfo  {journal} {Adv. Phys.}\ }\textbf {\bibinfo {volume} {52}},\
  \bibinfo {pages} {1} (\bibinfo {year} {2003})}\BibitemShut {NoStop}%
\bibitem [{\citenamefont {Bauer}\ \emph {et~al.}(2011)\citenamefont {Bauer},
  \citenamefont {Carr}, \citenamefont {Evertz}, \citenamefont {Feiguin},
  \citenamefont {Freire}, \citenamefont {Fuchs}, \citenamefont {Gamper},
  \citenamefont {Gukelberger}, \citenamefont {Gull}, \citenamefont {Guertler},
  \citenamefont {Hehn}, \citenamefont {Igarashi}, \citenamefont {Isakov},
  \citenamefont {Koop}, \citenamefont {Ma}, \citenamefont {Mates},
  \citenamefont {Matsuo}, \citenamefont {Parcollet}, \citenamefont
  {Paw{\l}owski}, \citenamefont {Picon}, \citenamefont {Pollet}, \citenamefont
  {Santos}, \citenamefont {Scarola}, \citenamefont {Schollwöck}, \citenamefont
  {Silva}, \citenamefont {Surer}, \citenamefont {Todo}, \citenamefont {Trebst},
  \citenamefont {Troyer}, \citenamefont {Wall}, \citenamefont {Werner},\ and\
  \citenamefont {Wessel}}]{ALPS}%
  \BibitemOpen
  \bibfield  {author} {\bibinfo {author} {\bibfnamefont {B.}~\bibnamefont
  {Bauer}}, \bibinfo {author} {\bibfnamefont {L.~D.}\ \bibnamefont {Carr}},
  \bibinfo {author} {\bibfnamefont {H.~G.}\ \bibnamefont {Evertz}}, \bibinfo
  {author} {\bibfnamefont {A.}~\bibnamefont {Feiguin}}, \bibinfo {author}
  {\bibfnamefont {J.}~\bibnamefont {Freire}}, \bibinfo {author} {\bibfnamefont
  {S.}~\bibnamefont {Fuchs}}, \bibinfo {author} {\bibfnamefont
  {L.}~\bibnamefont {Gamper}}, \bibinfo {author} {\bibfnamefont
  {J.}~\bibnamefont {Gukelberger}}, \bibinfo {author} {\bibfnamefont
  {E.}~\bibnamefont {Gull}}, \bibinfo {author} {\bibfnamefont {S.}~\bibnamefont
  {Guertler}}, \bibinfo {author} {\bibfnamefont {A.}~\bibnamefont {Hehn}},
  \bibinfo {author} {\bibfnamefont {R.}~\bibnamefont {Igarashi}}, \bibinfo
  {author} {\bibfnamefont {S.~V.}\ \bibnamefont {Isakov}}, \bibinfo {author}
  {\bibfnamefont {D.}~\bibnamefont {Koop}}, \bibinfo {author} {\bibfnamefont
  {P.~N.}\ \bibnamefont {Ma}}, \bibinfo {author} {\bibfnamefont
  {P.}~\bibnamefont {Mates}}, \bibinfo {author} {\bibfnamefont
  {H.}~\bibnamefont {Matsuo}}, \bibinfo {author} {\bibfnamefont
  {O.}~\bibnamefont {Parcollet}}, \bibinfo {author} {\bibfnamefont
  {G.}~\bibnamefont {Paw{\l}owski}}, \bibinfo {author} {\bibfnamefont {J.~D.}\
  \bibnamefont {Picon}}, \bibinfo {author} {\bibfnamefont {L.}~\bibnamefont
  {Pollet}}, \bibinfo {author} {\bibfnamefont {E.}~\bibnamefont {Santos}},
  \bibinfo {author} {\bibfnamefont {V.~W.}\ \bibnamefont {Scarola}}, \bibinfo
  {author} {\bibfnamefont {U.}~\bibnamefont {Schollwöck}}, \bibinfo {author}
  {\bibfnamefont {C.}~\bibnamefont {Silva}}, \bibinfo {author} {\bibfnamefont
  {B.}~\bibnamefont {Surer}}, \bibinfo {author} {\bibfnamefont
  {S.}~\bibnamefont {Todo}}, \bibinfo {author} {\bibfnamefont {S.}~\bibnamefont
  {Trebst}}, \bibinfo {author} {\bibfnamefont {M.}~\bibnamefont {Troyer}},
  \bibinfo {author} {\bibfnamefont {M.~L.}\ \bibnamefont {Wall}}, \bibinfo
  {author} {\bibfnamefont {P.}~\bibnamefont {Werner}},\ and\ \bibinfo {author}
  {\bibfnamefont {S.}~\bibnamefont {Wessel}},\ }\bibfield  {title} {\bibinfo
  {title} {{The {ALPS} project release 2.0: open source software for strongly
  correlated systems}},\ }\href
  {https://doi.org/10.1088/1742-5468/2011/05/p05001} {\bibfield  {journal}
  {\bibinfo  {journal} {J. Stat. Mech. Theory Exp.}\ }\textbf {\bibinfo
  {volume} {2011}},\ \bibinfo {pages} {P05001} (\bibinfo {year}
  {2011})}\BibitemShut {NoStop}%
\bibitem [{\citenamefont {Albuquerque}\ \emph {et~al.}(2007)\citenamefont
  {Albuquerque}, \citenamefont {Alet}, \citenamefont {Corboz}, \citenamefont
  {Dayal}, \citenamefont {Feiguin}, \citenamefont {Fuchs}, \citenamefont
  {Gamper}, \citenamefont {Gull}, \citenamefont {Gürtler}, \citenamefont
  {Honecker}, \citenamefont {Igarashi}, \citenamefont {Körner}, \citenamefont
  {Kozhevnikov}, \citenamefont {Läuchli}, \citenamefont {Manmana},
  \citenamefont {Matsumoto}, \citenamefont {McCulloch}, \citenamefont {Michel},
  \citenamefont {Noack}, \citenamefont {Pawłowski}, \citenamefont {Pollet},
  \citenamefont {Pruschke}, \citenamefont {Schollwöck}, \citenamefont {Todo},
  \citenamefont {Trebst}, \citenamefont {Troyer}, \citenamefont {Werner},\ and\
  \citenamefont {Wessel}}]{ALPS2}%
  \BibitemOpen
  \bibfield  {author} {\bibinfo {author} {\bibfnamefont {A.}~\bibnamefont
  {Albuquerque}}, \bibinfo {author} {\bibfnamefont {F.}~\bibnamefont {Alet}},
  \bibinfo {author} {\bibfnamefont {P.}~\bibnamefont {Corboz}}, \bibinfo
  {author} {\bibfnamefont {P.}~\bibnamefont {Dayal}}, \bibinfo {author}
  {\bibfnamefont {A.}~\bibnamefont {Feiguin}}, \bibinfo {author} {\bibfnamefont
  {S.}~\bibnamefont {Fuchs}}, \bibinfo {author} {\bibfnamefont
  {L.}~\bibnamefont {Gamper}}, \bibinfo {author} {\bibfnamefont
  {E.}~\bibnamefont {Gull}}, \bibinfo {author} {\bibfnamefont {S.}~\bibnamefont
  {Gürtler}}, \bibinfo {author} {\bibfnamefont {A.}~\bibnamefont {Honecker}},
  \bibinfo {author} {\bibfnamefont {R.}~\bibnamefont {Igarashi}}, \bibinfo
  {author} {\bibfnamefont {M.}~\bibnamefont {Körner}}, \bibinfo {author}
  {\bibfnamefont {A.}~\bibnamefont {Kozhevnikov}}, \bibinfo {author}
  {\bibfnamefont {A.}~\bibnamefont {Läuchli}}, \bibinfo {author}
  {\bibfnamefont {S.}~\bibnamefont {Manmana}}, \bibinfo {author} {\bibfnamefont
  {M.}~\bibnamefont {Matsumoto}}, \bibinfo {author} {\bibfnamefont
  {I.}~\bibnamefont {McCulloch}}, \bibinfo {author} {\bibfnamefont
  {F.}~\bibnamefont {Michel}}, \bibinfo {author} {\bibfnamefont
  {R.}~\bibnamefont {Noack}}, \bibinfo {author} {\bibfnamefont
  {G.}~\bibnamefont {Pawłowski}}, \bibinfo {author} {\bibfnamefont
  {L.}~\bibnamefont {Pollet}}, \bibinfo {author} {\bibfnamefont
  {T.}~\bibnamefont {Pruschke}}, \bibinfo {author} {\bibfnamefont
  {U.}~\bibnamefont {Schollwöck}}, \bibinfo {author} {\bibfnamefont
  {S.}~\bibnamefont {Todo}}, \bibinfo {author} {\bibfnamefont {S.}~\bibnamefont
  {Trebst}}, \bibinfo {author} {\bibfnamefont {M.}~\bibnamefont {Troyer}},
  \bibinfo {author} {\bibfnamefont {P.}~\bibnamefont {Werner}},\ and\ \bibinfo
  {author} {\bibfnamefont {S.}~\bibnamefont {Wessel}},\ }\bibfield  {title}
  {\bibinfo {title} {{The ALPS project release 1.3: open-source software for
  strongly correlated systems}},\ }\href
  {https://doi.org/https://doi.org/10.1016/j.jmmm.2006.10.304} {\bibfield
  {journal} {\bibinfo  {journal} {J. Magn. Magn. Mater}\ }\textbf {\bibinfo
  {volume} {310}},\ \bibinfo {pages} {1187 } (\bibinfo {year}
  {2007})}\BibitemShut {NoStop}%
\bibitem [{\citenamefont {Lohmann}\ \emph {et~al.}(2014)\citenamefont
  {Lohmann}, \citenamefont {Schmidt},\ and\ \citenamefont {Richter}}]{Lohmann}%
  \BibitemOpen
  \bibfield  {author} {\bibinfo {author} {\bibfnamefont {A.}~\bibnamefont
  {Lohmann}}, \bibinfo {author} {\bibfnamefont {H.-J.}\ \bibnamefont
  {Schmidt}},\ and\ \bibinfo {author} {\bibfnamefont {J.}~\bibnamefont
  {Richter}},\ }\bibfield  {title} {\bibinfo {title} {{Tenth-order
  high-temperature expansion for the susceptibility and the specific heat of
  spin-$s$ Heisenberg models with arbitrary exchange patterns: application to
  pyrochlore and kagome magnets}},\ }\href
  {https://doi.org/10.1103/PhysRevB.89.014415} {\bibfield  {journal} {\bibinfo
  {journal} {Phys. Rev. B}\ }\textbf {\bibinfo {volume} {89}},\ \bibinfo
  {pages} {014415} (\bibinfo {year} {2014})}\BibitemShut {NoStop}%
\bibitem [{\citenamefont {Koepernik}\ and\ \citenamefont
  {Eschrig}(1999)}]{Koepernik1999}%
  \BibitemOpen
  \bibfield  {author} {\bibinfo {author} {\bibfnamefont {K.}~\bibnamefont
  {Koepernik}}\ and\ \bibinfo {author} {\bibfnamefont {H.}~\bibnamefont
  {Eschrig}},\ }\bibfield  {title} {\bibinfo {title} {{Full-potential
  nonorthogonal local-orbital minimum-basis band-structure scheme}},\ }\href
  {https://doi.org/10.1103/PhysRevB.59.1743} {\bibfield  {journal} {\bibinfo
  {journal} {Phys. Rev. B}\ }\textbf {\bibinfo {volume} {59}},\ \bibinfo
  {pages} {1743} (\bibinfo {year} {1999})}\BibitemShut {NoStop}%
\bibitem [{\citenamefont {Perdew}\ and\ \citenamefont
  {Wang}(1992)}]{Perdew1992}%
  \BibitemOpen
  \bibfield  {author} {\bibinfo {author} {\bibfnamefont {J.~P.}\ \bibnamefont
  {Perdew}}\ and\ \bibinfo {author} {\bibfnamefont {Y.}~\bibnamefont {Wang}},\
  }\bibfield  {title} {\bibinfo {title} {{Accurate and simple analytic
  representation of the electron-gas correlation energy}},\ }\href
  {https://doi.org/10.1103/PhysRevB.45.13244} {\bibfield  {journal} {\bibinfo
  {journal} {Phys. Rev. B}\ }\textbf {\bibinfo {volume} {45}},\ \bibinfo
  {pages} {13244} (\bibinfo {year} {1992})}\BibitemShut {NoStop}%
\bibitem [{Note1()}]{Note1}%
  \BibitemOpen
  \bibinfo {note} {See Supplemental Material at
  http://link.aps.org/supplemental/ 10.1103/PhysRevMaterials.xx.xxxxxx for
  further details on DFT calculations and outputs, modeling of magnetic
  susceptibility and heat capacity data, and fitting results of all symmetry
  allowed magnetic models to powder neutron diffraction data.}\BibitemShut
  {Stop}%
\bibitem [{\citenamefont {Momma}\ and\ \citenamefont {Izumi}(2011)}]{Vesta}%
  \BibitemOpen
  \bibfield  {author} {\bibinfo {author} {\bibfnamefont {K.}~\bibnamefont
  {Momma}}\ and\ \bibinfo {author} {\bibfnamefont {F.}~\bibnamefont {Izumi}},\
  }\bibfield  {title} {\bibinfo {title} {{\textit{VESTA 3} for
  three-dimensional visualization of crystal, volumetric and morphology
  data}},\ }\href {https://doi.org/10.1107/S0021889811038970} {\bibfield
  {journal} {\bibinfo  {journal} {J. Appl. Cryst.}\ }\textbf {\bibinfo {volume}
  {44}},\ \bibinfo {pages} {1272} (\bibinfo {year} {2011})}\BibitemShut
  {NoStop}%
\bibitem [{\citenamefont {Jiang}\ \emph {et~al.}(2020)\citenamefont {Jiang},
  \citenamefont {Ramanathan}, \citenamefont {Bacsa},\ and\ \citenamefont {{La
  Pierre}}}]{Jiang}%
  \BibitemOpen
  \bibfield  {author} {\bibinfo {author} {\bibfnamefont {N.}~\bibnamefont
  {Jiang}}, \bibinfo {author} {\bibfnamefont {A.}~\bibnamefont {Ramanathan}},
  \bibinfo {author} {\bibfnamefont {J.}~\bibnamefont {Bacsa}},\ and\ \bibinfo
  {author} {\bibfnamefont {H.~S.}\ \bibnamefont {{La Pierre}}},\ }\bibfield
  {title} {\bibinfo {title} {{Synthesis of a \textit{d}$^1$-titanium fluoride
  kagome lattice antiferromagnet}},\ }\href
  {https://doi.org/10.1038/s41557-020-0490-8} {\bibfield  {journal} {\bibinfo
  {journal} {Nat. Chem.}\ }\textbf {\bibinfo {volume} {12}},\ \bibinfo {pages}
  {691} (\bibinfo {year} {2020})}\BibitemShut {NoStop}%
\bibitem [{\citenamefont {Aggarwal}\ \emph {et~al.}(1986)\citenamefont
  {Aggarwal}, \citenamefont {Sanchez}, \citenamefont {Fahey},\ and\
  \citenamefont {Strauss}}]{Aggarwal1986}%
  \BibitemOpen
  \bibfield  {author} {\bibinfo {author} {\bibfnamefont {R.~L.}\ \bibnamefont
  {Aggarwal}}, \bibinfo {author} {\bibfnamefont {A.}~\bibnamefont {Sanchez}},
  \bibinfo {author} {\bibfnamefont {R.~E.}\ \bibnamefont {Fahey}},\ and\
  \bibinfo {author} {\bibfnamefont {A.~J.}\ \bibnamefont {Strauss}},\
  }\bibfield  {title} {\bibinfo {title} {{Magnetic and optical measurements on
  Ti:Al$_2$O$_3$ crystals for laser applications: Concentration and absorption
  cross section of Ti$^{3+}$ ions}},\ }\href {https://doi.org/10.1063/1.96904}
  {\bibfield  {journal} {\bibinfo  {journal} {Appl. Phys. Lett.}\ }\textbf
  {\bibinfo {volume} {48}},\ \bibinfo {pages} {1345} (\bibinfo {year}
  {1986})}\BibitemShut {NoStop}%
\bibitem [{\citenamefont {Eitel}\ and\ \citenamefont
  {Greedan}(1986)}]{Eitel1986}%
  \BibitemOpen
  \bibfield  {author} {\bibinfo {author} {\bibfnamefont {M.}~\bibnamefont
  {Eitel}}\ and\ \bibinfo {author} {\bibfnamefont {J.}~\bibnamefont
  {Greedan}},\ }\bibfield  {title} {\bibinfo {title} {{A high resolution
  neutron diffraction study of the perovskite LaTiO$_3$}},\ }\href
  {https://doi.org/https://doi.org/10.1016/0022-5088(86)90220-1} {\bibfield
  {journal} {\bibinfo  {journal} {J. Less-Common Met.}\ }\textbf {\bibinfo
  {volume} {116}},\ \bibinfo {pages} {95 } (\bibinfo {year}
  {1986})}\BibitemShut {NoStop}%
\bibitem [{\citenamefont {Carr}\ and\ \citenamefont {Foner}(1960)}]{Carr1960}%
  \BibitemOpen
  \bibfield  {author} {\bibinfo {author} {\bibfnamefont {P.~H.}\ \bibnamefont
  {Carr}}\ and\ \bibinfo {author} {\bibfnamefont {S.}~\bibnamefont {Foner}},\
  }\bibfield  {title} {\bibinfo {title} {{Magnetic Transitions in Ti$_2$O$_3$
  and V$_2$O$_3$}},\ }\href {https://doi.org/10.1063/1.1984740} {\bibfield
  {journal} {\bibinfo  {journal} {Int. J. Appl. Phys.}\ }\textbf {\bibinfo
  {volume} {31}},\ \bibinfo {pages} {S344} (\bibinfo {year}
  {1960})}\BibitemShut {NoStop}%
\bibitem [{\citenamefont {Kasinathan}\ \emph {et~al.}(2013)\citenamefont
  {Kasinathan}, \citenamefont {Koepernik}, \citenamefont {Janson},
  \citenamefont {Nilsen}, \citenamefont {Piatek}, \citenamefont {R{\o}nnow},\
  and\ \citenamefont {Rosner}}]{Kasinathan2013}%
  \BibitemOpen
  \bibfield  {author} {\bibinfo {author} {\bibfnamefont {D.}~\bibnamefont
  {Kasinathan}}, \bibinfo {author} {\bibfnamefont {K.}~\bibnamefont
  {Koepernik}}, \bibinfo {author} {\bibfnamefont {O.}~\bibnamefont {Janson}},
  \bibinfo {author} {\bibfnamefont {G.~J.}\ \bibnamefont {Nilsen}}, \bibinfo
  {author} {\bibfnamefont {J.~O.}\ \bibnamefont {Piatek}}, \bibinfo {author}
  {\bibfnamefont {H.~M.}\ \bibnamefont {R{\o}nnow}},\ and\ \bibinfo {author}
  {\bibfnamefont {H.}~\bibnamefont {Rosner}},\ }\bibfield  {title} {\bibinfo
  {title} {{Electronic structure of KTi(SO$_4$)$_2\cdot$H$_2$O: an $S = 1/2$
  frustrated chain antiferromagnet}},\ }\href
  {https://link.aps.org/doi/10.1103/PhysRevB.88.224410} {\bibfield  {journal}
  {\bibinfo  {journal} {Phys. Rev. B}\ }\textbf {\bibinfo {volume} {88}}
  (\bibinfo {year} {2013})}\BibitemShut {NoStop}%
\bibitem [{\citenamefont {Nilsen}\ \emph {et~al.}(2015)\citenamefont {Nilsen},
  \citenamefont {Raja}, \citenamefont {Tsirlin}, \citenamefont {Mutka},
  \citenamefont {Kasinathan}, \citenamefont {Ritter},\ and\ \citenamefont
  {R{\o}nnow}}]{Nilsen2015}%
  \BibitemOpen
  \bibfield  {author} {\bibinfo {author} {\bibfnamefont {G.~J.}\ \bibnamefont
  {Nilsen}}, \bibinfo {author} {\bibfnamefont {A.}~\bibnamefont {Raja}},
  \bibinfo {author} {\bibfnamefont {A.~A.}\ \bibnamefont {Tsirlin}}, \bibinfo
  {author} {\bibfnamefont {H.}~\bibnamefont {Mutka}}, \bibinfo {author}
  {\bibfnamefont {D.}~\bibnamefont {Kasinathan}}, \bibinfo {author}
  {\bibfnamefont {C.}~\bibnamefont {Ritter}},\ and\ \bibinfo {author}
  {\bibfnamefont {H.~M.}\ \bibnamefont {R{\o}nnow}},\ }\bibfield  {title}
  {\bibinfo {title} {{One-dimensional quantum magnetism in the anhydrous alum
  KTi(SO$_4$)$_2$}},\ }\href {https://doi.org/10.1088/1367-2630/17/11/113035}
  {\bibfield  {journal} {\bibinfo  {journal} {New J. Phys}\ }\textbf {\bibinfo
  {volume} {17}},\ \bibinfo {pages} {113035} (\bibinfo {year}
  {2015})}\BibitemShut {NoStop}%
\bibitem [{\citenamefont {Bramwell}\ \emph {et~al.}(1996)\citenamefont
  {Bramwell}, \citenamefont {Carling}, \citenamefont {Harding}, \citenamefont
  {Harris}, \citenamefont {Kariuki}, \citenamefont {Nixon},\ and\ \citenamefont
  {Parkin}}]{Bramwell1996}%
  \BibitemOpen
  \bibfield  {author} {\bibinfo {author} {\bibfnamefont {S.~T.}\ \bibnamefont
  {Bramwell}}, \bibinfo {author} {\bibfnamefont {S.~G.}\ \bibnamefont
  {Carling}}, \bibinfo {author} {\bibfnamefont {C.~J.}\ \bibnamefont
  {Harding}}, \bibinfo {author} {\bibfnamefont {K.~D.~M.}\ \bibnamefont
  {Harris}}, \bibinfo {author} {\bibfnamefont {B.~M.}\ \bibnamefont {Kariuki}},
  \bibinfo {author} {\bibfnamefont {L.}~\bibnamefont {Nixon}},\ and\ \bibinfo
  {author} {\bibfnamefont {I.~P.}\ \bibnamefont {Parkin}},\ }\bibfield  {title}
  {\bibinfo {title} {{The anhydrous alums as model triangular-lattice
  magnets}},\ }\href {https://doi.org/10.1088/0953-8984/8/9/002} {\bibfield
  {journal} {\bibinfo  {journal} {J. Phys.: Condens. Matter}\ }\textbf
  {\bibinfo {volume} {8}},\ \bibinfo {pages} {L123} (\bibinfo {year}
  {1996})}\BibitemShut {NoStop}%
\bibitem [{\citenamefont {Nilsen}\ \emph {et~al.}(2008)\citenamefont {Nilsen},
  \citenamefont {R{\o}nnow}, \citenamefont {L{\"{a}}uchli}, \citenamefont
  {Fabbiani}, \citenamefont {Sanchez-Benitez}, \citenamefont {Kamenev},\ and\
  \citenamefont {Harrison}}]{Nilsen2008}%
  \BibitemOpen
  \bibfield  {author} {\bibinfo {author} {\bibfnamefont {G.~J.}\ \bibnamefont
  {Nilsen}}, \bibinfo {author} {\bibfnamefont {H.~M.}\ \bibnamefont
  {R{\o}nnow}}, \bibinfo {author} {\bibfnamefont {A.~M.}\ \bibnamefont
  {L{\"{a}}uchli}}, \bibinfo {author} {\bibfnamefont {F.~P.~A.}\ \bibnamefont
  {Fabbiani}}, \bibinfo {author} {\bibfnamefont {J.}~\bibnamefont
  {Sanchez-Benitez}}, \bibinfo {author} {\bibfnamefont {K.~V.}\ \bibnamefont
  {Kamenev}},\ and\ \bibinfo {author} {\bibfnamefont {A.}~\bibnamefont
  {Harrison}},\ }\bibfield  {title} {\bibinfo {title} {{A new realisation of
  the $S = 1/2$ frustrated chain antiferromagnet}},\ }\href
  {https://doi.org/10.1021/cm7023263} {\bibfield  {journal} {\bibinfo
  {journal} {Chem. Mater}\ }\textbf {\bibinfo {volume} {20}},\ \bibinfo {pages}
  {8} (\bibinfo {year} {2008})}\BibitemShut {NoStop}%
\bibitem [{\citenamefont {Lancaster}\ \emph {et~al.}(2007)\citenamefont
  {Lancaster}, \citenamefont {Blundell}, \citenamefont {Brooks}, \citenamefont
  {Baker}, \citenamefont {Pratt}, \citenamefont {Manson}, \citenamefont
  {Conner}, \citenamefont {Xiao}, \citenamefont {Landee}, \citenamefont
  {Chaves}, \citenamefont {Soriano}, \citenamefont {Novak}, \citenamefont
  {Papageorgiou}, \citenamefont {Bianchi}, \citenamefont
  {Herrmannsd{\"{o}}rfer}, \citenamefont {Wosnitza},\ and\ \citenamefont
  {Schlueter}}]{Lancaster2007}%
  \BibitemOpen
  \bibfield  {author} {\bibinfo {author} {\bibfnamefont {T.}~\bibnamefont
  {Lancaster}}, \bibinfo {author} {\bibfnamefont {S.~J.}\ \bibnamefont
  {Blundell}}, \bibinfo {author} {\bibfnamefont {M.~L.}\ \bibnamefont
  {Brooks}}, \bibinfo {author} {\bibfnamefont {P.~J.}\ \bibnamefont {Baker}},
  \bibinfo {author} {\bibfnamefont {F.~L.}\ \bibnamefont {Pratt}}, \bibinfo
  {author} {\bibfnamefont {J.~L.}\ \bibnamefont {Manson}}, \bibinfo {author}
  {\bibfnamefont {M.~M.}\ \bibnamefont {Conner}}, \bibinfo {author}
  {\bibfnamefont {F.}~\bibnamefont {Xiao}}, \bibinfo {author} {\bibfnamefont
  {C.~P.}\ \bibnamefont {Landee}}, \bibinfo {author} {\bibfnamefont {F.~A.}\
  \bibnamefont {Chaves}}, \bibinfo {author} {\bibfnamefont {S.}~\bibnamefont
  {Soriano}}, \bibinfo {author} {\bibfnamefont {M.~A.}\ \bibnamefont {Novak}},
  \bibinfo {author} {\bibfnamefont {T.~P.}\ \bibnamefont {Papageorgiou}},
  \bibinfo {author} {\bibfnamefont {A.~D.}\ \bibnamefont {Bianchi}}, \bibinfo
  {author} {\bibfnamefont {T.}~\bibnamefont {Herrmannsd{\"{o}}rfer}}, \bibinfo
  {author} {\bibfnamefont {J.}~\bibnamefont {Wosnitza}},\ and\ \bibinfo
  {author} {\bibfnamefont {J.~A.}\ \bibnamefont {Schlueter}},\ }\bibfield
  {title} {\bibinfo {title} {{Magnetic order in the $S = 1/2$ two-dimensional
  molecular antiferromagnet copper pyrazine perchlorate
  Cu(Pz)$_2$(ClO$_4$)$_2$}},\ }\href
  {https://doi.org/10.1103/PhysRevB.75.094421} {\bibfield  {journal} {\bibinfo
  {journal} {Phys. Rev. B}\ }\textbf {\bibinfo {volume} {75}},\ \bibinfo
  {pages} {94421} (\bibinfo {year} {2007})}\BibitemShut {NoStop}%
\bibitem [{\citenamefont {Nath}\ \emph {et~al.}(2015)\citenamefont {Nath},
  \citenamefont {Padmanabhan}, \citenamefont {Baby}, \citenamefont
  {Thirumurugan}, \citenamefont {Ehlers}, \citenamefont {Hemmida},
  \citenamefont {{Krug von Nidda}},\ and\ \citenamefont {Tsirlin}}]{Nath2015}%
  \BibitemOpen
  \bibfield  {author} {\bibinfo {author} {\bibfnamefont {R.}~\bibnamefont
  {Nath}}, \bibinfo {author} {\bibfnamefont {M.}~\bibnamefont {Padmanabhan}},
  \bibinfo {author} {\bibfnamefont {S.}~\bibnamefont {Baby}}, \bibinfo {author}
  {\bibfnamefont {A.}~\bibnamefont {Thirumurugan}}, \bibinfo {author}
  {\bibfnamefont {D.}~\bibnamefont {Ehlers}}, \bibinfo {author} {\bibfnamefont
  {M.}~\bibnamefont {Hemmida}}, \bibinfo {author} {\bibfnamefont {H.-A.}\
  \bibnamefont {{Krug von Nidda}}},\ and\ \bibinfo {author} {\bibfnamefont
  {A.~A.}\ \bibnamefont {Tsirlin}},\ }\bibfield  {title} {\bibinfo {title}
  {{Quasi-two-dimensional Cu[C$_6$H$_2$(COO)$_4$][C$_2$H$_5$NH$_3$]$_2$}},\
  }\href {https://doi.org/10.1103/PhysRevB.91.054409} {\bibfield  {journal}
  {\bibinfo  {journal} {Phys. Rev. B}\ }\textbf {\bibinfo {volume} {91}},\
  \bibinfo {pages} {54409} (\bibinfo {year} {2015})}\BibitemShut {NoStop}%
\bibitem [{\citenamefont {Yasuda}\ \emph {et~al.}(2005)\citenamefont {Yasuda},
  \citenamefont {Todo}, \citenamefont {Hukushima}, \citenamefont {Alet},
  \citenamefont {Keller}, \citenamefont {Troyer},\ and\ \citenamefont
  {Takayama}}]{Yasuda2005}%
  \BibitemOpen
  \bibfield  {author} {\bibinfo {author} {\bibfnamefont {C.}~\bibnamefont
  {Yasuda}}, \bibinfo {author} {\bibfnamefont {S.}~\bibnamefont {Todo}},
  \bibinfo {author} {\bibfnamefont {K.}~\bibnamefont {Hukushima}}, \bibinfo
  {author} {\bibfnamefont {F.}~\bibnamefont {Alet}}, \bibinfo {author}
  {\bibfnamefont {M.}~\bibnamefont {Keller}}, \bibinfo {author} {\bibfnamefont
  {M.}~\bibnamefont {Troyer}},\ and\ \bibinfo {author} {\bibfnamefont
  {H.}~\bibnamefont {Takayama}},\ }\bibfield  {title} {\bibinfo {title}
  {{N\'eel temperature of quasi-low-dimensional Heisenberg antiferromagnets}},\
  }\href {https://doi.org/10.1103/PhysRevLett.94.217201} {\bibfield  {journal}
  {\bibinfo  {journal} {Phys. Rev. Lett.}\ }\textbf {\bibinfo {volume} {94}},\
  \bibinfo {pages} {217201} (\bibinfo {year} {2005})}\BibitemShut {NoStop}%
\bibitem [{\citenamefont {Miller}\ and\ \citenamefont {Love}(1967)}]{Miller}%
  \BibitemOpen
  \bibfield  {author} {\bibinfo {author} {\bibfnamefont {S.~C.}\ \bibnamefont
  {Miller}}\ and\ \bibinfo {author} {\bibfnamefont {W.~F.}\ \bibnamefont
  {Love}},\ }\href@noop {} {\emph {\bibinfo {title} {{Tables of irreducible
  representations of space groups and co-representations of magnetic space
  groups}}}}\ (\bibinfo  {publisher} {Boulder, Colo: Pruett Press},\ \bibinfo
  {year} {1967})\BibitemShut {NoStop}%
\bibitem [{\citenamefont {Perez-Mato}\ \emph {et~al.}(2015)\citenamefont
  {Perez-Mato}, \citenamefont {Gallego}, \citenamefont {Tasci}, \citenamefont
  {Elcoro}, \citenamefont {de~la Flor},\ and\ \citenamefont {Aroyo}}]{maxmagn}%
  \BibitemOpen
  \bibfield  {author} {\bibinfo {author} {\bibfnamefont {J.}~\bibnamefont
  {Perez-Mato}}, \bibinfo {author} {\bibfnamefont {S.}~\bibnamefont {Gallego}},
  \bibinfo {author} {\bibfnamefont {E.}~\bibnamefont {Tasci}}, \bibinfo
  {author} {\bibfnamefont {L.}~\bibnamefont {Elcoro}}, \bibinfo {author}
  {\bibfnamefont {G.}~\bibnamefont {de~la Flor}},\ and\ \bibinfo {author}
  {\bibfnamefont {M.}~\bibnamefont {Aroyo}},\ }\bibfield  {title} {\bibinfo
  {title} {Symmetry-based computational tools for magnetic crystallography},\
  }\href {https://doi.org/10.1146/annurev-matsci-070214-021008} {\bibfield
  {journal} {\bibinfo  {journal} {Annu. Rev. Mater. Res.}\ }\textbf {\bibinfo
  {volume} {45}},\ \bibinfo {pages} {217} (\bibinfo {year} {2015})}\BibitemShut
  {NoStop}%
\bibitem [{\citenamefont {Belov}\ \emph {et~al.}(1957)\citenamefont {Belov},
  \citenamefont {Neronova},\ and\ \citenamefont {Smirnova}}]{belov}%
  \BibitemOpen
  \bibfield  {author} {\bibinfo {author} {\bibfnamefont {N.}~\bibnamefont
  {Belov}}, \bibinfo {author} {\bibfnamefont {N.}~\bibnamefont {Neronova}},\
  and\ \bibinfo {author} {\bibfnamefont {T.}~\bibnamefont {Smirnova}},\
  }\bibfield  {title} {\bibinfo {title} {Shubnikov groups},\ }\href@noop {}
  {\bibfield  {journal} {\bibinfo  {journal} {Kristallografiya}\ }\textbf
  {\bibinfo {volume} {2}},\ \bibinfo {pages} {315} (\bibinfo {year}
  {1957})}\BibitemShut {NoStop}%
\bibitem [{\citenamefont {Bettler}\ \emph {et~al.}()\citenamefont {Bettler},
  \citenamefont {Landolt}, \citenamefont {Aksoy}, \citenamefont {Yan},
  \citenamefont {Gvasaliya}, \citenamefont {Qiu}, \citenamefont {Ressouche},
  \citenamefont {Beauvois}, \citenamefont {Raymond}, \citenamefont
  {Ponomaryov}, \citenamefont {Zvyagin},\ and\ \citenamefont
  {Zheludev}}]{Bettler2019}%
  \BibitemOpen
  \bibfield  {author} {\bibinfo {author} {\bibfnamefont {S.}~\bibnamefont
  {Bettler}}, \bibinfo {author} {\bibfnamefont {F.}~\bibnamefont {Landolt}},
  \bibinfo {author} {\bibfnamefont {{\"{O}}.~M.}\ \bibnamefont {Aksoy}},
  \bibinfo {author} {\bibfnamefont {Z.}~\bibnamefont {Yan}}, \bibinfo {author}
  {\bibfnamefont {S.}~\bibnamefont {Gvasaliya}}, \bibinfo {author}
  {\bibfnamefont {Y.}~\bibnamefont {Qiu}}, \bibinfo {author} {\bibfnamefont
  {E.}~\bibnamefont {Ressouche}}, \bibinfo {author} {\bibfnamefont
  {K.}~\bibnamefont {Beauvois}}, \bibinfo {author} {\bibfnamefont
  {S.}~\bibnamefont {Raymond}}, \bibinfo {author} {\bibfnamefont {A.~N.}\
  \bibnamefont {Ponomaryov}}, \bibinfo {author} {\bibfnamefont {S.~A.}\
  \bibnamefont {Zvyagin}},\ and\ \bibinfo {author} {\bibfnamefont
  {A.}~\bibnamefont {Zheludev}},\ }\bibfield  {title} {\bibinfo {title}
  {{Magnetic structure and spin waves in the frustrated ferro-antiferromagnet
  Pb$_2$VO(PO$_4$)$_2$}},\ }\href {https://doi.org/10.1103/PhysRevB.99.184437}
  {\bibinfo  {journal} {Phys. Rev. B}\ ,\ \bibinfo {pages}
  {184437}}\BibitemShut {NoStop}%
\bibitem [{\citenamefont {Koga}\ \emph {et~al.}(2016)\citenamefont {Koga},
  \citenamefont {Kurita}, \citenamefont {Avdeev}, \citenamefont {Danilkin},
  \citenamefont {Sato},\ and\ \citenamefont {Tanaka}}]{Koga2016}%
  \BibitemOpen
\bibfield  {journal} {  }\bibfield  {author} {\bibinfo {author} {\bibfnamefont
  {T.}~\bibnamefont {Koga}}, \bibinfo {author} {\bibfnamefont {N.}~\bibnamefont
  {Kurita}}, \bibinfo {author} {\bibfnamefont {M.}~\bibnamefont {Avdeev}},
  \bibinfo {author} {\bibfnamefont {S.}~\bibnamefont {Danilkin}}, \bibinfo
  {author} {\bibfnamefont {T.~J.}\ \bibnamefont {Sato}},\ and\ \bibinfo
  {author} {\bibfnamefont {H.}~\bibnamefont {Tanaka}},\ }\bibfield  {title}
  {\bibinfo {title} {{Magnetic structure of the $S = 1/2$ quasi-two-dimensional
  square-lattice Heisenberg antiferromagnet Sr$_2$CuTeO$_6$}},\ }\href
  {https://doi.org/10.1103/PhysRevB.93.054426} {\bibfield  {journal} {\bibinfo
  {journal} {Phys. Rev. B}\ }\textbf {\bibinfo {volume} {93}},\ \bibinfo
  {pages} {054426} (\bibinfo {year} {2016})}\BibitemShut {NoStop}%
\bibitem [{\citenamefont {Manousakis}(1991)}]{Manousakis1991}%
  \BibitemOpen
  \bibfield  {author} {\bibinfo {author} {\bibfnamefont {E.}~\bibnamefont
  {Manousakis}},\ }\bibfield  {title} {\bibinfo {title} {{The spin-$1/2$
  Heisenberg antiferromagnet on a square lattice and its application to the
  cuprous oxides}},\ }\href {https://doi.org/10.1103/RevModPhys.63.1}
  {\bibfield  {journal} {\bibinfo  {journal} {Rev. Mod. Phys.}\ }\textbf
  {\bibinfo {volume} {63}},\ \bibinfo {pages} {1} (\bibinfo {year}
  {1991})}\BibitemShut {NoStop}%
\bibitem [{\citenamefont {Liu}(1990)}]{Liu1990}%
  \BibitemOpen
  \bibfield  {author} {\bibinfo {author} {\bibfnamefont {B.-G.}\ \bibnamefont
  {Liu}},\ }\bibfield  {title} {\bibinfo {title} {{Low-temperature properties
  of the quasi-two-dimensional antiferromagnetic Heisenberg model}},\ }\href
  {https://doi.org/10.1103/PhysRevB.41.9563} {\bibfield  {journal} {\bibinfo
  {journal} {Phys. Rev. B}\ }\textbf {\bibinfo {volume} {41}},\ \bibinfo
  {pages} {9563} (\bibinfo {year} {1990})}\BibitemShut {NoStop}%
\bibitem [{\citenamefont {Holt}\ \emph {et~al.}(2011)\citenamefont {Holt},
  \citenamefont {Sushkov}, \citenamefont {Stanek},\ and\ \citenamefont
  {Uhrig}}]{Holt2011}%
  \BibitemOpen
  \bibfield  {author} {\bibinfo {author} {\bibfnamefont {M.}~\bibnamefont
  {Holt}}, \bibinfo {author} {\bibfnamefont {O.}~\bibnamefont {Sushkov}},
  \bibinfo {author} {\bibfnamefont {D.}~\bibnamefont {Stanek}},\ and\ \bibinfo
  {author} {\bibfnamefont {G.}~\bibnamefont {Uhrig}},\ }\bibfield  {title}
  {\bibinfo {title} {{Three-dimensional generalization of the J$_1$-J$_2$
  Heisenberg model on a square lattice and role of the interlayer coupling
  J$_c$}},\ }\href {https://link.aps.org/doi/10.1103/PhysRevB.83.144528}
  {\bibfield  {journal} {\bibinfo  {journal} {Phys. Rev. B}\ }\textbf {\bibinfo
  {volume} {83}} (\bibinfo {year} {2011})}\BibitemShut {NoStop}%
\bibitem [{\citenamefont {Sheu}\ \emph {et~al.}(1996)\citenamefont {Sheu},
  \citenamefont {Wu}, \citenamefont {Wang},\ and\ \citenamefont
  {English}}]{Sheu1996}%
  \BibitemOpen
  \bibfield  {author} {\bibinfo {author} {\bibfnamefont {H.~S.}\ \bibnamefont
  {Sheu}}, \bibinfo {author} {\bibfnamefont {J.~C.}\ \bibnamefont {Wu}},
  \bibinfo {author} {\bibfnamefont {Y.}~\bibnamefont {Wang}},\ and\ \bibinfo
  {author} {\bibfnamefont {R.~B.}\ \bibnamefont {English}},\ }\bibfield
  {title} {\bibinfo {title} {{Charge density studies in
  NH$_4$[Ti(C$_2$O$_4$)$_2$].2H$_2$O crystals at two wavelengths}},\ }\href
  {https://doi.org/10.1107/S0108768195012900} {\bibfield  {journal} {\bibinfo
  {journal} {Acta Crystallogr. B}\ }\textbf {\bibinfo {volume} {52}},\ \bibinfo
  {pages} {458} (\bibinfo {year} {1996})}\BibitemShut {NoStop}%
\bibitem [{\citenamefont {Fujio}\ \emph {et~al.}(2007)\citenamefont {Fujio},
  \citenamefont {Tanaka},\ and\ \citenamefont {Inui}}]{Fujio2007}%
  \BibitemOpen
  \bibfield  {author} {\bibinfo {author} {\bibfnamefont {S.}~\bibnamefont
  {Fujio}}, \bibinfo {author} {\bibfnamefont {K.}~\bibnamefont {Tanaka}},\ and\
  \bibinfo {author} {\bibfnamefont {H.}~\bibnamefont {Inui}},\ }\bibfield
  {title} {\bibinfo {title} {{Formation probability for enantiomorphic crystals
  (with the space groups of $P6_222$ and $P6_422$) in transition-metal
  disilicides with the C40 structure as determined by convergent-beam electron
  diffraction}},\ }\href
  {https://doi.org/https://doi.org/10.1016/j.intermet.2006.05.010} {\bibfield
  {journal} {\bibinfo  {journal} {Intermetallics}\ }\textbf {\bibinfo {volume}
  {15}},\ \bibinfo {pages} {245} (\bibinfo {year} {2007})}\BibitemShut
  {NoStop}%
\bibitem [{\citenamefont {Kalinnikov}\ \emph {et~al.}(1969)\citenamefont
  {Kalinnikov}, \citenamefont {Zelentsov}, \citenamefont {Zharkikh},\ and\
  \citenamefont {Aminov}}]{Kalinnikov1969}%
  \BibitemOpen
  \bibfield  {author} {\bibinfo {author} {\bibfnamefont {V.~T.}\ \bibnamefont
  {Kalinnikov}}, \bibinfo {author} {\bibfnamefont {V.~V.}\ \bibnamefont
  {Zelentsov}}, \bibinfo {author} {\bibfnamefont {A.~A.}\ \bibnamefont
  {Zharkikh}},\ and\ \bibinfo {author} {\bibfnamefont {T.~G.}\ \bibnamefont
  {Aminov}},\ }\bibfield  {title} {\bibinfo {title} {{The magnetic
  susceptibilities of complex Ti(III) oxalates}},\ }\href
  {https://doi.org/10.1007/BF00912572} {\bibfield  {journal} {\bibinfo
  {journal} {Bull. Acad. Sci. USSR, Div. Chem. Sci.}\ }\textbf {\bibinfo
  {volume} {18}},\ \bibinfo {pages} {2670} (\bibinfo {year}
  {1969})}\BibitemShut {NoStop}%
\bibitem [{\citenamefont {Oitmaa}(2018)}]{Oitmaa2018}%
  \BibitemOpen
  \bibfield  {author} {\bibinfo {author} {\bibfnamefont {J.}~\bibnamefont
  {Oitmaa}},\ }\bibfield  {title} {\bibinfo {title} {{Diamond lattice
  Heisenberg antiferromagnet}},\ }\href
  {https://doi.org/10.1088/1361-648x/aab22c} {\bibfield  {journal} {\bibinfo
  {journal} {J. Phys. Condens. Matter}\ }\textbf {\bibinfo {volume} {30}},\
  \bibinfo {pages} {155801} (\bibinfo {year} {2018})}\BibitemShut {NoStop}%
\bibitem [{\citenamefont {Oitmaa}(2019)}]{Oitmaa2019}%
  \BibitemOpen
  \bibfield  {author} {\bibinfo {author} {\bibfnamefont {J.}~\bibnamefont
  {Oitmaa}},\ }\bibfield  {title} {\bibinfo {title} {{Frustrated diamond
  lattice antiferromagnet}},\ }\href
  {https://doi.org/10.1103/PhysRevB.99.134407} {\bibfield  {journal} {\bibinfo
  {journal} {Phys. Rev. B}\ }\textbf {\bibinfo {volume} {99}},\ \bibinfo
  {pages} {134407} (\bibinfo {year} {2019})}\BibitemShut {NoStop}%
\bibitem [{\citenamefont {Campbell}\ \emph {et~al.}(2006)\citenamefont
  {Campbell}, \citenamefont {Stokes}, \citenamefont {Tanner},\ and\
  \citenamefont {Hatch}}]{Campbell2006}%
  \BibitemOpen
  \bibfield  {author} {\bibinfo {author} {\bibfnamefont {B.~J.}\ \bibnamefont
  {Campbell}}, \bibinfo {author} {\bibfnamefont {H.~T.}\ \bibnamefont
  {Stokes}}, \bibinfo {author} {\bibfnamefont {D.~E.}\ \bibnamefont {Tanner}},\
  and\ \bibinfo {author} {\bibfnamefont {D.~M.}\ \bibnamefont {Hatch}},\
  }\bibfield  {title} {\bibinfo {title} {{ISODISPLACE: a web-based tool for
  exploring structural distortions}},\ }\href
  {https://doi.org/10.1107/S0021889806014075} {\bibfield  {journal} {\bibinfo
  {journal} {J. Appl. Crystallogr}\ }\textbf {\bibinfo {volume} {39}},\
  \bibinfo {pages} {607} (\bibinfo {year} {2006})}\BibitemShut {NoStop}%
\bibitem [{\citenamefont {Wills}(2000)}]{Wills2000}%
  \BibitemOpen
  \bibfield  {author} {\bibinfo {author} {\bibfnamefont {A.~S.}\ \bibnamefont
  {Wills}},\ }\bibfield  {title} {\bibinfo {title} {{A new protocol for the
  determination of magnetic structures using simulated annealing and
  representational analysis (SARAh)}},\ }\href
  {https://doi.org/https://doi.org/10.1016/S0921-4526(99)01722-6} {\bibfield
  {journal} {\bibinfo  {journal} {Physica B Condens. Matter}\ }\textbf
  {\bibinfo {volume} {276-278}},\ \bibinfo {pages} {680} (\bibinfo {year}
  {2000})}\BibitemShut {NoStop}%
\bibitem [{\citenamefont {Marjerrison}\ \emph {et~al.}(2016)\citenamefont
  {Marjerrison}, \citenamefont {Mauws}, \citenamefont {Sharma}, \citenamefont
  {Wiebe}, \citenamefont {Derakhshan}, \citenamefont {Boyer}, \citenamefont
  {Gaulin},\ and\ \citenamefont {Greedan}}]{Marjerrison2016}%
  \BibitemOpen
  \bibfield  {author} {\bibinfo {author} {\bibfnamefont {C.~A.}\ \bibnamefont
  {Marjerrison}}, \bibinfo {author} {\bibfnamefont {C.}~\bibnamefont {Mauws}},
  \bibinfo {author} {\bibfnamefont {A.~Z.}\ \bibnamefont {Sharma}}, \bibinfo
  {author} {\bibfnamefont {C.~R.}\ \bibnamefont {Wiebe}}, \bibinfo {author}
  {\bibfnamefont {S.}~\bibnamefont {Derakhshan}}, \bibinfo {author}
  {\bibfnamefont {C.}~\bibnamefont {Boyer}}, \bibinfo {author} {\bibfnamefont
  {B.~D.}\ \bibnamefont {Gaulin}},\ and\ \bibinfo {author} {\bibfnamefont
  {J.~E.}\ \bibnamefont {Greedan}},\ }\bibfield  {title} {\bibinfo {title}
  {{Structure and Magnetic Properties of KRuO$_4$}},\ }\href
  {https://doi.org/10.1021/acs.inorgchem.6b02284} {\bibfield  {journal}
  {\bibinfo  {journal} {Inorg. Chem.}\ }\textbf {\bibinfo {volume} {55}},\
  \bibinfo {pages} {12897} (\bibinfo {year} {2016})}\BibitemShut {NoStop}%
\bibitem [{\citenamefont {Injac}\ \emph {et~al.}(2019)\citenamefont {Injac},
  \citenamefont {Yuen}, \citenamefont {Avdeev}, \citenamefont {Orlandi},\ and\
  \citenamefont {Kennedy}}]{Injac2019}%
  \BibitemOpen
  \bibfield  {author} {\bibinfo {author} {\bibfnamefont {S.}~\bibnamefont
  {Injac}}, \bibinfo {author} {\bibfnamefont {A.~K.~L.}\ \bibnamefont {Yuen}},
  \bibinfo {author} {\bibfnamefont {M.}~\bibnamefont {Avdeev}}, \bibinfo
  {author} {\bibfnamefont {F.}~\bibnamefont {Orlandi}},\ and\ \bibinfo {author}
  {\bibfnamefont {B.~J.}\ \bibnamefont {Kennedy}},\ }\bibfield  {title}
  {\bibinfo {title} {{Structural and magnetic studies of KOsO$_4$, a
  5\textit{d}$^1$ quantum magnet oxide}},\ }\href
  {https://doi.org/10.1039/C9CP00448C} {\bibfield  {journal} {\bibinfo
  {journal} {Phys. Chem. Chem. Phys.}\ }\textbf {\bibinfo {volume} {21}},\
  \bibinfo {pages} {7261} (\bibinfo {year} {2019})}\BibitemShut {NoStop}%
\bibitem [{\citenamefont {Randi{\'{c}}}(1960)}]{Randic1960}%
  \BibitemOpen
  \bibfield  {author} {\bibinfo {author} {\bibfnamefont {M.}~\bibnamefont
  {Randi{\'{c}}}},\ }\bibfield  {title} {\bibinfo {title} {{Ligand field
  splitting of \textit{d}-orbitals in eight coordinated complexes of square
  antiprism structure}},\ }\href {https://hrcak.srce.hr/file/306218} {\bibfield
   {journal} {\bibinfo  {journal} {Croat. Chem. Acta}\ }\textbf {\bibinfo
  {volume} {32}},\ \bibinfo {pages} {189} (\bibinfo {year} {1960})}\BibitemShut
  {NoStop}%
\bibitem [{\citenamefont {Kahn}(1985)}]{Kahn1985}%
  \BibitemOpen
  \bibfield  {author} {\bibinfo {author} {\bibfnamefont {O.}~\bibnamefont
  {Kahn}},\ }\bibfield  {title} {\bibinfo {title} {{Dinuclear complexes with
  predictable magnetic properties}},\ }\href
  {https://doi.org/10.1002/anie.198508341} {\bibfield  {journal} {\bibinfo
  {journal} {Angew. Chem.}\ }\textbf {\bibinfo {volume} {24}},\ \bibinfo
  {pages} {834} (\bibinfo {year} {1985})}\BibitemShut {NoStop}%
\bibitem [{\citenamefont {Cano}\ \emph {et~al.}(1998)\citenamefont {Cano},
  \citenamefont {Alemany}, \citenamefont {Alvarez}, \citenamefont {Verdaguer},\
  and\ \citenamefont {Ruiz}}]{Cano1998}%
  \BibitemOpen
  \bibfield  {author} {\bibinfo {author} {\bibfnamefont {J.}~\bibnamefont
  {Cano}}, \bibinfo {author} {\bibfnamefont {P.}~\bibnamefont {Alemany}},
  \bibinfo {author} {\bibfnamefont {S.}~\bibnamefont {Alvarez}}, \bibinfo
  {author} {\bibfnamefont {M.}~\bibnamefont {Verdaguer}},\ and\ \bibinfo
  {author} {\bibfnamefont {E.}~\bibnamefont {Ruiz}},\ }\bibfield  {title}
  {\bibinfo {title} {{Exchange coupling in oxalato-bridged copper(II) binuclear
  compounds: a density functional study}},\ }\href
  {https://doi.org/10.1002/(SICI)1521-3765(19980310)4:3<476::AID-CHEM476>3.0.CO;2-8}
  {\bibfield  {journal} {\bibinfo  {journal} {Chem. Eur. J}\ }\textbf {\bibinfo
  {volume} {4}},\ \bibinfo {pages} {476} (\bibinfo {year} {1998})}\BibitemShut
  {NoStop}%
\bibitem [{\citenamefont {Carlin}(1986)}]{Carlin1986}%
  \BibitemOpen
  \bibfield  {author} {\bibinfo {author} {\bibfnamefont {R.~L.}\ \bibnamefont
  {Carlin}},\ }\href {https://doi.org/10.1007/978-3-642-70733-9} {\emph
  {\bibinfo {title} {{Magnetochemistry}}}}\ (\bibinfo  {publisher}
  {Springer-Verlag Berlin Heidelberg},\ \bibinfo {year} {1986})\BibitemShut
  {NoStop}%
\bibitem [{\citenamefont {Tsyrulin}\ \emph {et~al.}(2009)\citenamefont
  {Tsyrulin}, \citenamefont {Pardini}, \citenamefont {Singh}, \citenamefont
  {Xiao}, \citenamefont {Link}, \citenamefont {Schneidewind}, \citenamefont
  {Hiess}, \citenamefont {Landee}, \citenamefont {Turnbull},\ and\
  \citenamefont {Kenzelmann}}]{Tsyrulin2009}%
  \BibitemOpen
  \bibfield  {author} {\bibinfo {author} {\bibfnamefont {N.}~\bibnamefont
  {Tsyrulin}}, \bibinfo {author} {\bibfnamefont {T.}~\bibnamefont {Pardini}},
  \bibinfo {author} {\bibfnamefont {R.~R.~P.}\ \bibnamefont {Singh}}, \bibinfo
  {author} {\bibfnamefont {F.}~\bibnamefont {Xiao}}, \bibinfo {author}
  {\bibfnamefont {P.}~\bibnamefont {Link}}, \bibinfo {author} {\bibfnamefont
  {A.}~\bibnamefont {Schneidewind}}, \bibinfo {author} {\bibfnamefont
  {A.}~\bibnamefont {Hiess}}, \bibinfo {author} {\bibfnamefont {C.~P.}\
  \bibnamefont {Landee}}, \bibinfo {author} {\bibfnamefont {M.~M.}\
  \bibnamefont {Turnbull}},\ and\ \bibinfo {author} {\bibfnamefont
  {M.}~\bibnamefont {Kenzelmann}},\ }\bibfield  {title} {\bibinfo {title}
  {{Quantum effects in a weakly frustrated $S=1/2$ two-dimensional Heisenberg
  antiferromagnet in an applied magnetic field}},\ }\href
  {https://doi.org/10.1103/PhysRevLett.102.197201} {\bibfield  {journal}
  {\bibinfo  {journal} {Phys. Rev. Lett.}\ }\textbf {\bibinfo {volume} {102}},\
  \bibinfo {pages} {197201} (\bibinfo {year} {2009})}\BibitemShut {NoStop}%
\bibitem [{\citenamefont {Wrobleski}\ and\ \citenamefont
  {Brown}(1980)}]{WROBLESKI1980227}%
  \BibitemOpen
  \bibfield  {author} {\bibinfo {author} {\bibfnamefont {J.~T.}\ \bibnamefont
  {Wrobleski}}\ and\ \bibinfo {author} {\bibfnamefont {D.~B.}\ \bibnamefont
  {Brown}},\ }\bibfield  {title} {\bibinfo {title} {{A study of the
  variable-temperature magnetic susceptibility of two Ti(III) oxalate
  complexes}},\ }\href
  {https://doi.org/https://doi.org/10.1016/S0020-1693(00)91964-9} {\bibfield
  {journal} {\bibinfo  {journal} {Inorg. Chim. Acta}\ }\textbf {\bibinfo
  {volume} {38}},\ \bibinfo {pages} {227 } (\bibinfo {year}
  {1980})}\BibitemShut {NoStop}%
\bibitem [{\citenamefont {Zhitomirsky}\ and\ \citenamefont
  {Chernyshev}(1999)}]{Zhitomirsky1999}%
  \BibitemOpen
  \bibfield  {author} {\bibinfo {author} {\bibfnamefont {M.~E.}\ \bibnamefont
  {Zhitomirsky}}\ and\ \bibinfo {author} {\bibfnamefont {A.~L.}\ \bibnamefont
  {Chernyshev}},\ }\bibfield  {title} {\bibinfo {title} {{Instability of
  antiferromagnetic magnons in strong fields}},\ }\href
  {https://doi.org/10.1103/PhysRevLett.82.4536} {\bibfield  {journal} {\bibinfo
   {journal} {Phys. Rev. Lett.}\ }\textbf {\bibinfo {volume} {82}},\ \bibinfo
  {pages} {4536} (\bibinfo {year} {1999})}\BibitemShut {NoStop}%
\bibitem [{\citenamefont {Sylju{\aa}sen}(2008)}]{Syljuasen2008}%
  \BibitemOpen
  \bibfield  {author} {\bibinfo {author} {\bibfnamefont {O.~F.}\ \bibnamefont
  {Sylju{\aa}sen}},\ }\bibfield  {title} {\bibinfo {title} {{Numerical evidence
  for unstable magnons at high fields in the Heisenberg antiferromagnet on the
  square lattice}},\ }\href {https://doi.org/10.1103/PhysRevB.78.180413}
  {\bibfield  {journal} {\bibinfo  {journal} {Phys. Rev. B}\ }\textbf {\bibinfo
  {volume} {78}},\ \bibinfo {pages} {180413} (\bibinfo {year}
  {2008})}\BibitemShut {NoStop}%
\bibitem [{\citenamefont {Eve}\ and\ \citenamefont {Fowles}(1966)}]{Eve1966}%
  \BibitemOpen
  \bibfield  {author} {\bibinfo {author} {\bibfnamefont {D.~J.}\ \bibnamefont
  {Eve}}\ and\ \bibinfo {author} {\bibfnamefont {G.~W.~A.}\ \bibnamefont
  {Fowles}},\ }\bibfield  {title} {\bibinfo {title} {{Oxalate complexes of
  tervalent titanium}},\ }\href {https://doi.org/10.1039/J19660001183}
  {\bibfield  {journal} {\bibinfo  {journal} {J. Chem. Soc}\ ,\ \bibinfo
  {pages} {1183}} (\bibinfo {year} {1966})}\BibitemShut {NoStop}%
\bibitem [{\citenamefont {Bulc}\ \emph {et~al.}(1983)\citenamefont {Bulc},
  \citenamefont {Golic},\ and\ \citenamefont {Siftar}}]{Bulc1983}%
  \BibitemOpen
  \bibfield  {author} {\bibinfo {author} {\bibfnamefont {N.}~\bibnamefont
  {Bulc}}, \bibinfo {author} {\bibfnamefont {L.}~\bibnamefont {Golic}},\ and\
  \bibinfo {author} {\bibfnamefont {J.}~\bibnamefont {Siftar}},\ }\bibfield
  {title} {\bibinfo {title} {{Structure of ammonium and sodium
  bis(oxalato)indate(III) dihydrate, NH$_4$[In(C$_2$O$_4$)$_2$]$\cdot$2H$_2$O
  and Na[In(C$_2$O$_4$)$_2$]$\cdot$2H$_2$O}},\ }\href
  {https://doi.org/10.1107/S0108270183004011} {\bibfield  {journal} {\bibinfo
  {journal} {Acta Crystallogr. C}\ }\textbf {\bibinfo {volume} {39}},\ \bibinfo
  {pages} {176} (\bibinfo {year} {1983})}\BibitemShut {NoStop}%
\end{thebibliography}%

\end{document}